\documentclass[twocolumn]{aastex631}

\newcommand{\JWST}{JWST}
\newcommand{\jwst}{JWST}
\newcommand{\Mstar}{\ensuremath{M_{\star}}}
\newcommand{\tstar}{\ensuremath{t_{\star}}}
\newcommand{\Msun}{\ensuremath{M_{\sun}}}
\newcommand{\SFR}{SFR}
\newcommand{\Myr}{\mathrm{Myr}}
\newcommand{\yr}{\mathrm{yr}}
\newcommand{\pc}{\mathrm{pc}}
\newcommand{\kpc}{\mathrm{kpc}}
\newcommand{\nJy}{\mathrm{nJy}}

\newcommand{\figscale}{0.9}
\newcommand{\figscaleB}{\figscale}

\usepackage{amsmath}

\newcommand{\given}{\,|\,}

\newcommand{\like}{\mathcal{L}}

\begin{document}

\title{Earliest Galaxies in the JADES Origins Field: \\
Luminosity Function and Cosmic Star-Formation Rate Density 300 Myr after the Big Bang}

\author[0000-0002-4271-0364]{Brant Robertson}
\affiliation{Department of Astronomy and Astrophysics, University of California, Santa Cruz, 1156 High Street, Santa Cruz, CA 96054, USA}

\author[0000-0002-9280-7594]{Benjamin D.\ Johnson}
\affiliation{Center for Astrophysics $|$ Harvard \& Smithsonian, 60 Garden St., Cambridge MA 02138 USA}

\author[0000-0002-8224-4505]{Sandro Tacchella}
\affiliation{Kavli Institute for Cosmology, University of Cambridge, Madingley Road, Cambridge, CB3 0HA, UK}
\affiliation{Cavendish Laboratory, University of Cambridge, 19 JJ Thomson Avenue, Cambridge, CB3 0HE, UK}

\author[0000-0002-2929-3121]{Daniel J.\ Eisenstein}
\affiliation{Center for Astrophysics $|$ Harvard \& Smithsonian, 60 Garden St., Cambridge MA 02138 USA}

\author[0000-0003-4565-8239]{Kevin Hainline}
\affiliation{Steward Observatory, University of Arizona, 933 N. Cherry Avenue, Tucson, AZ 85721, USA}

\author[0000-0001-7997-1640]{Santiago Arribas }
\affiliation{Centro de Astrobiolog\'ia (CAB), CSIC–INTA, Cra. de Ajalvir Km.~4, 28850- Torrej\'on de Ardoz, Madrid, Spain}

\author[0000-0003-0215-1104]{William M.\ Baker}
\affiliation{Kavli Institute for Cosmology, University of Cambridge, Madingley Road, Cambridge, CB3 0HA, UK}
\affiliation{Cavendish Laboratory, University of Cambridge, 19 JJ Thomson Avenue, Cambridge, CB3 0HE, UK}

\author[0000-0002-8651-9879]{Andrew J.\ Bunker}
\affiliation{Department of Physics, University of Oxford, Denys Wilkinson Building, Keble Road, Oxford OX1 3RH, UK}

\author[0000-0002-6719-380X]{Stefano Carniani}
\affiliation{Scuola Normale Superiore, Piazza dei Cavalieri 7, I-56126 Pisa, Italy}

\author[0000-0001-6301-3667]{Courtney Carreira}
\affiliation{Department of Astronomy and Astrophysics, University of California, Santa Cruz, 1156 High Street, Santa Cruz, CA 96054, USA}

\author[0000-0002-1617-8917]{Phillip A.\ Cargile}
\affiliation{Center for Astrophysics $|$ Harvard \& Smithsonian, 60 Garden St., Cambridge MA 02138 USA}

\author[0000-0003-3458-2275]{Stephane Charlot}
\affiliation{Sorbonne Universit\'e, CNRS, UMR 7095, Institut d'Astrophysique de Paris, 98 bis bd Arago, 75014 Paris, France}

\author[0000-0002-7636-0534]{Jacopo Chevallard}
\affiliation{Department of Physics, University of Oxford, Denys Wilkinson Building, Keble Road, Oxford OX1 3RH, UK}

\author[0000-0002-2678-2560]{Mirko Curti}
\affiliation{European Southern Observatory, Karl-Schwarzschild-Strasse 2, 85748 Garching, Germany}

\author[0000-0002-9551-0534]{Emma Curtis-Lake}
\affiliation{Centre for Astrophysics Research, Department of Physics, Astronomy and Mathematics, University of Hertfordshire, Hatfield AL10 9AB, UK}

\author[0000-0003-2388-8172]{Francesco D'Eugenio}
\affiliation{Kavli Institute for Cosmology, University of Cambridge, Madingley Road, Cambridge, CB3 0HA, UK}
\affiliation{Cavendish Laboratory, University of Cambridge, 19 JJ Thomson Avenue, Cambridge, CB3 0HE, UK}

\author[0000-0003-1344-9475]{Eiichi Egami}
\affiliation{Steward Observatory, University of Arizona, 933 N. Cherry Avenue, Tucson, AZ 85721, USA}

\author[0000-0002-8543-761X]{Ryan Hausen}
\affiliation{Department of Physics and Astronomy, The Johns Hopkins University, 3400 N. Charles St., Baltimore, MD 21218}

\author[0000-0003-4337-6211]{Jakob M.\ Helton}
\affiliation{Steward Observatory, University of Arizona, 933 N. Cherry Avenue, Tucson, AZ 85721, USA}

\author[0000-0002-6780-2441]{Peter Jakobsen}
\affiliation{Cosmic Dawn Center (DAWN), Copenhagen, Denmark}
\affiliation{Niels Bohr Institute, University of Copenhagen, Jagtvej 128, DK-2200, Copenhagen, Denmark}

\author[0000-0001-7673-2257]{Zhiyuan Ji}
\affiliation{Steward Observatory, University of Arizona, 933 N. Cherry Avenue, Tucson, AZ 85721, USA}

\author[0000-0002-0267-9024]{Gareth C.\ Jones}
\affiliation{Department of Physics, University of Oxford, Denys Wilkinson Building, Keble Road, Oxford OX1 3RH, UK}

\author[0000-0002-4985-3819]{Roberto Maiolino}
\affiliation{Kavli Institute for Cosmology, University of Cambridge, Madingley Road, Cambridge, CB3 0HA, UK}
\affiliation{Cavendish Laboratory, University of Cambridge, 19 JJ Thomson Avenue, Cambridge, CB3 0HE, UK}
\affiliation{Department of Physics and Astronomy, University College London, Gower Street, London WC1E 6BT, UK}

\author[0000-0003-0695-4414]{Michael V.\ Maseda}
\affiliation{Department of Astronomy, University of Wisconsin-Madison, 475 N. Charter St., Madison, WI 53706 USA}

\author[0000-0002-7524-374X]{Erica Nelson}
\affiliation{Department for Astrophysical and Planetary Science, University of Colorado, Boulder, CO 80309, USA}

\author[0000-0003-4528-5639]{Pablo G. P\'erez-Gonz\'alez}
\affiliation{Centro de Astrobiolog\'ia (CAB), CSIC–INTA, Cra. de Ajalvir Km.~4, 28850- Torrej\'on de Ardoz, Madrid, Spain}

\author[0000-0001-8630-2031]{D\'avid Pusk\'as}
\affiliation{Kavli Institute for Cosmology, University of Cambridge, Madingley Road, Cambridge, CB3 0HA, UK}
\affiliation{Cavendish Laboratory, University of Cambridge, 19 JJ Thomson Avenue, Cambridge, CB3 0HE, UK}

\author[0000-0002-7893-6170]{Marcia Rieke}
\affiliation{Steward Observatory, University of Arizona, 933 N. Cherry Avenue, Tucson, AZ 85721, USA}

\author[0000-0001-8034-7802]{Renske Smit}
\affiliation{Astrophysics Research Institute, Liverpool John Moores University, 146 Brownlow Hill, Liverpool L3 5RF, UK}

\author[0000-0002-4622-6617]{Fengwu Sun}
\affiliation{Steward Observatory, University of Arizona, 933 N. Cherry Avenue, Tucson, AZ 85721, USA}

\author[0000-0003-4891-0794]{Hannah \"Ubler}
\affiliation{Kavli Institute for Cosmology, University of Cambridge, Madingley Road, Cambridge, CB3 0HA, UK}
\affiliation{Cavendish Laboratory, University of Cambridge, 19 JJ Thomson Avenue, Cambridge, CB3 0HE, UK}

\author[0000-0003-1432-7744]{Lily Whitler}
\affiliation{Steward Observatory, University of Arizona, 933 N. Cherry Avenue, Tucson, AZ 85721, USA}

\author[0000-0003-2919-7495]{Christina C. Williams}
\affiliation{NSF’s National Optical-Infrared Astronomy Research Laboratory, 950 North Cherry Avenue, Tucson, AZ 85719, USA}

\author[0000-0001-9262-9997]{Christopher N.\ A.\ Willmer}
\affiliation{Steward Observatory, University of Arizona, 933 N. Cherry Avenue, Tucson, AZ 85721, USA}

\author[0000-0002-4201-7367]{Chris Willott}
\affiliation{NRC Herzberg, 5071 West Saanich Rd, Victoria, BC V9E 2E7, Canada}

\author[0000-0002-7595-121X]{Joris Witstok}
\affiliation{Kavli Institute for Cosmology, University of Cambridge, Madingley Road, Cambridge, CB3 0HA, UK}
\affiliation{Cavendish Laboratory, University of Cambridge, 19 JJ Thomson Avenue, Cambridge, CB3 0HE, UK}

\begin{abstract}
We characterize the earliest galaxy population in the JADES Origins Field (JOF), the deepest imaging field observed with JWST.
We make use of the ancillary Hubble optical images (5 filters spanning $0.4-0.9\mu\mathrm{m}$) and novel JWST images with
14 filters spanning $0.8-5\mu\mathrm{m}$, including 7 medium-band filters, and reaching total exposure times of up to
46 hours per filter.
We combine all our data at $>2.3\mu\mathrm{m}$ to construct
 an ultradeep image, reaching as deep as $\approx31.4$ AB mag in the stack and
30.3-31.0 AB mag ($5\sigma$, $r=0.1"$ circular aperture) in individual filters.
We measure photometric redshifts and use robust selection criteria to identify a sample of
eight galaxy candidates at redshifts $z=11.5-15$.
These objects show compact half-light radii of $R_{1/2}\sim50-200$pc,
stellar masses of $\Mstar\sim10^7-10^8\Msun$, and star-formation rates of $\mathrm{SFR}\sim0.1-1~\Msun~\yr^{-1}$.
Our search finds no candidates at $15<z<20$, placing upper limits at these redshifts.
We develop a forward modeling approach to infer the properties of the evolving luminosity function without
binning in redshift or luminosity that marginalizes over the photometric redshift uncertainty of our
candidate galaxies and incorporates the impact of non-detections. We find a $z=12$ luminosity function in good agreement with prior results, and that the luminosity function normalization and UV luminosity density decline by a factor of $\sim2.5$ from $z=12$ to $z=14$. We discuss the possible implications of our results in the context of
theoretical models for evolution of the dark matter halo mass function.
\end{abstract}


\section{Introduction} \label{sec:intro}

\jwst\ has pushed the
forefront of our knowledge of galaxies in the distant universe
to the first 350 million years of cosmic time.
Within the first weeks of operations, surveys
with \jwst\ unveiled galaxy candidates beyond
redshift $z\sim12$ in an epoch when only the
most optimistic models of the cosmic star
formation rate density predicted that galaxies
would be easily discoverable \citep{naidu22_highz, castellano22, finkelstein23, adams23, atek23, donnan23, harikane23_uvlf, morishita2023a, bouwens2023a}. The identification
and spectroscopic confirmation by the
\jwst\ Advanced Deep Extragalactic Survey 
(JADES; PIs Rieke and Lutzgendorf; \citealt{eisenstein23_jades})
of the 
galaxies JADES-GS-z12-0
at $z=12.6$ and JADES-GS-z13-0 at $z=13.2$
affirmatively established for the first
time the reality of galaxies at $z>12$ \citep{curtis-lake23, robertson23, deugenio23}.
Subsequently, other galaxy candidates have
been confirmed at $z\sim12-13$ in other
surveys \citep{fujimoto23_zspec, wang23}
and many additional high-redshift candidates
identified photometrically \citep[e.g.,][]{hainline23, perez-gonzalez2023a,leung23}.

The discovery of these distant sources raises
substantial questions about the nature of
galaxy formation in the early universe \citep{ferrara23, mason23, dekel23, li23, lovell23, shen23, yung24}. The
earliest known galaxies appear relatively bright \citep[e.g.,][]{naidu22_highz, castellano22, treu23, finkelstein23_ceers},
show a range of stellar masses $\Mstar\sim10^7-10^9\Msun$,
and have young stellar ages of $\tstar\sim10^7-10^8~\yr$
\citep{robertson23}.
Structurally, these galaxies show
physical sizes of $r\sim0.1-1~\kpc$ and star formation
rate surface densities of $\dot{\Sigma}_\star\sim50-100~\Msun~\yr^{-1}\kpc^{-2}$ \citep{robertson23, arrabal-haro23, wang23}.
They are compact star forming galaxies undergoing
rapid star formation on a timescale comparable to their
local dynamical times. Individually, the properties
of these objects are not extreme given the densities
and dynamics of the early universe. Collectively,
the apparent, albeit uncertain, abundance of such
objects in the context of structure formation 
may be unexpectedly high. Resolving this essential
quandary requires statistical constraints on the
abundance of $z>12$ galaxies and 
information on their possible origins through
higher-redshift searches.

To answer these questions, 
this work presents first results on the search 
for distant galaxies in the JADES Origin Field 
(JOF; Program ID 3215, PIs Eisenstein and Maiolino; \citealt{eisenstein23_jof}).
The JOF observations were designed to
use \JWST{} medium bands, including NIRCam F162M, to isolate 
the Lyman-$\alpha$ break at $z\gtrsim12$ and
simultaneously control for contamination
by lower-redshift line emitters that 
can mimic the broad-band spectral
energy distributions (SEDs) of distant
galaxies \citep{naidu22, zavala23, arrabal-haro23, perez-gonzalez2023b}. 
In concert with ultra-deep
broad-band observations from JADES,
the $9.05$ arcmin$^2$ JOF provides the best current dataset
for finding and characterizing $z\gtrsim12$
galaxies.
We search the JOF for
objects to an effective limiting 
depth of $f_\nu\sim2-3~\nJy$, performing
SED fitting analyses to select the highest
redshift candidates.
We then use a forward-modeling approach
to infer the character of the
evolving luminosity function given the
properties of our sample of 
high-redshift candidate galaxies.
Our method accounts
for the photometric redshift posterior
constraints of our sample's galaxies without
binning in redshift or luminosity. 
We employ our method to study
the behavior of the evolving luminosity
function beyond $z\sim12$ and the
abundance of galaxies at earlier times.

This paper is organized as follows.
In \S \ref{sec:data} we review the
JOF data, the observations, 
data reduction procedure, source
detection, and photometry.
In \S \ref{sec:selection},
we describe our selection
procedure based on SED
template fitting.
Forward modeling constraints
on the galaxy candidate
structural properties
and
inference of the distant
stellar
population properties
are described in \S \ref{sec:completeness}.
We characterize the galaxy
luminosity functions at $z\sim12-15$ and our constraints on the UV luminosity density at $z\sim12-20$
in \S \ref{sec:uvlf}, and report the inferred physical 
properties of the high-redshift candidates in
\S \ref{sec:props}.
We
interpret the observational
results in the context of
galaxy formation theory in 
\S \ref{sec:discussion}.
We summarize our conclusions and preview future
work in \S \ref{sec:conclusion}.
Throughout this work, we use the AB magnitude system  \citep{oke-gunn1983}  and assume 
a flat Lambda cold dark matter ($\Lambda$CDM) cosmology with 
$\Omega_m= 0.3$ and $H_0 = 70$~km~s$^{-1}$~Mpc$^{-1}$.

\section{Data} \label{sec:data}

This work uses \jwst{} observations in the
JOF to discover and constrain the
abundance and properties of $z>12$
galaxies. In \S \ref{sec:obs}
we review the JOF and accompanying
JADES and Hubble Space Telescope (HST)
observations. 
In \S \ref{sec:reduction}, we
present the data reduction methods
used to process the imagery.
The detection and photometric
methods used to discover the
objects are described
in \S \ref{sec:detphot}.

\subsection{Observations}
\label{sec:obs}

\citet{eisenstein23_jof} presents the JOF, a single \jwst\ NIRCam pointing of exceptional depth, with about 7 days of exposure time spread between 14 filters covering an $A\sim9$ arcmin$^{2}$ area.  The JOF began with the parallel imaging of deep JADES spectroscopy (Program ID 1210, presented in \citealt{bunker23}) that produced long F090W, F115W, F150W, F200W, F277W, F335M, F356W, F410M, and F444W exposures in a field adjacent to the Hubble Ultra Deep Field within the GOODS-S field.
This campaign continued in Cycle 2 Program ID 3215, which observed in 6 
\jwst\ NIRCam medium
bands---F162M, F182M, F210M, F250M, F300M, and F335M---again
acquired
in parallel to deep NIRSpec
observations.
We also include all JADES GOODS-S
medium-depth imaging (Program ID 1180)
that overlaps with the JOF.
This
area of GOODS-S partially overlaps
with the FRESCO (Program ID 1895)
F182M, F210M, and F444W data,
which we incorporate. The field also has
partial coverage of
HST ACS F435W, F606W, F775W, F814W
and F850LP images reduced and released
through the Hubble Legacy Field
program \citep{illingworth16} reductions of the
Great Observatories Origins Deep Survey
\citep[GOODS;][]{giavalisco2004a} and Cosmic Assembly
 Near-infrared Deep Extragalactic Legacy Survey \citep[CANDELS;][]{grogin2011a,koekemoer2011a}
 images.
In total, these data provide nineteen
\JWST{} and HST photometric bands that we
use to constrain the galaxy SEDs and particularly the 
Lyman-$\alpha$ break.

\subsection{Data reduction}
\label{sec:reduction}

Our image reduction methods were outlined
in \citet{rieke23} and \citet{eisenstein23_jades},
detailed in Tacchella et al., (in prep),
and we provide a summary here. We process 
the images with the \emph{jwst} Calibration 
Pipeline (version 1.11.4) and Calibration 
Reference Data System pipeline mapping 
(CRDS pmap) 1130, which includes in-flight 
NIRCam dark, distortion, bad pixel mask, 
read noise, superbias and flat reference files.

We use \emph{jwst} Stage 1 to perform the detector-level
corrections and ramp fitting. We run this 
stage with the default parameters, except 
for the correction of ``snowball'' artifacts
from cosmic rays. The identification 
and correction of snowballs represent a big 
challenge. Heuristically, we find that the following 
parameters provide reasonable snowball amelioration:
$\texttt{max\_jump\_to\_flag\_neighbors}=1$, 
$\texttt{min\_jump\_to\_flag\_neighbors}=100000$, 
$\texttt{min\_jump\_area}=5$,
$\texttt{min\_sat\_area}=1$,
$\texttt{expand\_factor}=2$,
$\texttt{min\_sat\_radius\_extend}=2.5$, and
$\texttt{max\_extended\_radius}=200$. 

As detailed in \citet{rieke23},
we run \emph{jwst} Stage 2 with the default parameters, 
but replace the STScI flats for all LW bands except F250M and F300M with custom
super-sky 
flats. When we do not have sufficient images to 
produce a robust flat field, we 
interpolated the flat-field images from the 
bands adjacent in wavelength. 
Following Stage 2, we perform custom corrections for 
all additive effects including
$1/f$ noise, scattered light effects (``wisps'' and 
``claws''), and the large-scale background.
Furthermore, we also updated 
the \texttt{DQ} data quality array 
to mask additional features 
imprinted visually onto the mosaics, including 
persistence, uncorrected wisp features, and 
unflagged hot pixels.

Before running \emph{jwst} Stage 3, we perform astrometric 
registration to Gaia DR2 (G. Brammer priv. comm., 
Gaia Collaboration et al. 2018) with a modified 
\emph{jwst}-pipeline \texttt{tweakreg} code.
We apply both a 
rotation and offset to the individual level-2 
images.
For images taken in the A module with the medium bands
F182M, F210M, and F335M,
we replace the default distortion maps with the nearest
(in effective wavelength) wide-band distortion map for that detector.

We construct the mosaics using \emph{jwst} Stage 3. We create 
single mosaics for each filter by combining 
exposures from all observations, and run \emph{jwst} Stage 3 
with the default parameter values while 
setting the 
pixel scale to 0.03 "/pixel and a drizzle 
parameter of $\texttt{pixfrac}=1$ for the SW 
and LW images. Finally, we perform a custom 
background subtraction, following the procedure 
outlined in \citep{bagley23}. For F090W, 
F115W, and F150W, hot pixels that
pass median rejection are replaced with
median filtered values from the local
flux image.

\subsection{Detection and Photometry}
\label{sec:detphot}

The detection and photometry methods are
discussed in \citet{rieke23}
and \citet{eisenstein23_jades} and will be 
detailed in Robertson et al. (in prep).

To perform source detection, an inverse variance-weighted
stack of the long-wavelength NIRCam 
F250M, F277W, F300M, F335M, F356W, F410M, and
F444W SCI and ERR channels are constructed.
Small-scale noise residuals from incomplete masking
in the \emph{jwst} pipeline are median filtered from
the ERR images. The signal-to-noise ratio (SNR) image
created from the ratio of these images is used
as the detection image. An initial source significance
threshold of
$SNR>1.5$ is used to select regions of interest, and
a series of custom computational morphology algorithms
inspired by NoiseChisel \citep{akhlaghi15,akhlaghi19}
are applied to refine the segmentations.
Stars and diffraction spikes are masked by constructing
segmentations from stacks of all available filters and
integrated into the detection segmentation map.
The detection image segmentations
are deblended using a logarithmic scaling of the F200W
image. High-pass filtering is applied to the outer regions
of large segmentations to isolate proximate satellite galaxies.
After these refinements of the segmentation map,
a final pass to detect potentially missed compact, faint sources is applied.
The completeness as a function of flux and size for this detection
algorithm has been calculated using source injection simulations and is
presented in Section \ref{sec:det_comp}.

After engineering the segmentation map, we perform a 
set of customized photometric measurements based on the
\emph{photutils} \citep{bradley23} analysis package.
Object centroids are computed using the 
``windowed positions'' used by 
Source Extractor \citep{bertin96}. 
Apertures for measuring \citet{kron80}
fluxes are determined based
on the stacked signal image (the numerator of the SNR detection
image) using a Kron parameter of $2.5$.
We limit the area of the Kron aperture
to be less than twice an object's segmentation area.
In addition to Kron fluxes, we measure circular aperture 
photometry with aperture radii of $r=\{0.1",0.15",0.25",0.3",0.35",0.5"\}$.
To provide aperture corrections, we 
produce a model point spread
function (mPSF) following the method 
of \citet{ji23}, where we inject WebbPSF models 
into \emph{jwst} level-2 images and
mosaic them using the same exposure pattern as
 the JOF observations to provide a composite star field.
An mPSF for each
band (and observing program) is then constructed
from these PSF-mosaics. 
The circular aperture corrections are measured and tabulated,
and the Kron aperture corrections computed by integrating
within the corresponding elliptical apertures placed on the mPSF.
For HST, we measure empirical (e)PSFs using the \emph{photutils} \citep{bradley23}
ePSF Builder with visually inspected stars in the field.

We perform a bevy of photometric validation tests. Cross
validation against the CANDELS survey HST photometry
using completely independent HST reductions from the Hubble Legacy Field
program are presented in \citet{rieke23} for the broader
JADES/GOODS-S field. We also
compute median photometric offsets from SED templates 
using EAZY \citep{brammer08}, following the method described by \citet{hainline23}. We find these 
zeropoint offsets to be within $5.2\%$, and typically within $1\%$, for \JWST{} filters. 

\subsubsection{Surface Brightness Profile Modeling}
\label{sec:forcepho}
We also forward model each galaxy's surface brightness profile using
the \texttt{Forcepho} code (B. Johnson, in prep). We
use \texttt{Forcepho} with custom model point-spread
functions to model the surface brightness profile
of each galaxy in our survey simultaneously with any nearby objects in each individual
exposure where pixel covariance is minimized.
We restrict the modeling to the F200W and F277W bands, to minimize the chance of any PSF mismatch or astrometric errors while maximizing S/N and resolution.
The surface brightness profile is assumed to be
a \citet{sersic68} model, with a fast
Gaussian-based factorization
of the model.
\texttt{Forcepho} provides a Bayesian estimate
of the surface brightness profile parameters, 
including the galaxy half-light radius.
We have used \texttt{Forcepho} to study the
structure of other extremely high-redshift
galaxies \citep[e.g.,][]{robertson23,tacchella23}, and we refer
the reader to \citet{baker23} for more details on our 
morphological analysis methods.

\subsection{Image Depths}
\label{sec:depths}

With the construction of our broad- and medium-band NIRCam mosaics and
the long-wavelength ($\lambda>2.3\mu$m) detection image, we can use the
photometry method described in \S \ref{sec:detphot} to measure our
image depths. In Table \ref{tab:depth}, we
report the median aperture corrected $5-\sigma$ point-source
depth in each filter and the stack
(using the F277W PSF to estimate the stack's aperture correction). When measuring
the depth in each image, we use a dilated version of the segmentation map created
by the detection algorithm to mask source pixels.
We note that the single-band images depths listed in Table \ref{tab:depth}
are all within 10$-$25\% of the $5-\sigma$
point-source depths we reported in \citet{eisenstein23_jof} that were computed from the
JWST exposure time calculator, with the longest wavelength filters showing the most
improved depth.
Our single-band images reach 30.3$-$31.0 AB, and the combined $\lambda>2.3\mu$m stack
reaches 31.4 AB depth.
For comparison, we also list the $5-\sigma$ point-source depth the corresponding
$\lambda>2.3\mu$m stacks from available NIRCam long-wavelength images in NGDEEP
\citep[F277W+F356W+F444W;][]{bagley2023b},
the MIRI-UDF NIRCam parallel \citep[F277W+F356W;][]{perez-gonzalez2023a}, 
and the JADES GOODS-S Deep region that covers the Hubble
Ultra Deep Field \citep[][]{rieke23}. To measure their depths, we
processed these fields using identical
methods and used the same F277W PSF model to aperture correct them.
We report depths for each program separately, and note
that where the MIRI-UDF parallel and NGDEEP NIRCam imaging overlap the
combined depths will be even more sensitive than listed in Table \ref{tab:depth}.

\begin{deluxetable}{ccc}
\tabletypesize{\footnotesize}
\tablecaption{Depths of the JADES Origins Field\label{tab:depth}}
\tablehead{
	\colhead{Band} & \colhead{Median Depth$^a$} & \colhead{Median Depth}\\
	\colhead{} & \colhead{[nJy]} & \colhead{[AB]}
}
\startdata
\multicolumn{3}{c}{JWST/NIRCam Filters} \\
F090W & 2.80 & 30.28\\
F115W & 2.33 & 30.48\\
F150W & 2.19 & 30.55\\
F162M & 2.76 & 30.30\\
F182M & 1.78 & 30.77\\
F200W & 2.27 & 30.51\\
F210M & 2.29 & 30.50\\
F250M & 2.58 & 30.37\\
F277W & 1.42 & 31.02\\
F300M & 1.80 & 30.76\\
F335M & 1.70 & 30.82\\
F356W & 1.58 & 30.90\\
F410M & 2.65 & 30.34\\
F444W & 2.26 & 30.52\\
\multicolumn{3}{c}{Stacked Depth at $\lambda>2.3\mu$m} \\
JOF & 0.96 & 31.44\\
NGDEEP$^b$ & 0.82 & 31.61 \\
MIRI-UDF$^c$ & 1.28 & 31.13 \\
JADES GOODS-S Deep &1.39& 31.04 \\
\enddata
\tablecomments{$^a$ Median $r=0.1''$ aperture corrected $5\sigma$ point-source depth.
$^b$ This depth reflects our independent processing of the NGDEEP data, and we
refer the reader to \citet{bagley2023b} for their depth measurements.
$^c$ This depth reflects our independent processing of the MIRI-UDF data, and we
refer the reader to \citet{perez-gonzalez2023a} for their depth measurements.}
\end{deluxetable}

\section{Selection of Redshift \texorpdfstring{$z\gtrsim12$}{z>12} Galaxies}
\label{sec:selection}

The photometric selection of high-redshift galaxies
relies on identifying a strong Lyman-$\alpha$ break in the
rest-frame UV of a galaxy's SED
\citep[e.g.,][]{guhathakurta90,steidel95}.
Below, we detail our selection of $z\gtrsim12$
galaxies based on this feature.

\subsection{Photometric Redshift Estimation}
\label{sec:photoz}

To infer the photometric redshifts of galaxies in the JOF, we apply the techniques detailed in \citet{hainline23} to fit templates of galaxy SEDs to our \JWST{} and HST photometry, varying the redshift to assess the relative goodness of fit. To perform the SED fits, we use the \texttt{EAZY} code \citep{brammer08} to compute rapidly the photometric redshift posterior distributions for each galaxy in the JOF survey. When fitting SED templates, we use the template suite described in \citet{hainline23} that includes models with strong line emission and a range of UV continua. The photometric redshifts estimated from fits to these templates were shown to have an outlier fraction (defined as the fraction of sources with $|z_{phot}-z_{spec}| / (1 + z_{spec}) > 0.15$) of $f_{\mathrm{out}}=0.05$ in \citet{rieke23}, and $f_{\mathrm{out}}=0$ for 42 sources at $z > 8$ in \citet{hainline23}. A range of potential redshifts $z=0.01-22$ in $\Delta z=0.01$ increments were considered, and for selection, we adopt the use of the redshift corresponding to the minimum $\chi^2$ from the fit, $z_a$. For each nominal redshift, we use the \citet{inoue14} model for attenuation from the intergalactic medium \citep[see also][]{madau95}. We do not adopt any magnitude priors, we impose an error floor of 5\% on the photometry, and allow for negative fluxes. When fitting the SED models to determine a photometric redshift, to maximize signal-to-noise ratios we use aperture-corrected $r=0.1"$ circular aperture fluxes on the native resolution JOF images without convolution to a common PSF, multiplied by the photometric offsets discussed in Section \ref{sec:detphot}. We have checked that we obtain comparably high-redshift solutions when using common-PSF Kron aperture photometry with lower SNR, except where noted below.
We note that for some objects, the
best-fit SED model has Lyman-$\alpha$ line emission. This feature arises as an artifact
of the optimization process in EAZY that mixes templates
with and without Lyman-$\alpha$ emission.  
We do not claim this line emission to be real. 
The equivalent width of Lyman-$\alpha$ is degenerate
with the redshift of the break, which can contribute to a
photometric redshift offset of $\Delta z\approx0.2-0.4$
relative to a spectroscopic redshift. Local attenuation
from the galaxy interstellar medium or circumgalactic medium
can shift the photometric redshift by a similar amount
\citep[e.g.,][]{deugenio23,heintz2023a}

\subsection{Selection Criteria}
\label{sec:selection_criteria}

In the JOF, we apply the following 
criteria to identify our high-redshift
sample. These criteria have been
adapted from \citet{hainline23} but
further tailored to a $12\lesssim z\lesssim20$
selection. We note that these
criteria both select objects
previously discovered, notably by \citet{hainline23}, and identify
new objects. We provide the provenance of each object when discussing our samples below. 
Our selection criteria are:

\begin{enumerate}
    \item The redshift at the \texttt{EAZY} fit $\chi^2$ minimum must be $z_a\ge11.5$.
    \item Two of F277W, F356W, and F444W \JWST{} NIRcam filters must show $>5\sigma$ detections.
    \item All the long-wavelength NIRCam fluxes (F250M, F277W, F300M, F335M, F356W, F410M, F444W) must exceed $1.5\sigma$ significance.
    \item The redshift posterior distribution must have an integral probability of $P(z_a>11)>0.68$, where we take $P(z)\propto \exp(-\chi^2/2)$.
    \item The goodness of fit difference between the best high redshift ($z>11$) and low redshift ($z<7$) solutions must satisfy $\Delta \chi^2>4$, and for the best fit we require $\chi^2<100$ summed over all 19 filters.
    \item The flux in F090W and F115W each must be below $2.5\sigma$ significance, as we expect no robust detection of flux blueward of the Lyman-$\alpha$ break.
    \item To avoid objects redder than the typically blue high-redshift objects \citep[e.g.,][]{topping23}, we require that sources cannot have both (F277W-F356W)$>0.125$ and (F356W-F444W)$>0.25$.
    \item Each object must have F150W, F162M, F182M, F210M, and F277W coverage. This criterion limits our survey area to the F162M JOF footprint.
    \item The NIRCam short and long wavelength local exposure time must be within a factor of four, which avoids edge effects from the mosaic pattern.
    \item To avoid variable sources, the flux in the NIRCam Medium bands acquired in the second year of \JWST{} operations must not exceed the broad band NIRCam fluxes acquired in the first year by more than $1\sigma$ in all bands simultaneously. In practice, we treat overlapping medium and broad band filters as random samples of the same flux density, and then flag when the difference between such pairs of flux estimates exceeds the quadrature sum of each pair's errors when taken in different epochs.
    \item We require that the source not be covered by another galaxy as determined from the segmentation map, which lowers the available area by 22\%. The final available area after accounting for foreground sources is approximately $A'=7.06$ square arcmin.
\end{enumerate}
\noindent
We note that without the data quality (criteria 9-10), minimum long-wavelength SNR threshold (criteria 2-3), or color criteria (criteria 6-7), fifteen objects would be selected. 
However, of these sources,
one (JADES+53.05101-27.89787) sits in an oversubtracted area of a distant star diffraction spike and three more are covered
by a stray light ``wisp'' feature in F162M (JADES+53.08317-27.86572, JADES+53.07681-27.86286, and JADES+53.04964-27.88605).
For a discussion of wisp features in JWST, please see \citet[e.g.,][]{rigby2023a}. 

\begin{deluxetable*}{cccccccc}
\centerwidetable
\tabletypesize{\footnotesize}
\tablecaption{High-redshift candidates in the JADES Origins Field\label{tab:properties}}
\tablehead{
	\colhead{Name} & \colhead{NIRCam ID} & \colhead{RA} & \colhead{Dec}& \colhead{$z_{\rm phot}^{b}$}  & \colhead{$M_{UV}$} &  \colhead{$R_{1/2}$ [mas]$^{a}$} & \colhead{$P(z<7)^{c}$}  
}
\startdata
\multicolumn{8}{c}{Main Sample $z>11.5$} \\
JADES+53.09731-27.84714 & 74977 & 53.09731 & --27.84714 & $11.53_{-0.78}^{+0.27}$ & $-17.66\pm0.14$ &  $12_{-7}^{+8}$ & $3.72\times10^{-5}$\\
JADES+53.02618-27.88716 & 16699 & 53.02618 & --27.88716 & $11.56_{-0.46}^{+0.41}$ & $-17.94\pm0.15$ &  $35_{-7}^{+7}$ & $9.12\times10^{-4}$\\
JADES+53.04017-27.87603 & 33309 & 53.04017 & --27.87603 & $12.10_{-0.16}^{+0.37}$ & $-17.73\pm0.10$ &  $12_{-3}^{+3}$ & $4.02\times10^{-5}$\\
JADES+53.03547-27.90037 & 160071 & 53.03547 & --27.90037 & $12.38_{-0.40}^{+0.17}$ & $-18.16\pm0.11$ &  $33_{-4}^{+4}$ & $7.87\times10^{-4}$\\
JADES+53.06475-27.89024 & 13731 & 53.06475 & --27.89024 & $12.93_{-0.16}^{+0.08}$ & $-18.78\pm0.04$ &  $4_{-2}^{+4}$ & $5.12\times10^{-24}$\\
JADES+53.02868-27.89301 & 11457 & 53.02868 & --27.89301 & $13.52_{-0.82}^{+0.26}$ & $-18.55\pm0.11$ &  $19_{-4}^{+4}$ & $7.75\times10^{-5}$\\
JADES+53.07557-27.87268 & 376946 & 53.07557 & --27.87268 & $14.38_{-0.37}^{+1.05}$ & $-18.28\pm0.22$ &  {$6_{-3}^{+6}$} & $7.63\times10^{-2}$\\
JADES+53.08294-27.85563$^{d}$ & 183348 & 53.08294 & --27.85563 & $14.39_{-0.09}^{+0.23}$ & $-21.00\pm0.05$ &{$76_{-2}^{+2}$} & 0\\ 
JADES+53.10762-27.86013 & 55733 & 53.10762 & --27.86013 & $14.63_{-0.75}^{+0.06}$ & $-18.54\pm0.13$ &  {$45_{-5}^{+6}$} & $2.26\times10^{-2}$\\
\multicolumn{8}{c}{Contributing Sample $z<11.5$} \\
JADES+53.03139-27.87219 & 172510 & 53.03139 & --27.87219 & $10.76_{-0.36}^{+0.66}$ & $-17.85\pm0.10$ & $32_{-7}^{+8}$ & $6.49\times10^{-5}$\\
JADES+53.09292-27.84607 & 76035 & 53.09292 & --27.84607 & $11.05_{-0.42}^{+0.49}$ & $-17.83\pm0.15$ & {$6_{-4}^{+5}$} & $4.06\times10^{-4}$\\
JADES+53.06857-27.85093 & 70836 & 53.06857 & --27.85093 & $11.17_{-0.31}^{+0.26}$ & $-18.02\pm0.10$ & {$5_{-3}^{+4}$} & $2.38\times10^{-3}$\\ 
\multicolumn{8}{c}{Auxiliary Sample $z>11.5$} \\
JADES+53.08468-27.86666$^{e}$ & 44962 & 53.08468 & --27.86666 & $12.9_{-0.25}^{+1.20}$ & $-18.16\pm0.10$ & {$56_{-7}^{+9}$} & $1.09\times10^{-2}$\\
JADES+53.07385-27.86072$^{f}$ & 54586 & 53.07385 & --27.86072 & $13.06_{-0.49}^{+0.97}$ & $-17.08\pm0.12$ & {$40_{-11}^{+16}$} & $6.16\times10^{-2}$\\ 
\enddata
\tablecomments{$^{a}$ The half-light size refers to the intrinsic, PSF-deconvolved size of each source, in milliarcseconds. $^{b}$ Best fit photometric redshift with 16- and 84-percentile uncertainties from the inferred photometric redshift distribution. $^{c}$ The posterior probability density for the photometric redshift of the candidate to lie at redshift $z<7$, given the SED fitting method described in \S \ref{sec:photoz}. $^{d}$ Spectroscopically confirmed at $z=14.32$ by Carniani et al. (submitted). $^{e}$ Fails red color limit. $^{f}$ Fails minimum SNR criterion.}
\end{deluxetable*}

\begin{figure*}[p]
\noindent
\includegraphics[width=\linewidth]{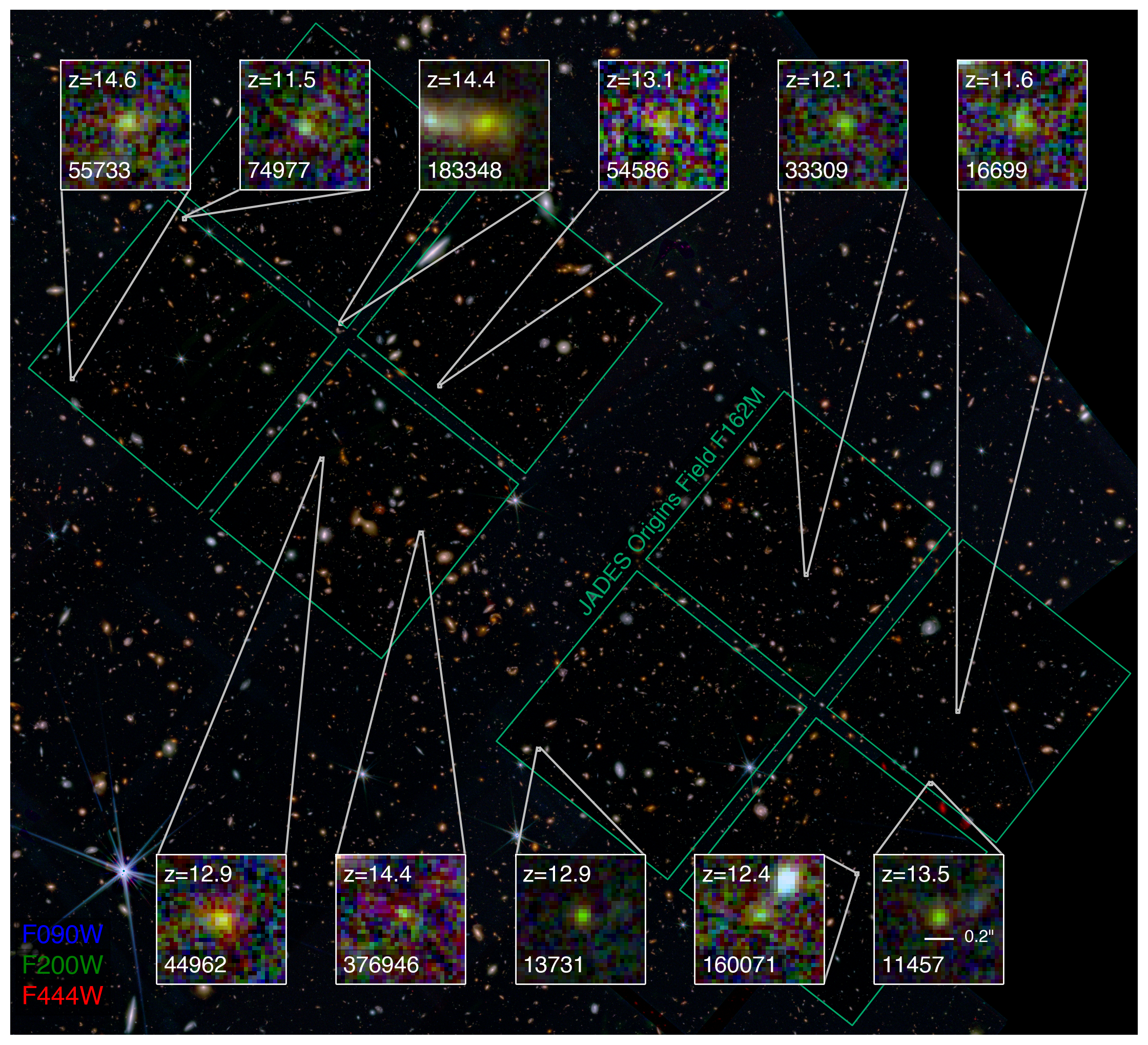}
\caption{F444W/F200W/F090W false color red/green/blue image of the JADES Origin Field (background image; 27.5 arcmin$^{2}$), the JOF F162M footprint (jade outline) and
F356W+F410M+F444W/F200W+F210M/F090W+F115W false color red/green/blue thumbnail images (each 0.86 arcsec$^2$) for $z\gtrsim12$ high-redshift
galaxy candidates. The RGB images of the galaxy candidates typically appear to have a green hue in this color space,
as they are all detected in the filters used for both the green and red channels, but not the blue channel.
Each inset thumbnail lists the best-fit \texttt{EAZY} photometric redshift and the JADES NIRCam ID, and we indicate the shared angular scale of the thumbnails with a scale bar showing 0.2". Table \ref{tab:properties}
lists the designations of the objects based on [RA, Dec]. NIRCam ID 183348 was spectroscopically confirmed as JADES-GS-z14-0 by Carniani et al. (submitted) at $z=14.32$.\label{fig:composite}}
\end{figure*}

Of the remaining eleven, one fails the minimum SNR threshold (for JADES+53.07385-27.86072, all filters redward of F335M have $f_{\nu}<2$nJy) and  one fails the color criteria (JADES+53.08468-27.86666 is red).
In total nine objects pass the selection; these comprise our Main Sample (see Table \ref{tab:properties}). 
While
we will comment on the additional two interesting objects that nearly satisfy our selection, we do not consider them in our fiducial luminosity function analyses. We call this collection of two objects the ``Auxiliary Sample'' at $z>11.5$.
There are also five sources in \citet{hainline23} sample in the vicinity of the JOF with previously reported photometric redshifts $z>11$ that are not in our sample. 
Of these, with the additional JOF data we find four
objects to have revised photometric redshifts $z < 7$ or fail other selection criteria
(JADES+53.02700-27.89808, JADES+53.03696-27.89422, JADES+53.07901-27.87154, JADES+53.10469-27.86187). The fifth falls in the F162M gap (JADES+53.07076-27.86544; NIRCam ID 176151) and therefore does not reside in our effective area.

We note that the F250M SNR criterion fixes the upper redshift limit of our selection. If we remove this criterion and the $z>11.5$ limit, we find one additional $z\sim11.4$ candidate (JADES+53.10131-27.85696) detected in all filters F150W and redder with $f_\nu\approx2$nJy, excepting F250M which is about $1.5\sigma$ low. In other words, we would find no $z\gtrsim20$ candidates by removing the weak F250M detection criterion.

The luminosity function analysis discussed below in \S \ref{sec:uvlf} enables the accounting of potential
contributions to the inferred galaxy abundance from galaxies with maximum-likelihood photometric redshifts below the putative
redshift of interest. We identified galaxies with maximum
likelihood redshifts $z>8$ and $P(z>12)>0.01$ that otherwise satisfy the above selection
criteria. There are three such galaxies, which fall in the photometric
redshift $z\approx10.5-11.2$ range, which will be referred to as the ``Contributing Sample'' at $z<11.5$.  Two of these (NIRCam IDs 76035 and 172510) were previously found in \citet{hainline23}. A third galaxy, NIRCam ID 64312 with photometric redshift $z\approx10.6$ 
and $P(z>12)\approx0.05$ from photometry, was subsequently
confirmed to lie at slightly lower redshift with $P(z<12)<0.01$ and was not considered further.

\begin{deluxetable*}{cccccc}
\centerwidetable
\tabletypesize{\footnotesize}
\tablecaption{Aperture-corrected HST/ACS photometry$^{a}$ in $r=0.1''$ circular apertures.\label{tab:acs}}
\tablehead{
	\colhead{Name} &  \colhead{F435W [nJy]} & \colhead{F606W [nJy]} & \colhead{F775W [nJy]} & \colhead{F814W [nJy]} & \colhead{F850LP [nJy]}
}
\startdata
\multicolumn{6}{c}{Main Sample $z>11.5$} \\
JADES+53.09731-27.84714 &  $-1.06\pm2.49$ & $-0.58\pm1.35$ & $2.25\pm3.52$ & $-3.58\pm1.40$ & $-11.28\pm4.63$\\
JADES+53.02618-27.88716 &  $-2.47\pm3.10$ & $-0.79\pm3.71$ & $-5.52\pm5.67$ & $2.57\pm2.33$ & $7.01\pm7.62$\\
JADES+53.04017-27.87603 &  $0.15\pm1.70$ & $-1.38\pm3.37$ & $8.59\pm5.73$ & $2.04\pm1.98$ & $-2.17\pm7.90$\\
JADES+53.03547-27.90037 & $-3.50\pm1.73$ & $-2.86\pm2.40$ & $-2.08\pm3.28$ & $1.55\pm2.24$ & $8.51\pm8.26$\\
JADES+53.06475-27.89024 &  $1.60\pm2.97$ & $0.47\pm1.79$ & $-3.36\pm2.67$ & $4.45\pm2.30$ & $0.20\pm5.04$\\
JADES+53.02868-27.89301 &  $-4.05\pm3.02$ & $-1.82\pm3.33$ & $4.03\pm3.64$ & $2.84\pm2.22$ & $3.77\pm8.78$\\
JADES+53.07557-27.87268 &  $-0.99\pm1.88$ & $1.83\pm1.73$ & $0.99\pm3.72$ & $0.73\pm1.80$ & $3.17\pm4.94$\\
JADES+53.08294-27.85563 &  $-2.80\pm3.47$ & $0.54\pm1.36$ & $3.87\pm3.94$ & $3.67\pm1.47$ & $1.66\pm4.45$\\
JADES+53.10762-27.86013 &  $-3.64\pm2.62$ & $-0.29\pm1.56$ & $3.47\pm3.38$ & $-1.35\pm1.89$ & $-0.35\pm4.72$\\
\multicolumn{6}{c}{Contributing Sample $z<11.5$} \\
JADES+53.03139-27.87219 &  $-1.30\pm1.69$ & $3.82\pm4.85$ & $-1.02\pm5.67$ & $-1.37\pm1.46$ & $5.42\pm8.00$\\
JADES+53.09292-27.84607 &  $-0.95\pm2.41$ & $0.69\pm1.27$ & $0.76\pm3.56$ & $-2.21\pm1.41$ & $-2.95\pm4.35$\\
JADES+53.06857-27.85093 &  $4.54\pm2.64$ & $2.21\pm1.33$ & $-7.54\pm2.83$ & $3.70\pm1.40$ & $-1.87\pm3.75$\\
\multicolumn{6}{c}{Auxiliary Sample $z>11.5$} \\
JADES+53.08468-27.86666 &  $-0.01\pm2.19$ & $-1.82\pm1.37$ & $3.51\pm3.72$ & $-0.03\pm1.58$ & $-6.87\pm4.91$\\
JADES+53.07385-27.86072 &  $7.83\pm2.55$ & $-0.31\pm1.38$ & $0.98\pm3.71$ & $-0.17\pm1.34$ & $0.43\pm4.35$\\
\enddata
\tablecomments{$^{a}$ These photometric measurements were made on the native-resolution Hubble Legacy Field images \citep{illingworth2016a}.}
\end{deluxetable*}

\begin{deluxetable*}{cccccccc}
\centerwidetable
\tabletypesize{\footnotesize}
\tablecaption{Aperture-corrected short-wavelength JWST/NIRCam photometry in $r=0.1''$ circular apertures.\label{tab:nircam_sw}}
\tablehead{
	\colhead{Name} &  \colhead{F090W [nJy]} & \colhead{F115W [nJy]} & \colhead{F150W [nJy]} & \colhead{F162M [nJy]} & \colhead{F182M [nJy]} & \colhead{F200W [nJy]} & \colhead{F210M [nJy]}
}
\startdata
\multicolumn{8}{c}{Main Sample $z>11.5$} \\
JADES+53.09731-27.84714 &  $0.67\pm0.53$ & $-0.01\pm0.46$ & $2.13\pm0.46$ & $4.14\pm0.56$ & $2.93\pm0.36$ & $3.49\pm0.52$ & $2.62\pm0.47$\\
JADES+53.02618-27.88716 &  $0.62\pm0.61$ & $-0.49\pm0.50$ & $1.80\pm0.47$ & $4.10\pm0.63$ & $3.85\pm0.45$ & $3.83\pm0.49$ & $2.83\pm0.56$\\
JADES+53.04017-27.87603 & $-0.20\pm0.59$ & $-0.32\pm0.49$ & $0.56\pm0.45$ & $2.81\pm0.58$ & $3.14\pm0.42$ & $3.71\pm0.45$ & $3.43\pm0.49$\\
JADES+53.03547-27.90037 &  $0.65\pm0.52$ & $-0.69\pm0.42$ & $0.65\pm0.41$ & $2.06\pm0.53$ & $3.06\pm0.38$ & $2.49\pm0.41$ & $2.82\pm0.48$\\
JADES+53.06475-27.89024 & $-0.08\pm0.50$ & $0.29\pm0.40$ & $0.10\pm0.37$ & $1.16\pm0.46$ & $7.80\pm0.36$ & $6.24\pm0.42$ & $7.48\pm0.41$\\
JADES+53.02868-27.89301 &  $-0.82\pm0.66$ & $0.77\pm0.54$ & $0.54\pm0.49$ & $1.46\pm0.64$ & $4.97\pm0.46$ & $5.90\pm0.51$ & $7.46\pm0.58$\\
JADES+53.07557-27.87268 & $0.21\pm0.52$ & $0.21\pm0.43$ & $-0.82\pm0.42$ & $0.17\pm0.52$ & $0.75\pm0.38$ & $2.18\pm0.41$ & $0.60\pm0.46$\\
JADES+53.08294-27.85563 &  $-1.12\pm0.68$ & $0.73\pm0.66$ & $1.10\pm0.55$ & $-$ & $9.73\pm0.90$ & $20.78\pm0.58$ & $29.66\pm1.14$\\
JADES+53.10762-27.86013 &  $0.36\pm0.56$ & $0.60\pm0.47$ & $0.16\pm0.44$ & $0.74\pm0.61$ & $1.87\pm0.38$ & $3.37\pm0.45$ & $3.72\pm0.53$\\
\multicolumn{8}{c}{Contributing Sample $z<11.5$} \\
JADES+53.03139-27.87219 & $-0.41\pm0.67$ & $0.03\pm0.56$ & $3.07\pm0.52$ & $4.70\pm0.65$ & $4.02\pm0.47$ & $3.70\pm0.56$ & $2.95\pm0.57$\\
JADES+53.09292-27.84607 &  $0.16\pm0.51$ & $0.26\pm0.44$ & $2.36\pm0.50$ & $4.03\pm0.63$ & $3.43\pm0.37$ & $3.43\pm0.54$ & $3.29\pm0.57$\\
JADES+53.06857-27.85093 &  $0.49\pm0.48$ & $0.16\pm0.39$ & $2.36\pm0.40$ & $3.95\pm0.49$ & $3.70\pm0.31$ & $3.56\pm0.39$ & $3.86\pm0.43$\\
\multicolumn{8}{c}{Auxiliary Sample $z>11.5$} \\
JADES+53.08468-27.86666 &  $0.05\pm0.48$ & $-0.47\pm0.43$ & $0.38\pm0.42$ & $0.60\pm0.46$ & $3.02\pm0.39$ & $2.78\pm0.42$ & $3.82\pm0.44$\\
JADES+53.07385-27.86072 &  $-0.52\pm0.46$ & $0.26\pm0.40$ & $0.59\pm0.38$ & $-0.01\pm0.48$ & $2.03\pm0.37$ & $1.55\pm0.39$ & $1.81\pm0.44$\\
\enddata
\end{deluxetable*}

\begin{deluxetable*}{cccccccc}
\centerwidetable
\tabletypesize{\footnotesize}
\tablecaption{Aperture-corrected long-wavelength JWST/NIRCam photometry in $r=0.1''$ circular apertures.\label{tab:nircam_lw}}
\tablehead{
	\colhead{Name} & \colhead{F250M [nJy]} & \colhead{F277W [nJy]} & \colhead{F300M [nJy]} & \colhead{F335M [nJy]} & \colhead{F356W [nJy]} & \colhead{F410M [nJy]} & \colhead{F444W [nJy]}
}
\startdata
\multicolumn{8}{c}{Main Sample $z>11.5$} \\
JADES+53.09731-27.84714 & $1.68\pm0.73$ & $2.13\pm0.33$ & $2.49\pm0.52$ & $2.25\pm0.45$ & $2.04\pm0.35$ & $2.96\pm0.60$ & $3.07\pm0.49$\\
JADES+53.02618-27.88716 &  $1.74\pm0.66$ & $2.15\pm0.35$ & $2.32\pm0.46$ & $2.57\pm0.36$ & $2.17\pm0.38$ & $1.76\pm0.63$ & $2.50\pm0.51$\\
JADES+53.04017-27.87603 & $2.81\pm0.75$ & $2.63\pm0.34$ & $2.70\pm0.47$ & $2.80\pm0.37$ & $2.52\pm0.37$ & $1.71\pm0.63$ & $1.20\pm0.51$\\
JADES+53.03547-27.90037 &  $3.16\pm0.65$ & $3.56\pm0.34$ & $3.58\pm0.44$ & $2.73\pm0.38$ & $2.32\pm0.39$ & $2.79\pm0.66$ & $2.47\pm0.52$\\
JADES+53.06475-27.89024 &  $7.22\pm0.65$ & $5.34\pm0.30$ & $5.20\pm0.46$ & $6.21\pm0.37$ & $5.47\pm0.34$ & $5.01\pm0.58$ & $4.30\pm0.46$\\
JADES+53.02868-27.89301 & $5.63\pm0.69$ & $4.71\pm0.36$ & $4.55\pm0.46$ & $3.75\pm0.37$ & $5.50\pm0.38$ & $5.79\pm0.65$ & $5.24\pm0.51$\\
JADES+53.07557-27.87268 &  $1.80\pm0.60$ & $2.22\pm0.33$ & $2.77\pm0.45$ & $2.27\pm0.39$ & $2.41\pm0.35$ & $2.18\pm0.57$ & $2.20\pm0.46$\\
JADES+53.08294-27.85563 &  $32.02\pm0.75$ & $32.83\pm0.44$ & $31.30\pm0.55$ & $27.43\pm0.48$ & $28.21\pm0.44$ & $28.56\pm0.71$ & $29.58\pm0.55$\\
JADES+53.10762-27.86013 & $3.90\pm0.70$ & $3.75\pm0.38$ & $4.25\pm0.50$ & $4.36\pm0.45$ & $3.76\pm0.42$ & $2.68\pm0.65$ & $4.07\pm0.49$\\
\multicolumn{8}{c}{Contributing Sample $z<11.5$} \\
JADES+53.03139-27.87219 &  $2.34\pm0.69$ & $2.72\pm0.33$ & $3.51\pm0.47$ & $2.90\pm0.38$ & $2.26\pm0.37$ & $2.21\pm0.62$ & $3.33\pm0.51$\\
JADES+53.09292-27.84607 &  $3.96\pm0.71$ & $2.41\pm0.34$ & $3.26\pm0.52$ & $2.51\pm0.46$ & $2.43\pm0.37$ & $2.90\pm0.60$ & $1.69\pm0.49$\\
JADES+53.06857-27.85093 &  $4.49\pm0.68$ & $3.63\pm0.37$ & $3.08\pm0.49$ & $2.87\pm0.43$ & $2.98\pm0.37$ & $3.21\pm0.59$ & $2.25\pm0.48$\\
\multicolumn{8}{c}{Auxiliary Sample $z>11.5$} \\
JADES+53.08468-27.86666 &  $4.31\pm0.73$ & $4.21\pm0.36$ & $5.08\pm0.50$ & $5.14\pm0.45$ & $4.78\pm0.39$ & $5.30\pm0.62$ & $6.14\pm0.51$\\
JADES+53.07385-27.86072 &  $1.62\pm0.69$ & $2.12\pm0.39$ & $2.60\pm0.51$ & $2.28\pm0.44$ & $1.84\pm0.39$ & $1.41\pm0.64$ & $1.74\pm0.51$\\
\enddata
\end{deluxetable*}

\subsection{Sample}
\label{sec:sample}

Given the selection criteria presented
in \S \ref{sec:selection_criteria}, our
entire $11.5<z<15$ sample consists of eight galaxy candidates (our ``Main Sample'').
Table \ref{tab:properties} lists their
designations based on [RA, Dec], the internal JADES NIRCam ID, and the 
best-fit redshift $z_a$.  
Five of these objects (ID 16699, 33309, 13731, 11457, and 55733) were previously identified in \citet{hainline23}; the other three are new here.
We also record galaxy sizes measured from our \texttt{Forcepho}
modeling in Table \ref{tab:properties}.
For each object, we provide
 $r=0.1"$ circular aperture photometry
for HST ACS bands in
Table \ref{tab:acs}, and \JWST{} $0.1"$-radius circular aperture photometry for the NIRCam short-wavelength and long-wavelength
filters appear in Tables \ref{tab:nircam_sw} and \ref{tab:nircam_lw}, respectively. Note the fluxes we report in Tables \ref{tab:acs}-\ref{tab:nircam_lw} are measured on the native resolution images and are not convolved to a common PSF.

\begin{figure*}
\noindent
\centerline{\includegraphics[width=\figscale\linewidth]{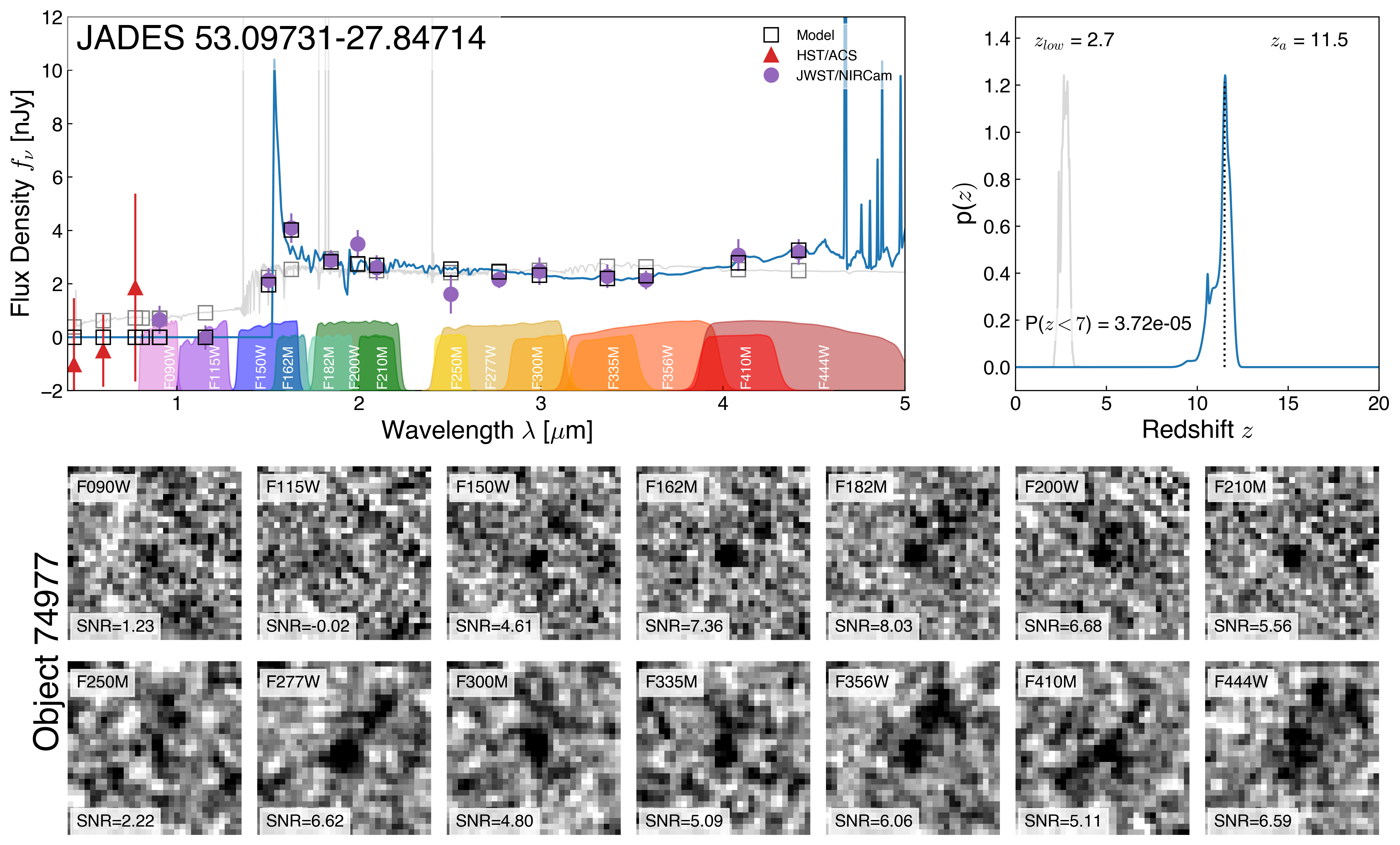}}
\caption{SED model, photometric redshift posterior distributions, and \JWST{} NIRCam image thumbnails
for galaxy candidate JADES+53.09731-27.84714 (NIRcam ID 74977). The upper left panel shows the aperture-corrected $r=0.1"$ flux density $f_\nu$ in
nJy of the NIRCam (purple points with $1\sigma$ uncertainties)
and HST/ACS (red points with $1\sigma$ uncertainties) photometry for the object, with median photometric
offset corrections applied. The best-fit
SED is shown in blue, while the best fit low-redshift solution is shown in gray. The synthetic
model photometry for both models are shown as open squares, and the \JWST{} NIRCam filter
transmission curves are shown as colored regions.
The upper
right panel shows the posterior distribution of photometric redshifts for the object (blue),
the best-fit redshift (vertical dashed line), the photo-$z$ posterior if only redshifts $z<7$
are considered (light gray), and the best-fit redshifts provided as an annotation, as is the posterior probability density at redshifts below $z\sim7$. The
bottom panel shows inverted grayscale thumbnails of the fourteen NIRCam filters in a $0.93\times0.93$ arcsec$^{2}$
region around each object, the stretch applied to each filter scaled with the mean value in the thumbnail. The signal-to-noise ratio of the aperture-corrected $r=0.1"$ circular aperture photometry for each band is noted in the corresponding thumbnail. The JADES NIRCam ID is
also provided on the left side of the image.\label{fig:sed_74977}}
\end{figure*}

Figure \ref{fig:composite} shows an F444W/F200W/F090W red/green/blue false color mosaic of the
JOF region. Of the 27.5 arcmin$^{2}$ area shown, about 9.05 arcmin$^{2}$ has acceptable F162M
coverage. The inset thumbnail images for each galaxy candidate show $0.86$ arcsec$^2$ regions
with red/green/blue colors provided by F356W+F410M+F444W/F200W+F210M/F090W+F115W, along
with the best fit redshift $z_a$ and the JADES NIRCam ID referenced in Table \ref{tab:properties}.
We plot the eight galaxy candidates in our Main Sample, and the two 
auxiliary objects that have 
photometric redshifts $z>11.5$ but that fail data quality or redness cuts.

Next, in order of increasing photometric redshift, we introduce each galaxy candidate with some summary discussion
and a figure of the SED fits to the $0.1"$-radius circular aperture photometry. 
We show the photometric
redshift posterior distribution and best-fit
redshift for each and the redshift posterior distribution limited to $z<7$, as well as the best-fit SED and most likely
low-redshift SEDs. We also show the \JWST{} filter transmission curves and the fourteen \JWST{} filter cutouts
for each galaxy.

\subsubsection{JADES+53.09731-27.84714; NIRCam ID 74977}
\label{sec:74977}

Figure \ref{fig:sed_74977} shows the
best-fit SED for object JADES+53.09731-27.84714
(NIRCam ID 74977). The object
is remarkably faint with $m_{AB}\approx30.5$
redward of Lyman-$\alpha$, and has a best-fit
redshift of $z_a=11.5$. The best
low-redshift solution has $z_{low}=2.7$, but
exceeds the observed F115W constraint.

\subsubsection{JADES+53.02618-27.88716; NIRCam ID 16699}
\label{sec:16699}

\begin{figure*}
\noindent
\centerline{\includegraphics[width=\figscale\linewidth]{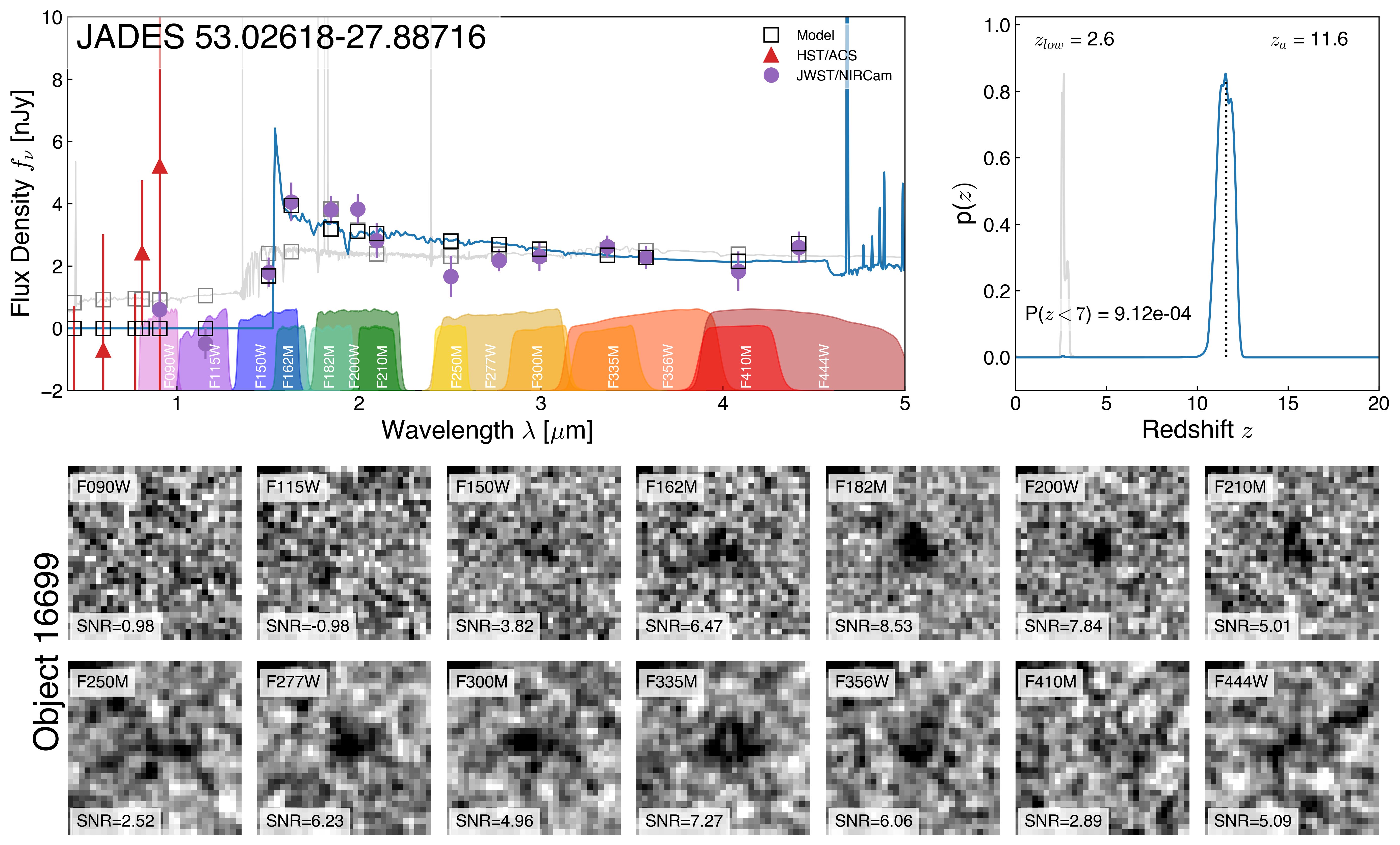}}
\caption{Same as Figure \ref{fig:sed_74977}, but for galaxy candidate JADES+53.02618-27.88716 (NIRCam ID 16699).\label{fig:sed_16699}}
\end{figure*}

Figure \ref{fig:sed_16699} shows the
best-fit SED for object JADES+53.02618-27.88716
\citep[NIRCam ID 16699,][]{hainline23}. The best-fit redshift
is $z_a=11.6$ for this faint source,
which has $m_{AB}\approx30.2-30.5$ in
the NIRCam long-wavelength channels.
The best low-redshift solution
has $z_{low}=2.6$. We note that using the
BAGPIPES SED-fitting code \citep{carnall18} to
constrain the photometric redshift of this galaxy candidate
provides
a slightly lower redshift of $z\approx11.3$, without
Lyman-$\alpha$ emission and
with still non-zero $P(z>11)$.

\subsubsection{JADES+53.04017-27.87603; NIRCam ID 33309}
\label{sec:33309}

\begin{figure*}
\noindent
\centerline{\includegraphics[width=\figscale\linewidth]{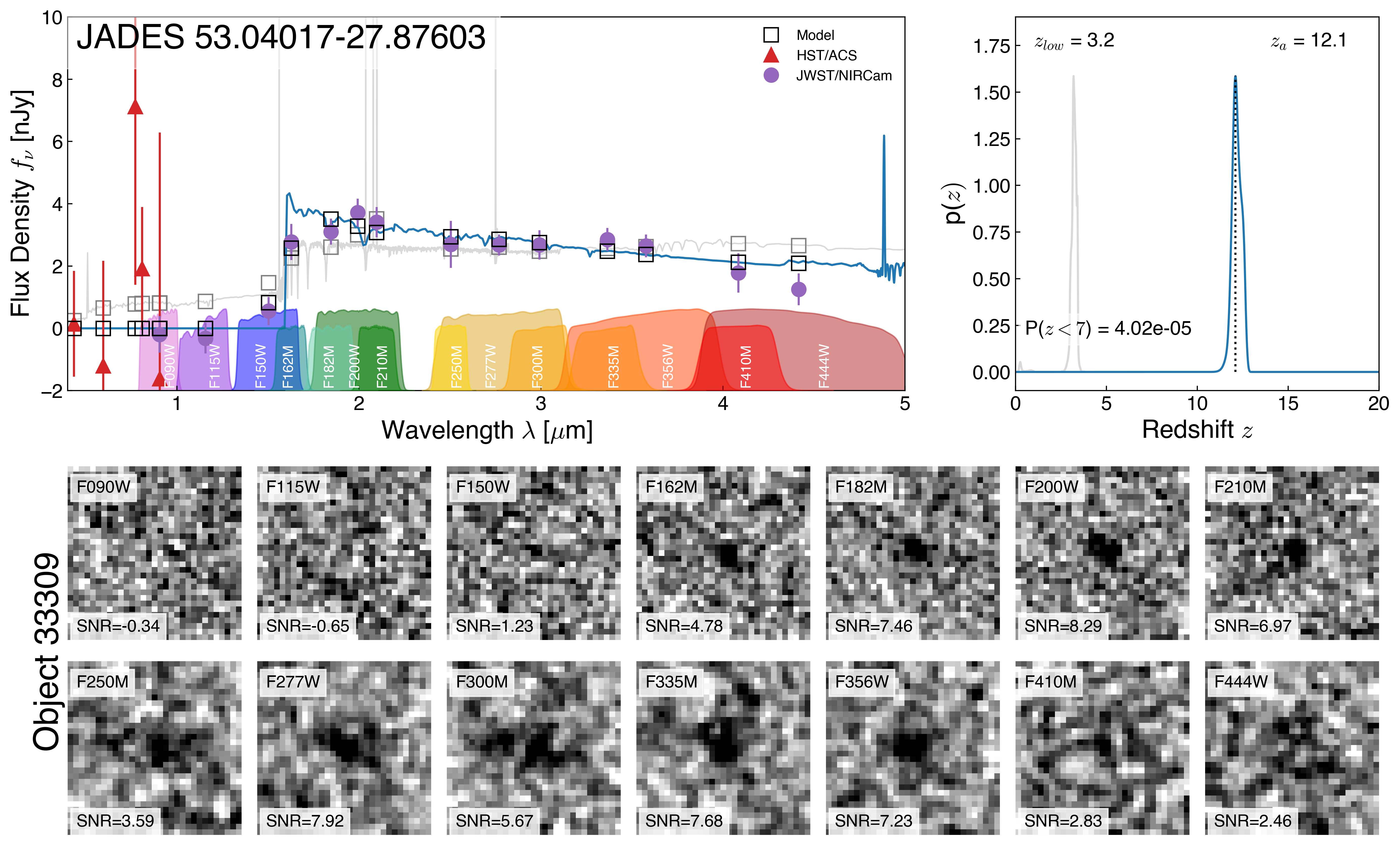}}
\caption{Same as Figure \ref{fig:sed_74977}, but for galaxy candidate JADES+53.04017-27.87603 (NIRCam ID 33309).\label{fig:sed_33309}}
\end{figure*}

Figure \ref{fig:sed_33309} shows the
best-fit SED for object JADES+53.04017-27.87603
\citep[NIRCam ID 33309,][]{hainline23}. The best-fit
SED model has $z_a=12.1$, while the best
low-redshift solution has $z_{low}=3.2$.
The source is also remarkably faint, with 
$m_{AB}\approx 30.2$ in the NIRCam long-wavelength
filters.

\subsubsection{JADES+53.03547-27.90037; NIRCam ID 160071}
\label{sec:160071}

\begin{figure*}
\noindent
\centerline{\includegraphics[width=\figscale\linewidth]{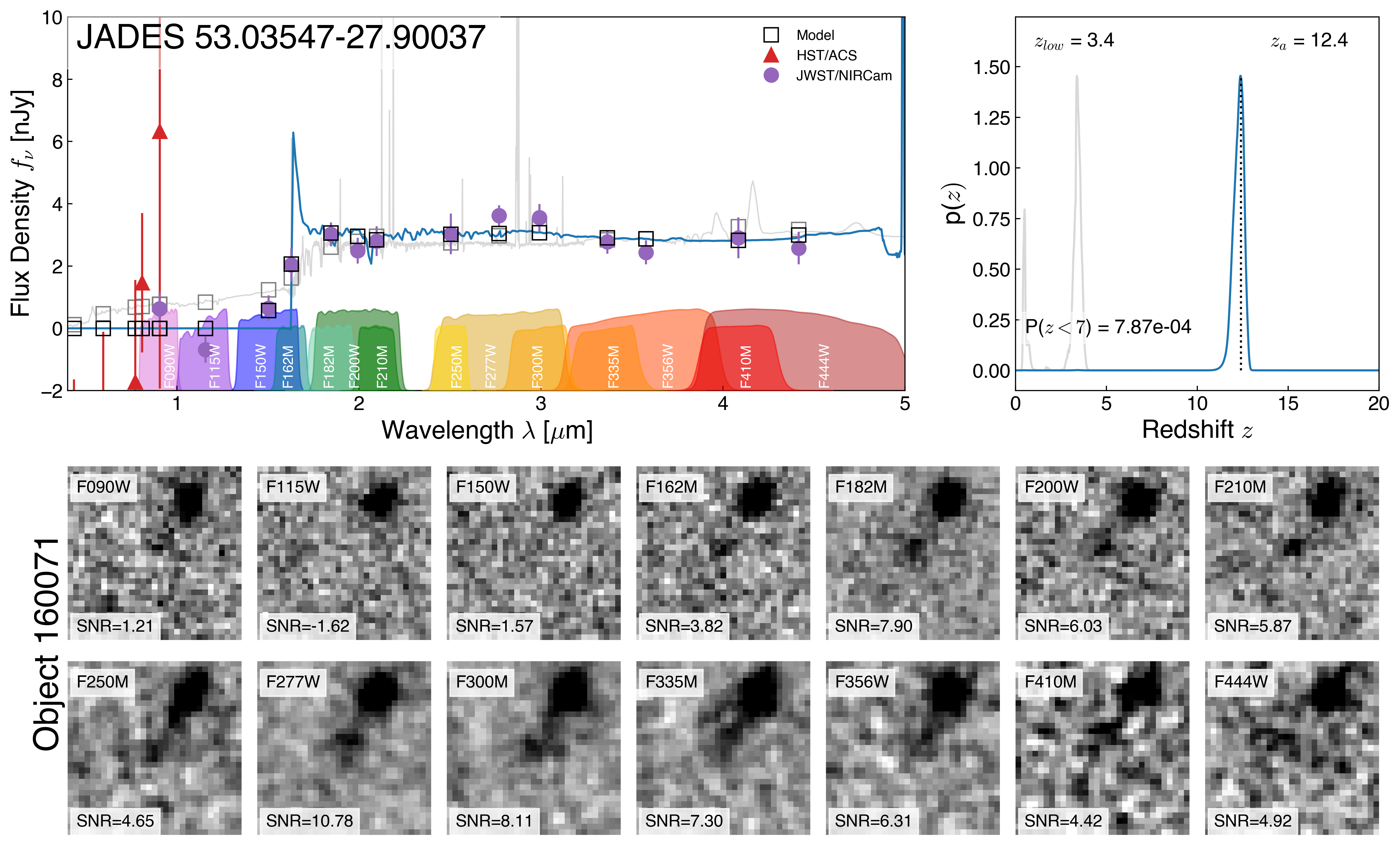}}
\caption{Same as Figure \ref{fig:sed_74977}, but for galaxy candidate JADES+53.03547-27.90037 (NIRCam ID 160071).\label{fig:sed_160071}}
\end{figure*}

Figure \ref{fig:sed_160071} shows the
best-fit SED for object JADES+53.03547-27.90037
(NIRcam ID 160071). The flux densities
of this object are $f_\nu\approx3.5$ nJy ($m_{AB}\approx30$).
When fitting an SED model, the observed photometry, including
the strong break in F150W, constrain the
redshift to be $z_a = 12.4$.
The best
solution at low redshift has $z_{low} = 3.4$.

\subsubsection{JADES+53.06475-27.89024; NIRCam ID 13731}
\label{sec:13731}

\begin{figure*}
\noindent
\centerline{\includegraphics[width=\figscale\linewidth]{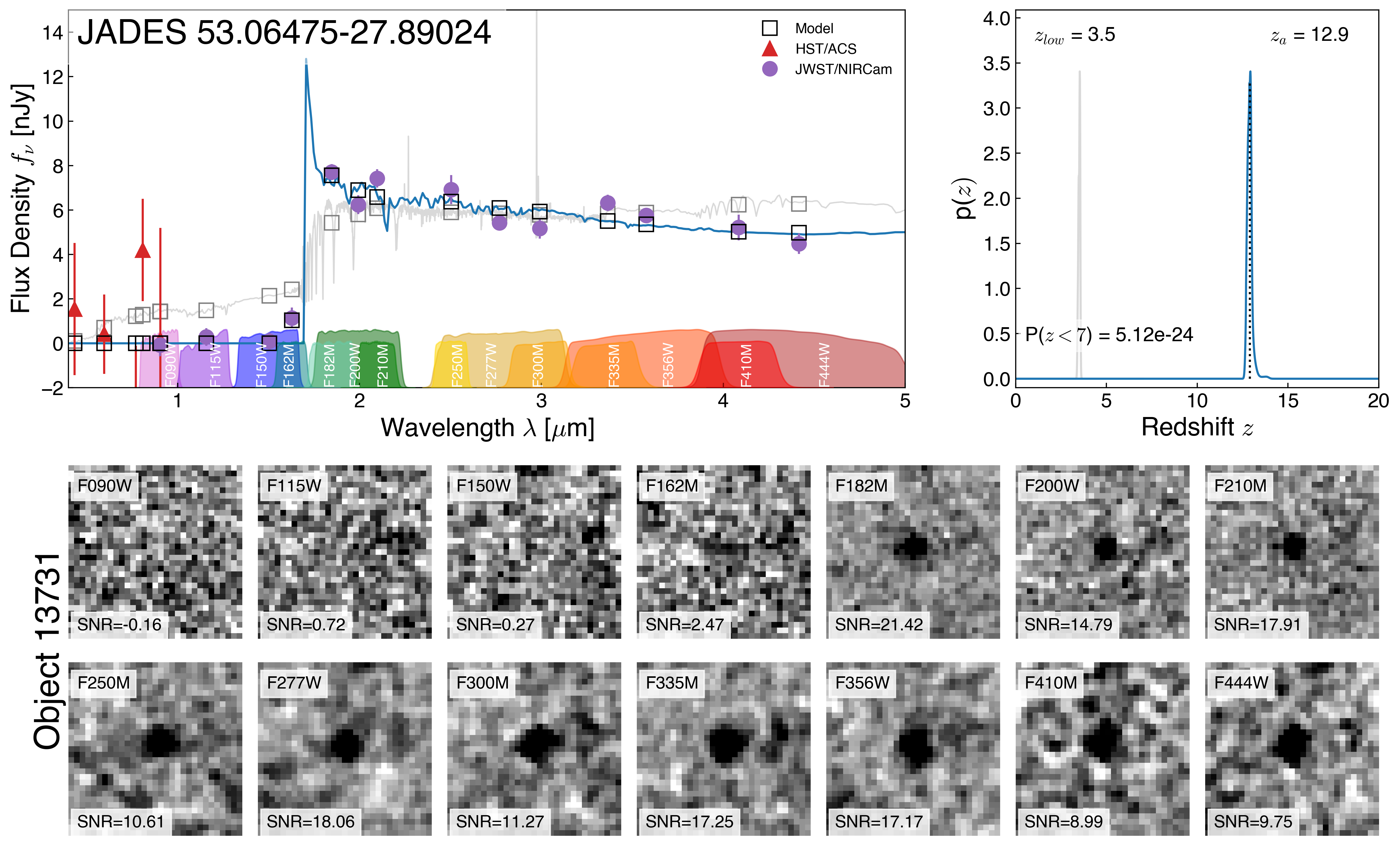}}
\caption{Same as Figure \ref{fig:sed_74977}, but for galaxy candidate JADES+53.06475-27.89024 (NIRCam ID 13731).\label{fig:sed_13731}}
\end{figure*}

Figure \ref{fig:sed_13731} shows the
best-fit SED for object JADES+53.06475-27.89024
\citep[NIRCam ID 13731,][]{hainline23}. The long-wavelength JWST/NIRCam photometry
shows $m_{AB}\approx29.5$, and constrains the
posterior photometric redshift distribution to be
peaked strongly near $z_a=12.9$. The best-fit
low-redshift solution at $z_{low}=3.5$ would
exceed the F090W, F115W, F150W, and F162M 
photometry by several standard deviations.

\begin{figure*}
\noindent
\centerline{\includegraphics[width=\figscale\linewidth]{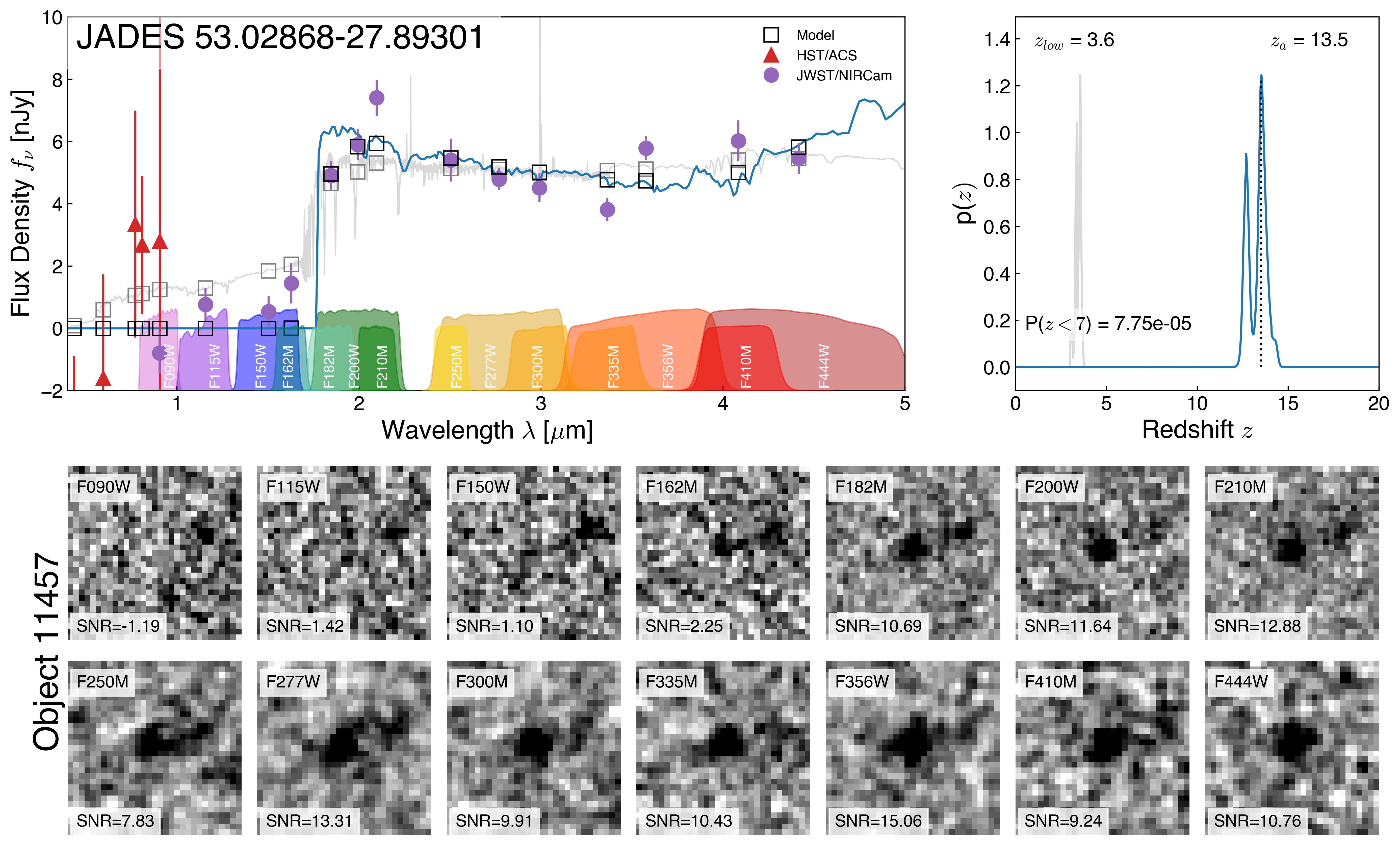}}
\caption{Same as Figure \ref{fig:sed_74977}, but for galaxy candidate JADES+53.02868-27.89301 (NIRCam ID 11457).\label{fig:sed_11457}}
\end{figure*}

\begin{figure*}
\noindent
\centerline{\includegraphics[width=\figscale\linewidth]{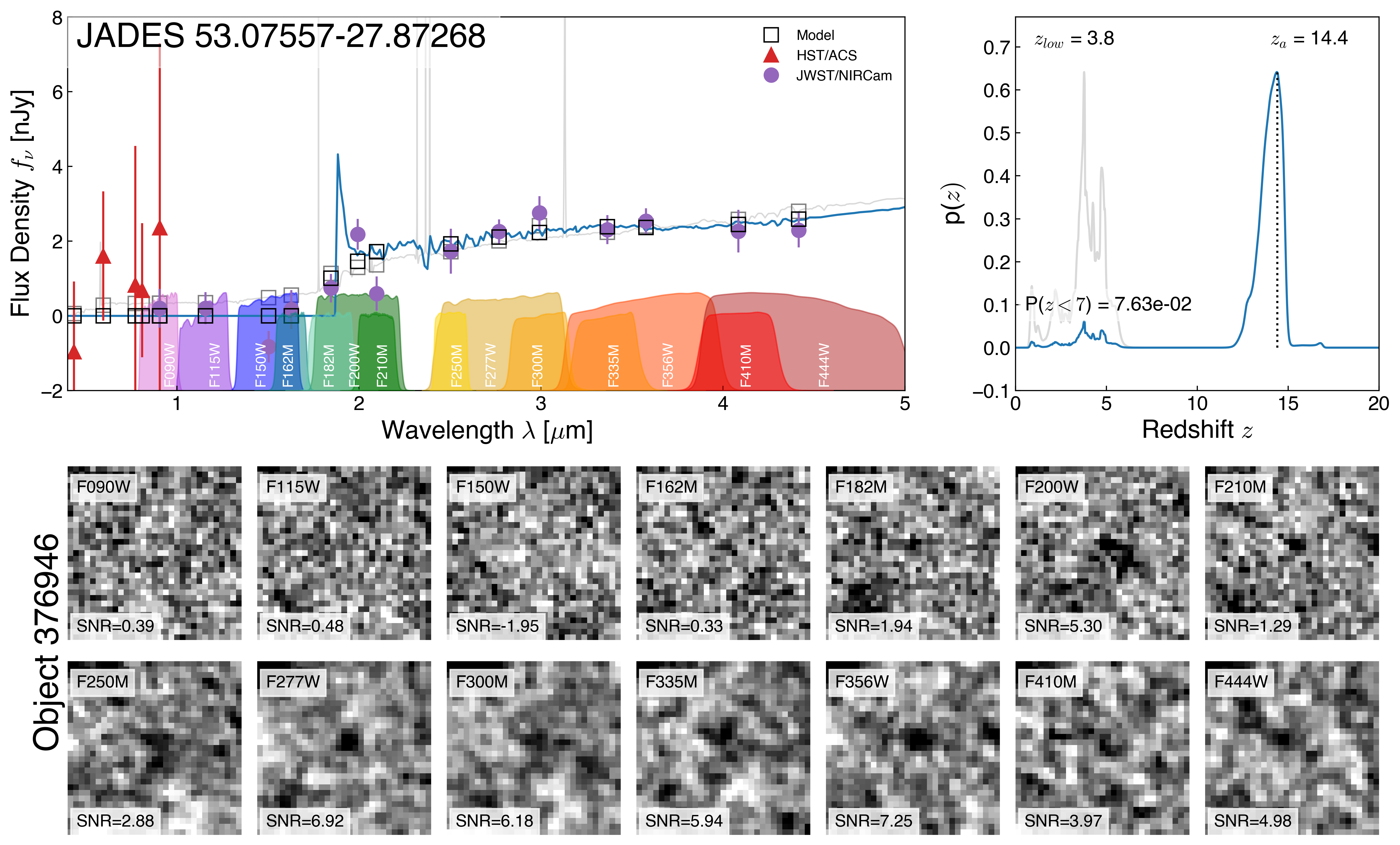}}
\caption{Same as Figure \ref{fig:sed_74977}, but for galaxy candidate JADES+53.07557-27.87268 (NIRCam ID 376946).\label{fig:sed_376946}}
\end{figure*}

\begin{figure*}
\noindent
\centerline{\includegraphics[width=\figscale\linewidth]{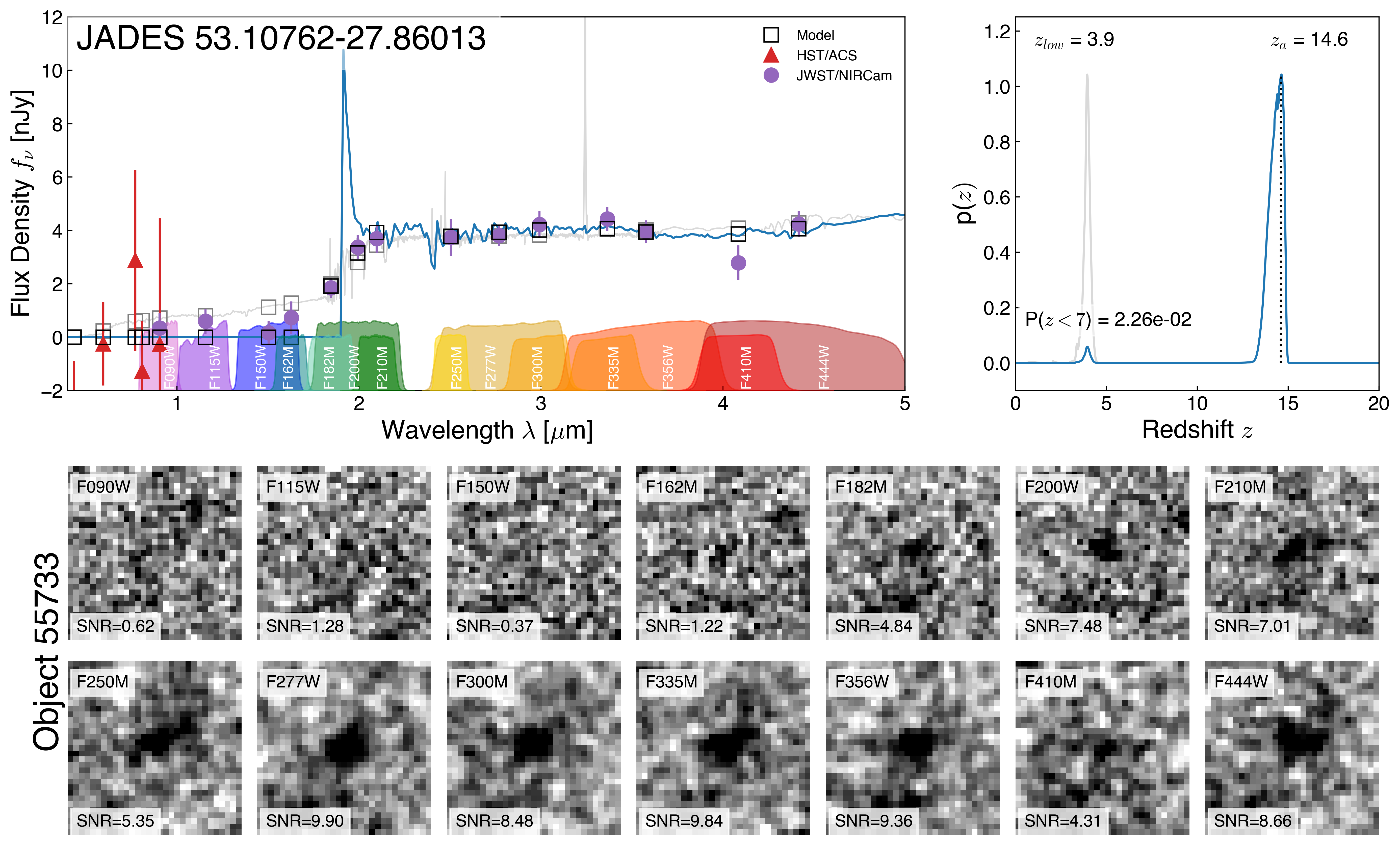}}
\caption{Same as Figure \ref{fig:sed_74977}, but for galaxy candidate JADES+53.10762-27.86013 (NIRCam ID 55733).\label{fig:sed_55733}}
\end{figure*}

\begin{figure*}
\noindent
\centerline{\includegraphics[width=\figscaleB\linewidth]{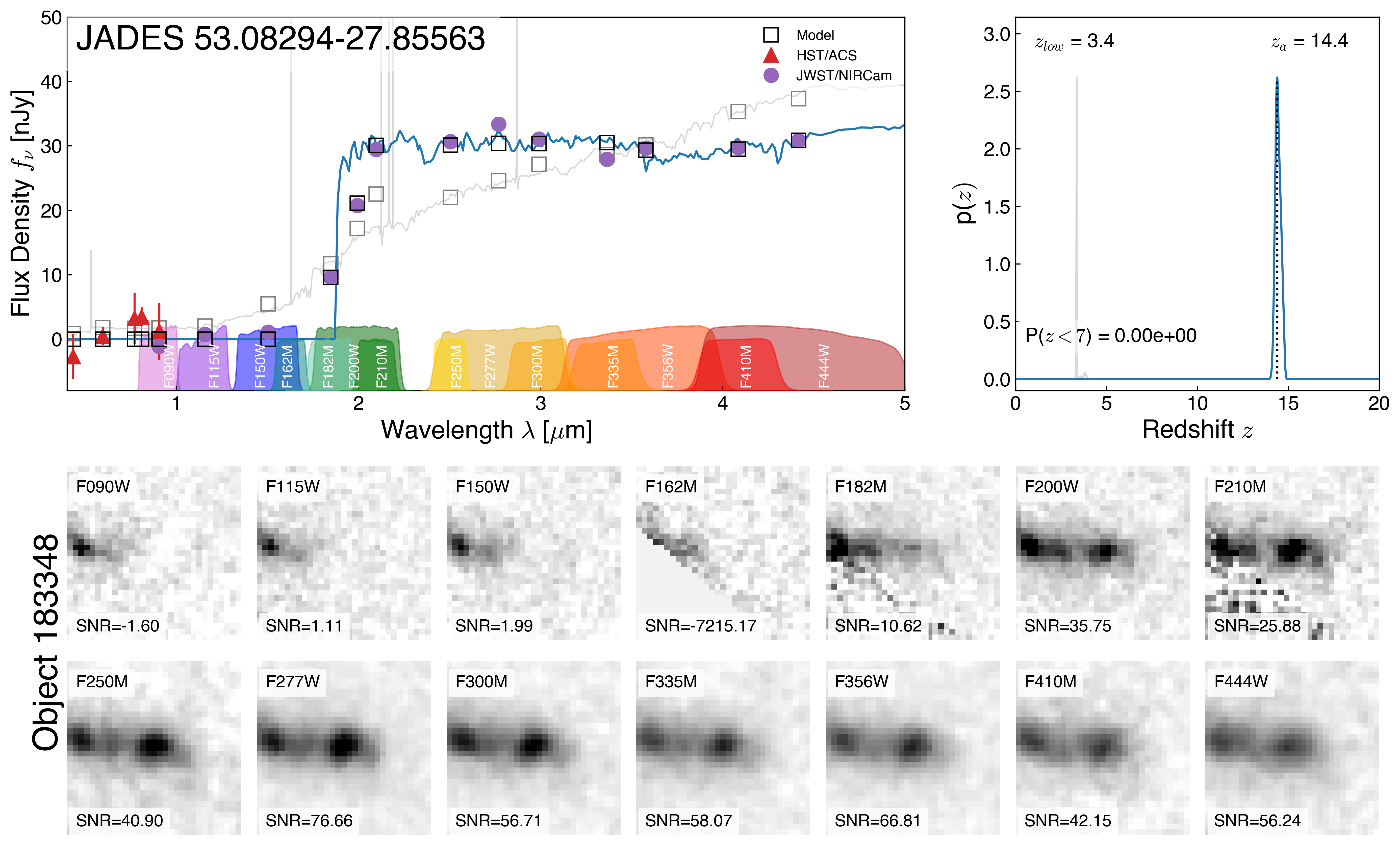}}
\caption{Same as Figure \ref{fig:sed_74977}, but for galaxy JADES+53.08294-27.85563 (NIRCam ID 183348).
We note this object has been discussed previously in \citet{hainline23} and \citet{williams23}, and
spectroscopically confirmed by Carniari et al. (submitted).
The F162M data for this object has been 
omitted because of data quality issues.\label{fig:sed_183348}}
\end{figure*}

\subsubsection{JADES+53.02868-27.89301; NIRCam ID 11457}
\label{sec:11457}

Figure \ref{fig:sed_11457} shows the
best-fit SED for object JADES+53.02868-27.89301
\citep[NIRCam ID 11457,][]{hainline23}. The object has NIRCam flux
densities redward of the break 
of $f_\nu\approx4-6$ nJy ($m_{AB}\approx29.5-29.9$),
that constrain SED models to yield a best-fit
redshift $z_a=13.5$. The best-fit low-redshift solution SED
at $z_{low}=3.6$ exceeds the F090W and F150W constraints by
several standard deviations. The second peak in the high-redshift
$p(z)$ at $z\approx12.7$ is driven by the marginal ($2.3\sigma$) detection in F162M, which if real would
prefer a slightly lower redshift than the mode but still within our selection 
criteria. However, we caution that the F162M detection, the F182M-F210M color, and the rising SED shape longward of $3.5\mu$m
could indicate a potential low-redshift contaminant not well-modeled by our
SED template set. We therefore proceed with caution while including this candidate
in our sample.

\subsubsection{JADES+53.07557-27.87268; NIRCam ID 376946}
\label{sec:376946}

Figure \ref{fig:sed_376946} shows the
best-fit SED for object JADES+53.07557-27.87268
(NIRcam ID 376946). This faint ($m_{AB}=30.5$) object at redshift $z_a=14.4$
is slightly redder
than most of the other candidates.
JADES+53.07557-27.87268 displays an unusual SED in that either the F182M and F210M fluxes must be biased low by several sigma to be consistent with the F200W flux, and  the high-redshift solution does not match
well the observed F182M, F200W, and F210M data.
The best
solution at low redshift has $z_{low} = 3.8$ with nearly 8\% of the EAZY probability, although it overpredicts the observed F150W flux.
We note that when using the
BAGPIPES SED-fitting code \citep{carnall18} with a broad log-uniform prior ($\Mstar\in[10^5,10^{13}]\Msun$) on stellar 
mass to
constrain the photometric redshift of this galaxy candidate,
we find a yet larger low-redshift probability density
than with EAZY. The best fit redshift is still $z>14$
and most of its photometric redshift posterior probability
is at very high-redshift. We also note that this
object has the largest increase in the low-redshift
probability density when using common-PSF Kron
aperture fluxes to fit a photometric redshift, but, 
given the loss in SNR for this exceedingly faint object, the photometric SED become much noisier.

\subsubsection{JADES+53.08294-27.8556; NIRCam ID 183348}
\label{sec:183348}
JADES+53.08294-27.8556 (NIRcam ID 183348) with redshift $z=14.4$
is the most remarkable
object in our sample, with a best-fit SED shown in 
Figure \ref{fig:sed_183348}.
The object appears relatively bright ($f_\nu\approx30$nJy; $r=0.1"$ radius aperture)
but shows strong break from F210M to F182M and no significant flux at shorter wavelengths.
Before the JOF ultradeep JWST/NIRCam medium band data was acquired,
based on JADES JWST/NIRCam broadband data
\citet{hainline23} first discussed this source
with a photometric redshift of $z_{phot} = 14.51$.
Owing to the observed brightness of the source and its
close proximity to another lower-redshift source,
183348 was rejected from their main sample.
Subsequently, \cite{williams23} determined a lower
photometric redshift $z_{phot} = 3.38$,
and found the source was detected by JWST/MIRI at 7$\mu$m
from the SMILES program (PID 1207; PI Rieke).
Given the addition of our ultradeep JOF JWST/NIRCam medium band
data, we find the photometric redshift posterior of 183348
distribution is sharply peaked at $z\sim14.4$. 
This high redshift peak is now much more strongly favored than
low redshift solutions as the new JOF
medium band measurements better constrain the shape and depth of the
break at $\sim1.8\mu\mathrm{m}$ while placing limits on strong emission lines
redward of the break.
While low-redshift solutions have low probability, 
the low-redshift photometric redshift posterior distribution 
is very sharply
peaked near $z_{low}=3.4$ and requires a very red object with
strong emission lines in F200W and F277W. 
A principal concern regarding 183348 is the close proximity of a
neighboring galaxy (NIRCam ID 183349) that has a best-fit photometric
redshift of $z_a\approx3.4$. 
This alignment obviously supported the previous
suspicion that 183348 was also at the lower redshift.  
However, our analysis of the initial JOF NIRCam medium-band photometry as well as
JWST/MIRI photometry (Helton et al., submitted)
further supported the higher redshift, and on that basis,
the galaxy was selected for spectroscopic followup.
Carniani et al. (submitted) present a spectroscopic redshift confirmation
of $z=14.32$, and we refer the reader to that work for a detailed
analysis of the properties of this intriguing galaxy. Here, we do compare the
properties inferred for this galaxy along with other objects in the Main
Sample measured in the same manner. We note that the photometric and
spectroscopic redshift distributions are very similar, and 
our choice to adopt its photometric redshift distribution during the
luminosity function inference has little impact on our results.
We also note that gravitational lensing by the neighbor is considered by
Carniani et al. (submitted), but find the magnification to be small.

\subsubsection{JADES+53.10762-27.86013; NIRCam ID 55733}
\label{sec:55733}

Figure \ref{fig:sed_55733} shows the
best-fit SED for object JADES+53.10762-27.86013
\citep[NIRCam ID 55733,][]{hainline23}. The galaxy candidate
has NIRCam long-wavelength fluxes of 
$m_{AB}\approx29.9$ and a best-fit
redshift of $z_a\approx14.6$.
The best low-redshift solution has $z_{low}=3.9$ with 2\% of the EAZY probability,
although the corresponding SED model would substantially exceed the 
observed F150W.
We note that this object shows F162M flux
at $1.1\sigma$ significance, and confirmation
of this hint of a signal would negate a
possible high-redshift solution.

\subsubsection{Auxiliary Objects}
\label{sec:Auxiliary}

\begin{figure*}
\noindent
\centerline{\includegraphics[width=\figscaleB\linewidth]{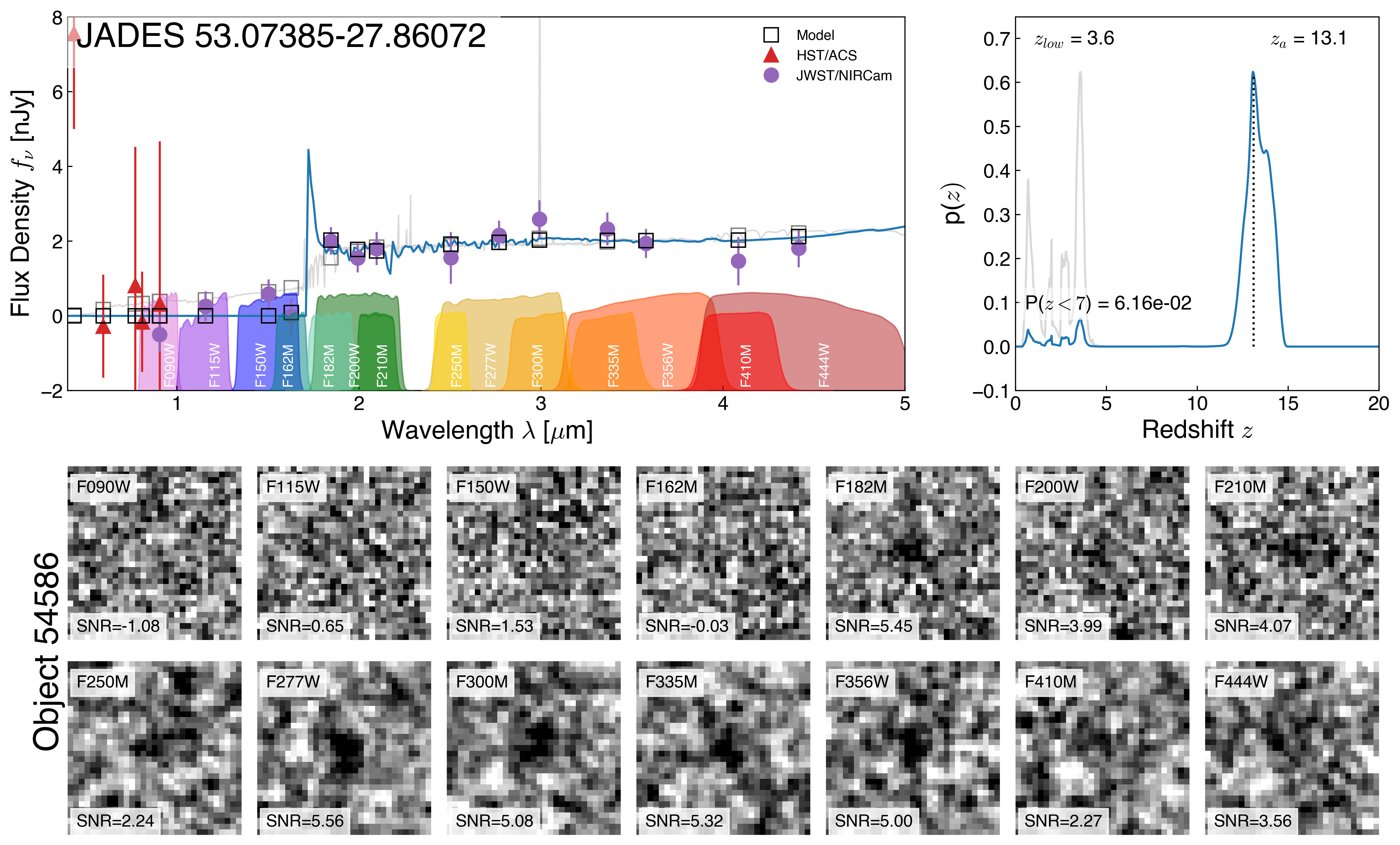}}
\caption{Same as Figure \ref{fig:sed_74977}, but for Auxiliary galaxy candidate JADES+53.07385-27.86072 (NIRCam ID 54586).\label{fig:sed_54586}}
\end{figure*}

\begin{figure*}
\noindent
\centerline{\includegraphics[width=\figscaleB\linewidth]{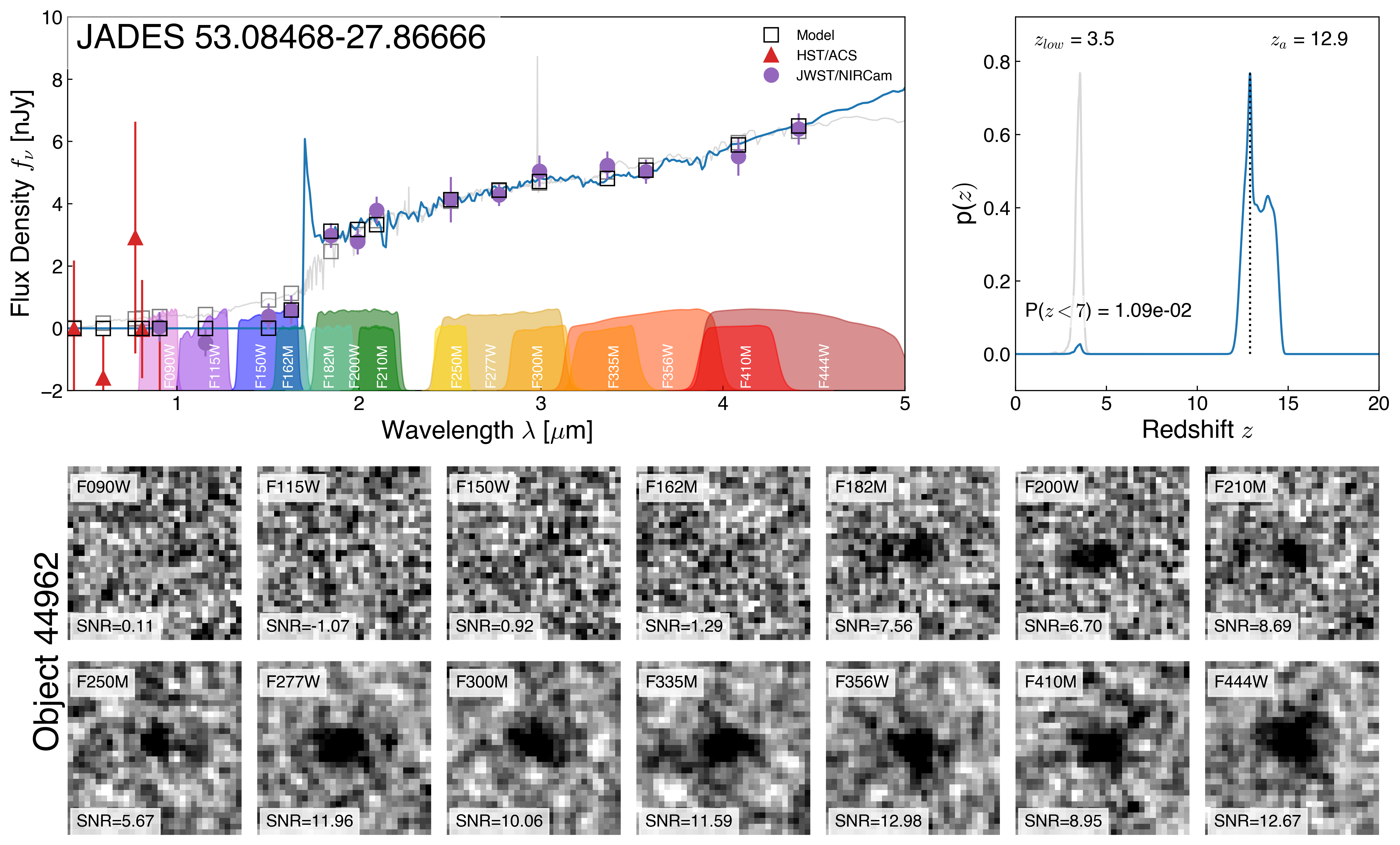}}
\caption{Same as Figure \ref{fig:sed_74977}, but for Auxiliary galaxy candidate JADES+53.08468-27.86666 (NIRCam ID 44962).\label{fig:sed_44962}}
\end{figure*}

We also provide SED fits for the Auxiliary
candidates
JADES+53.07385-27.86072 (Figure \ref{fig:sed_54586}),
and JADES+53.08468-27.86666. (Figure \ref{fig:sed_44962}).

JADES+53.07385-27.86072 (NIRcam ID 54586) is exceedingly faint and is relegated to our Auxiliary sample by failing the minimum SNR criteria, with
some long-wavelength NIRCam filters showing $m_{AB}>30.5$
flux levels. The high-redshift posterior distribution for this object
is correspondingly broader, with a peak at $z_a=13.1$. The best
low redshift solution has $z_{low}=3.6$.

Finally, JADES+53.08468-27.86666 
\citep[NIRCam ID 44962,][]{hainline23} is in our
Auxiliary sample owing to its 
red SED that increases from
$f_\nu\approx3$ nJy in F182M to $f_\nu\approx6$nJy in F444W.
The redshift posterior distribution is double-valued, with a peak
at $z_a=12.9$. The best low-redshift solution has $z_{low}=3.5$.

\section{Completeness Simulations}
\label{sec:completeness}

The detection and selection of high-redshift galaxy
candidates impose limitations that reduce the
completeness of a sample. To convert the number
of observed galaxies satisfying the selection
criteria into a measurement of the galaxy
number density, the completeness of the detection
and selection process can be computed and 
incorporated. Below, in \S \ref{sec:det_comp}
we use simulations to
characterize our detection completeness and 
in \S \ref{sec:sel_comp} we simulate our
selection completeness. These calculations 
are used in \S \ref{sec:uvlf} to include
completeness corrections in the rest-UV
luminosity function.

We note that the requirement to compute the
completeness suggests that the detection and
selection process should be algorithmic and automatable. We
therefore do not apply any cuts based on
visual inspection or judgment beyond 
crafting the detection method described 
in \S \ref{sec:detphot} or the selection criteria
presented in \S \ref{sec:selection}. This
restriction allows us to simulate both the
detection and selection completeness.

\subsection{Detection Completeness}
\label{sec:det_comp}

To compute the detection completeness
of our photometric pipeline, we
performed detailed source injection
simulations using a wide range of input
sources. First, we create a mock input
galaxy catalog by drawing from randomized
distributions of galaxy physical properties
including redshift, star formation
rate, stellar mass, size, \citet{sersic68}
surface brightness profile index, position
angle, and
axis ratio. The objects are selected to have
properties comparable to the $z>8$ sources
reported by \citet{hainline23}.
We use the \texttt{Prospector}
code \citep{johnson21} to compute the
object fluxes given their physical properties
and redshift. With this mock catalog,
we use the GalSim \citep{rowe2015a} image
simulation software to create simulated
\citet{sersic68} profile objects distributed
across a grid on the sky. We compute the
overlap of the JOF mosaics in each filter
with this grid of objects, and then add the
randomized objects as injected sources
in the JOF images. The result is a large
set of synthetic
JOF mosaics with injected sources.
We can then process the images identically
to the real data and attempt to discover
sources.

With the injected images, we combine the
long-wavelength NIRCam images as for
the real data, creating an ultradeep
stack. Our pipeline detection algorithm
is applied to the injected mosaic
stack to create a new detection catalog
with simulated sources. We can then characterize
the completeness of our detection method
as a function of the source properties.
We repeat the simulations with ten separate realizations, such that a total of 115,000 injected sources with widely-ranging
intrinsic properties are used.

\begin{figure}[t]
\noindent
\includegraphics[width=\linewidth]{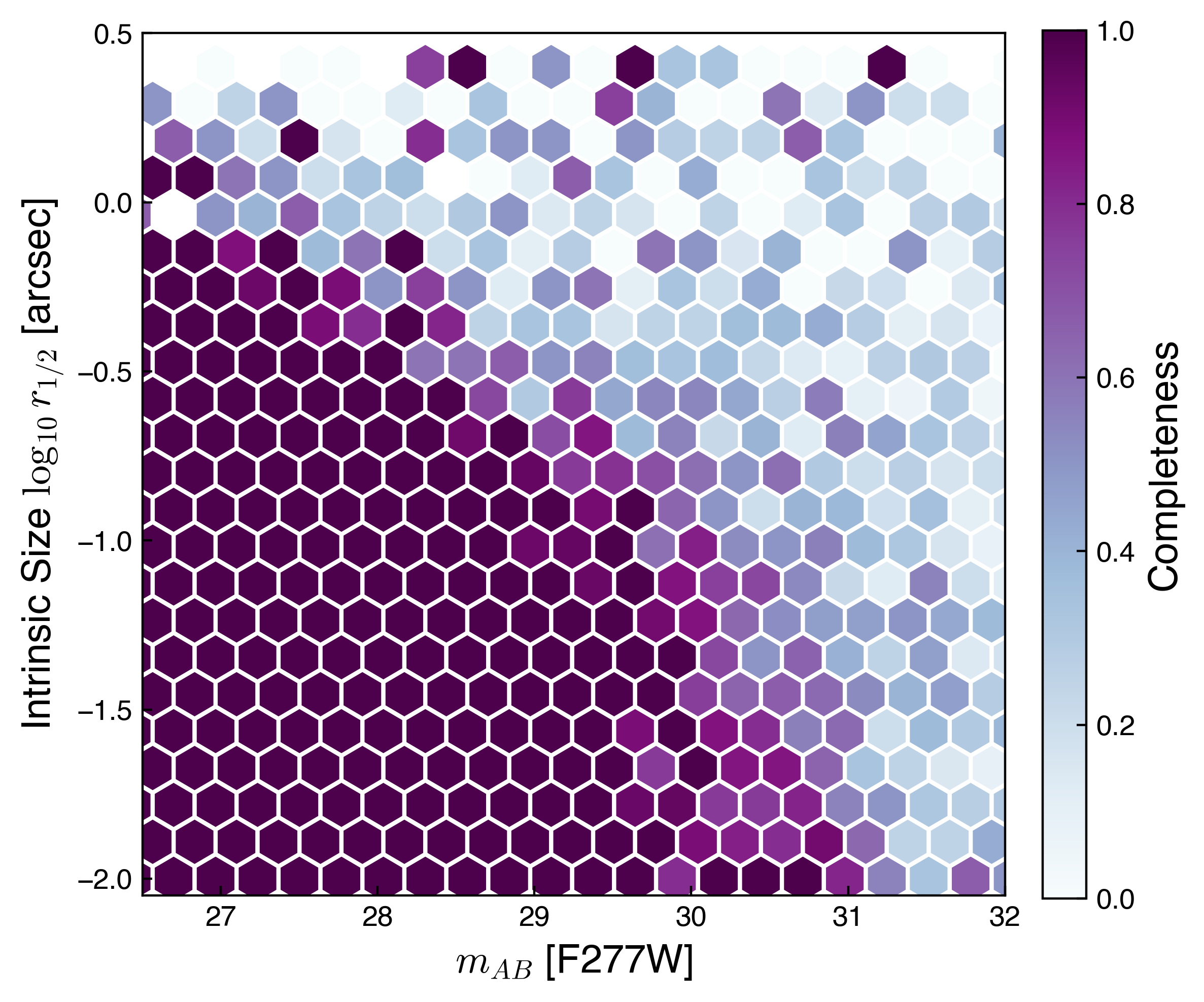}
\caption{Detection completeness in our JOF analysis as a function of intrinsic
half-light radius and F277W apparent magnitude.
The detection method is complete for small objects and bright magnitudes,
and the differential completeness reaches about 90\% at F277W$\approx30.2$AB for small objects.
Shown is a two-dimensional normalized histogram of object size and flux 
indicating the fraction of sources with such properties detected by the pipeline.
The method becomes highly incomplete fainter than $m_{AB}\sim31$ or for half-light
radii above about half an arcsec. Owing to pixels covered by foreground sources,
the maximum detection completeness will be reduced to $\sim78$\% of that shown here.
\label{fig:detection_completeness}}
\end{figure}

Figure \ref{fig:detection_completeness}
shows the detection completeness as a function
of the two main factors affecting this
completeness. The apparent brightness
of the objects influence their signal-to-noise
ratio in the stacked detection image.
The size of the object affects the surface
brightness, which in turn determines the
per pixel SNR that governs the contrast
an object of a given luminosity relative
to the sky background. The
detection algorithm reaches 90\%
completeness at around $m_{AB}\sim30.2$
for small objects ($R_{1/2}\lesssim0.1$ arcsec).
This completeness function can be integrated
into an interpolator to allow for the
object completeness as a function of
apparent magnitude and size to be utilized
in inferring the UV luminosity function.
We note that through this simulation for
the JOF we find that only about
78\% of the pixels are not impacted by foreground
objects, which we account for in computing
our effective survey volume. Given that the
objects of interest are small, only several pixels across,
and our detection
method reaches fairly low significance (SNR$\sim1.5$) per pixel such that the 
segmentations reach low surface brightnesses, we find
this number to be representative of the impact of
foreground sources on our detection completeness.

\subsection{Selection Completeness}
\label{sec:sel_comp}

To simulate the selection completeness, we can use the spectral energy
distributions in our mock galaxy catalog and the photometric
uncertainty measured for our galaxy sample to simulate the effects of
photometric noise on our selection and consequently the inferred UV
luminosity function. We create a sample of two million mock galaxies
with model SEDs, induce photometric noise with a normal scatter
in each HST and \JWST{} filter of the magnitude of our measured
sky background. Our measurement uncertainties are sky-dominated, so only include sky noise in our simulated fluxes. These two million noisy simulated SEDs are then
provided to \texttt{EAZY} exactly in the same manner as our 
real catalog, and SED fitting is performed to each object. This
enables us to estimate how the photometric noise can disrupt the
mapping between true redshift and photometric redshift, and
identify which redshift windows could provide non-negligible
contamination for our selection criteria. For reference, we note that in our simulations, the fraction of
objects with F200W SNR$>5$ that are photometric redshift outliers with $(|z_a-z_\mathrm{true}|/(1+z_{\mathrm{true}}))>0.1$ is 3.8\%.

\begin{figure}[t]
\noindent
\includegraphics[width=\linewidth]{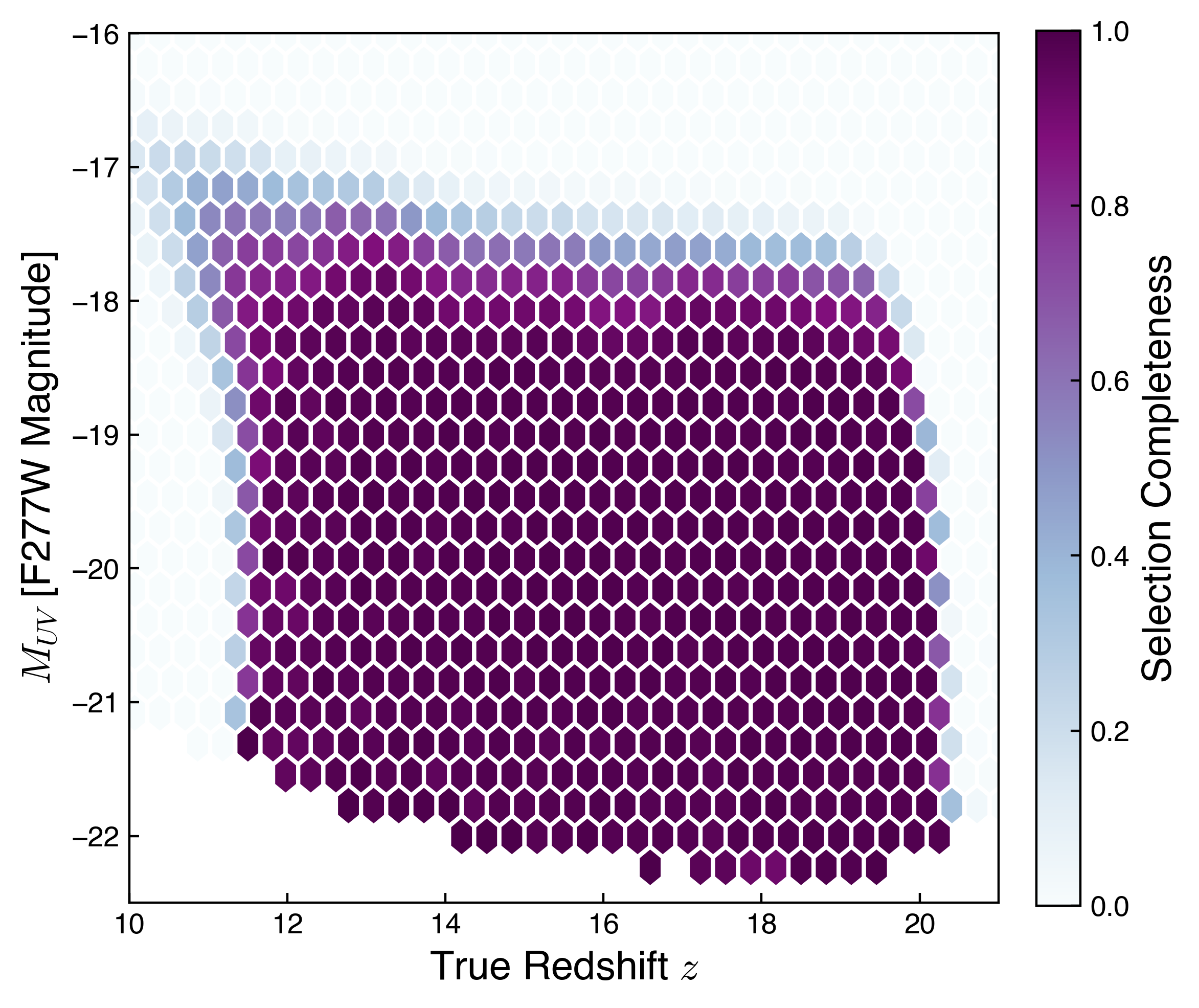}
\caption{Completeness of our selection criteria as a function of galaxy
redshift and absolute magnitude. For bright objects, the selection criteria
described in \S \ref{sec:selection} produce a substantially complete sample.
For fainter objects, the $\Delta \chi^2$ criterion fails as the photometric
noise prevents the SED fitting procedure from distinguishing robustly between
high and low photometric redshifts.
\label{fig:selection_completeness}}
\end{figure}

Figure \ref{fig:selection_completeness} shows the 
completeness of selection criteria as applied to our mock
galaxy catalog, as a function of the true object redshift  
and absolute UV magnitude. The selection proves highly
complete at $M_{UV}<-18$ for redshifts $z\gtrsim12$.
At magnitudes fainter than $M_{UV}>-17.5$, the photometric
noise prevents the strict elimination of low-redshift
solutions such that the objects fail the $\Delta \chi^2$
selection described in \ref{sec:selection}. At the high-redshift
end, the selection declines at $z\approx20$ when the Lyman-$\alpha$ break
affects F250M and our SNR requirement in that filter becomes
limiting. As with the
detection completeness, an interpolator can be constructed
from the selection completeness and then used to correct
the galaxy number counts for the lossy selection process.
We note here that we define $M_{UV}$ as the rest-frame
1500\AA\ UV luminosity density, computed by fitting a power-law to rest-frame UV photometry and marginalizing over any covariance with the spectral slope (for more details, see \S \ref{sec:muv_beta}).

\section{Rest-Frame UV Luminosity Function at \texorpdfstring{$z\gtrsim12$}{z>12}}
\label{sec:uvlf}

To compute the rest-frame UV luminosity function from our sample
of galaxy candidates and our completeness calculations, we can
construct multiple measures of the galaxy abundance. We wish to
account for several confounding effects. 

First, galaxies with a range of intrinsic redshifts will contribute to the
the observed number
counts of galaxies at a given photometric redshift. The
degree of this contamination will depend on the abundance of galaxies
at other proximate intrinsic redshifts whose photometric redshifts
overlap with the epoch of our measurement. We must therefore account
for the evolving luminosity function and mixing between populations
at different redshifts.

Second, each individual galaxy has a posterior distribution for
its photometric redshift. Rather than assign each galaxy to a 
specific redshift bin and absolute magnitude, we can allow for a
posterior distribution on the photometric redshift to represent
a track of inferred absolute magnitude and redshift. Each
galaxy can make a fractional contribution to the UV luminosity
function at redshifts where its posterior has support.

Given these considerations, we want to allow
for flexibility in our representation of the UV luminosity function.
We can either infer a parameterized luminosity function by
computing the likelihood of observing each galaxy, given the evolving
distribution of galaxy counts with luminosity and redshift, 
fully without binning, or we could bin in magnitude and
redshift but account for the photometric redshift
posterior distributions of each object. In either case, with
the known individual properties of each object, we
want to treat the completeness of our detection and selection
methods at the per-object level rather than through binning. Below, we
present both methods, where we expand on the methods used
by \citet{leja20_mf} to infer the evolving stellar mass function
at low redshift but now applied to the UV luminosity function
evolution at high redshifts. We have tested both methods
using mock galaxy samples constructed from specified
luminosity functions and posterior photometric redshift distributions.

\subsection{Inferring Evolving Luminosity Function Parameters}
\label{sec:lf_inference}

The probability of observing an object with a given true luminosity and redshift is given by the product of the redshift dependent luminosity function $\Phi(L, z | \theta)$, the selection function $S(L, z)$, and the differential comoving volume element probed $V(z)$.  We can assume the luminosity function depends on some parameters $\theta$.
Unfortunately, we do not know the true luminosity and redshift of each galaxy $i$, but instead estimate it from photometric data $D_i$, by using SED models to construct the likelihood function $\like(D_i \given L, z)$.  The likelihood of observing a galaxy with $D_i$ must then be marginalized over the unknown true parameters
\begin{eqnarray}
\like(D_i \given \theta) &\propto \int \, dL \int \, dz \, \like(D_i \given L, z) \, \lambda(L, z \given \theta) \label{eqn:obj_marg}\\
\lambda(L, z | \theta) &= \Phi(L, z | \theta) \, S(L, z) \, V(z) \\
\Phi(L, z | \theta) &= \phi(z) \, (L/L_*(z))^{\alpha(z)} e^{-\frac{L}{L_*(z)}}
\end{eqnarray}
Here $\lambda(L, z)$ is the differential number of objects expected to be selected from the survey, as a function of the true $L$ and $z$.
We have parameterized the luminosity function as a single Schechter function.
The redshift evolution of the luminosity function can be treated with a dependence of the parameters on $(z-z_{\rm ref})$ where $z_{\rm ref}$ is some reference redshift, e.g. the midpoint of the redshift range of interest. For our purposes, we will adopt either simple log-linear or log-exponential evolution with redshift.
To compute the likelihood of each object marginalized over the true object redshift and luminosity we numerically integrate the marginalization integrals using samples from the probability distribution provided by EAZY.
\begin{eqnarray}
\like(D_i \given \theta) &\sim \sum_j w_{i,j} \, \lambda(L_{i,j}, z_{i, j} | \theta) / \sum_j w_{i,j} 
\end{eqnarray}
By drawing fair samples from the probability distributions provided by EAZY, and noting that the effective priors on $z$ and $L$ were uniform, each sample has equal weight $w_{i,j}$.
With the ability to compute the likelihood of each object given the model,  
the likelihood for an ensemble of objects is then the product of the individual likelihoods.
However, we must include the overall constraint given by the number of observed objects. 
The total expected number of selected objects is given by the integral of the product of the luminosity
function and the effective volume, and the observational constraint is given by the Poisson
likelihood of the actual number of observed objects\footnote{This can be derived from the treatment
of the luminosity function as an inhomogeneous Poisson process; in the case that the effective rate $\lambda$
is constant this reduces to the typical Poisson likelihood.}
\begin{eqnarray}
\like(D \given \theta) &= e^{-N_{\theta}} \prod_i \like(D_i \given \theta) \\
N_{\theta} &=  \int dL \int \, dz \, \lambda(L, z)
\end{eqnarray}
Here $N_{\theta}$ is the total number of observed objects. Note that for redshifts and luminosities for which
our observations are complete, the method accounts for the likelihood of non-detections given the chosen
luminosity function parameter values.

\subsection{Estimating a Step-Wise Luminosity Function}
\label{sec:lf_stepwise}

While the method in \S \ref{sec:lf_inference} does not bin in redshift or
luminosity, the observed candidate galaxies could be
assigned to specific redshift and luminosity bins. If nothing else,
binning allows for the measured galaxy abundance to be usefully
plotted and compared with other measurements. The binned luminosity
function summarizes the information retained by the 
unbinned parameterized LF for which representing constraints on
the galaxy abundance requires access to samples of the posterior
distribution of LF parameters.
 
Consider the photometric redshift posterior distribution $p_i(z)$
of a candidate galaxy $i$ with observed apparent magnitude
$m_i$. In the absence of photometric noise,
the absolute magnitude of the object is $M_i = m_i - DM(z)$,
where $DM(z)$ is the cosmological distance modulus including $K$-correction.
Accounting for photometric noise, we will instead have some distribution of
absolute magnitudes $p(M_i|m_i,z)$ for each object at a given
photometric redshift.
The
distribution of inferred absolute magnitudes in some redshift
bin $z_1$ to $z_2$ is
\begin{equation}
p(M_i| z_1, z_2) = \int_{z_1}^{z_2} dz \int d m_i p(M_i | m_i, z) p(z).
\end{equation}
\noindent
The contribution of a galaxy to an absolute magnitude
bin would then be
\begin{equation}
N_i(M_1,M_2,z_1,z_2) = \int_{M_1}^{M_2} p(M_i|z_1,z_2) dM_i.
\end{equation}
\noindent
The total number density per magnitude $n_j$ in 
a magnitude bin $M_1 < M_j < M_2$
would then be
\begin{equation}
n_j(M_1,M_2,z_1,z_2) = \frac{\sum_{i} N_i(M_1,M_2,z_1,z_2)}{(M_2-M_1)V_j}
\end{equation}
\noindent
where $V_j$ is the average effective volume in the bin, allowing
for the completeness to vary for each object $i$. 
In practice, evaluating these
equations involves summing over samples from the photometric
posterior distributions of the galaxies while accounting
for samples that lie outside the redshift bin to enforce
the posterior normalization constraint $\int p(z) dz = 1$.
We note that when computing the samples in $M_{UV}$ and $z$,
to compute $M_{UV}$ we use
the 1500\AA\ rest-frame flux computed in the appropriate
\JWST{} filter given a putative redshift $z$. When
computing $M_{UV}$, we use the total fluxes computed from the \emph{Forcepho} morphological decompositions.

Procedurally, for each redshift bin we take all ordered 
$M_{UV}$ samples and separate them into
bins whose edges are set to maintain a comparable number
of samples per bin. We sum the number of samples
in each bin and divide by the total number of samples
across all galaxies, which provides the (non-integer)
number of galaxies per bin. The average completeness
in the bin is computed from the per-object selection
and detection completeness based on the object 
properties and the fraction of pixels in the
image not covered by foreground sources. We then
divide the number of galaxies in each bin by the
bin width, the completeness, and the volume to
get the number density. The uncertainties for each
bin are estimated from number count statistics.

While we report our step-wise estimate, which
accounts for photometric scatter between
magnitude bins and variable completeness, we
consider these measurements estimated checks
on the inferred LF constraints described in 
\S \ref{sec:lf_inference} that do not bin 
in either redshift or magnitude and additionally
account for potential contamination from
proximate redshifts and the evolving shape
of the luminosity function with redshift.
We emphasize here that our formal derived
constraints on the luminosity function are
provided through our inference procedure
in the form of the computed
posterior distributions of the parameters
of our model evolving luminosity functions.

\subsection{Luminosity Function Constraints}
\label{sec:lf_constraints}

\begin{deluxetable}{cc}
\tabletypesize{\footnotesize}
\tablecaption{Step-wise Luminosity Function.\label{tab:lf_stepwise}}
\tablehead{
	\colhead{$M_{UV}$} & \colhead{$\phi_{UV}$ [10$^{-4}$ mag$^{-1}$ Mpc$^{-3}$]}
}
\startdata
\multicolumn{2}{c}{$11.5<z<13.5$} \\
\hline
$-18.5_{-0.48}^{+0.18}$ & $1.22\pm0.94$ \\
$-18.0_{-0.18}^{+0.14}$ & $3.20\pm2.46$ \\
$-17.6_{-0.19}^{+0.65}$ & $1.54\pm1.18$ \\
\hline
\multicolumn{2}{c}{$13.5<z<15$} \\
\hline
$-20.8_{-0.32}^{+2.12}$ & $0.371\pm0.357$ \\
$-18.4_{-0.50}^{+0.16}$ & $2.56\pm2.46$ \\
$-18.1_{-0.23}^{+1.13}$ & $0.783\pm0.754$ \\
\enddata
\tablecomments{The ranges listed for each $M_{UV}$ reflect the widths of the magnitude bins, which are determined by the distribution of photometric redshift posterior samples for the objects contributing to each bin.}
\end{deluxetable}
\begin{deluxetable}{ccccc}
\tabletypesize{\footnotesize}
\tablecaption{Luminosity Function Marginalized Parameter Constraints\label{tab:lf_model}}
\tablehead{
	\colhead{Parameter} & \colhead{Prior} & \multicolumn{3}{c}{Constraint}
}
\startdata
$\log_{10} \phi_{\star,0}^a$   & $\mathcal{U}(-8,-2)$& $-6.39$& $-5.22$& $-4.24$\\
$M^{\star b}$ & $\mathcal{U}(-17,-24)$ & $-24.95$ & $-22.80$ & $-20.71$\\
$\eta^c$ & $\mathcal{U}(-3, 3)$& $-0.29$& $-0.20$& $-0.13$\\
$\alpha$  & $\mathcal{U}(-3,-1)$  & $-2.16$  & $-1.79$  & $-1.43$ \\
\enddata
\tablecomments{
$^a$ The lower limit on the LF normalization is not well constrained,
but the 95\% upper limit is $\log_{10} \phi_{\star,0} < -3.84$.
$^b$ The 95\% upper limit on the characteristic magnitude is $M^{\star}<-19.9$.
$^c$ We constrain the evolution parameter to be $\eta<-0.08$ at 95\%.}
\end{deluxetable}

Given the measured properties of 
our sample galaxies, their photometric
redshift distributions $p(z)$, and the method
described in \S \ref{sec:lf_inference}, we
can compute marginalized constraints of an 
evolving UV luminosity function once we adopt
a parameterized form.

\begin{figure*}[t]
\noindent
\includegraphics[width=\linewidth]{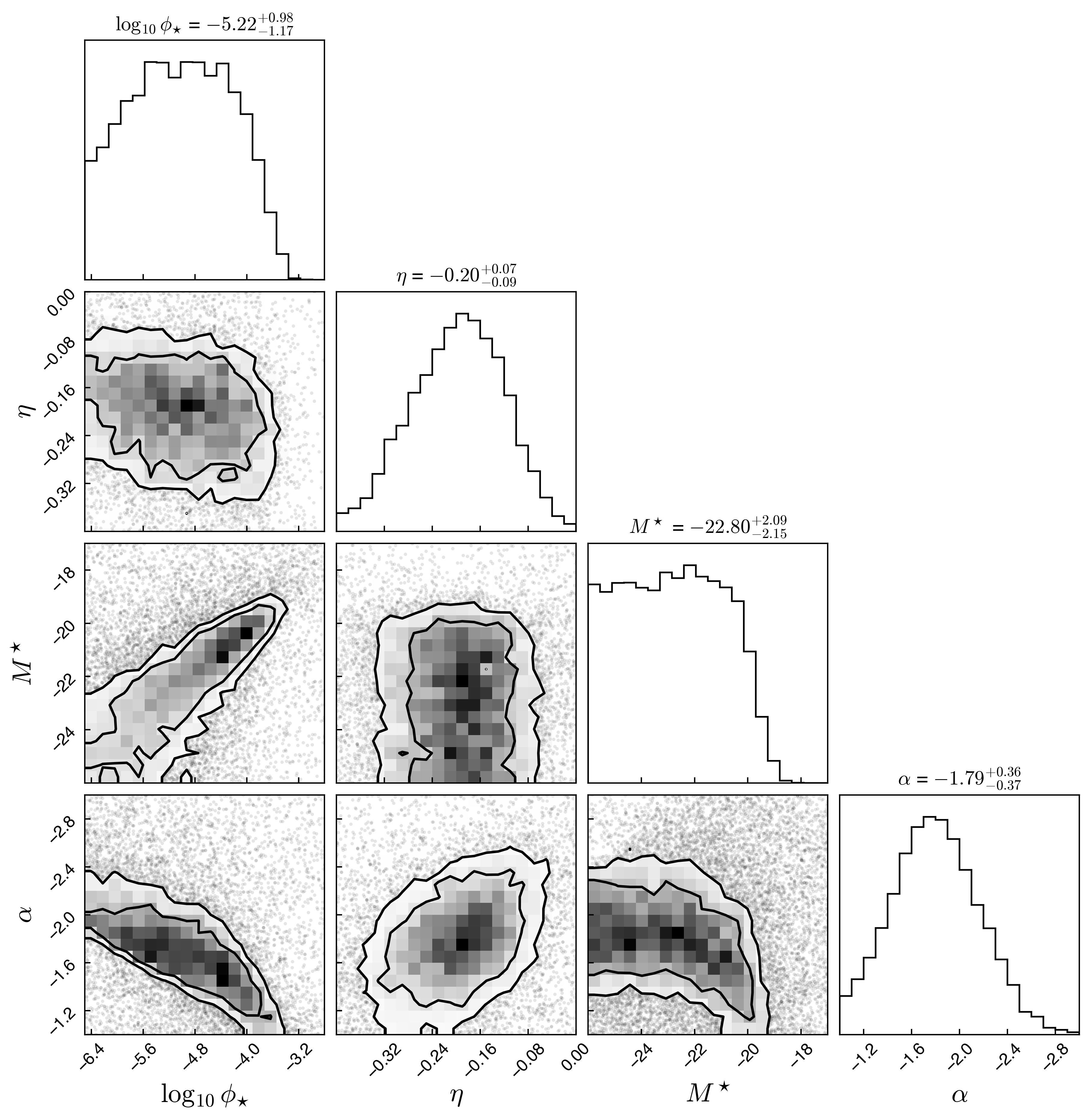}
\caption{Posterior distributions of the evolving luminosity function parameters. Shown are the posterior distributions for the luminosity function normalization $\log_{10} \phi_\star$ [Mpc$^{-3}$ mag$^{-1}$], the normalization evolution parameter $\eta$, the characteristic magnitude $M^\star$ in absolute magnitude, and the faint-end slope $\alpha$. Contours represent the 68\% and 90\% enclosed probabilities for each parameter. The marginalized posterior distributions for each parameter are shown at the top of each column, along with the 16\%, 50\% and 84\% marginal constraints (see also Table \ref{tab:lf_model}). 
The lower limits on $\phi_\star$ and $M_\star$ are
not well constrained, but we constrain at $95\%$ probability that
$\log_{10} \phi_\star<-3.84$ and $M_\star<-19.9$.
We note that $\eta<0$ with $>95\%$ probability, indicating that we infer a declining luminosity density at $z>12$.\label{fig:posteriors}}
\end{figure*}

For the luminosity function, we adopt a
redshift-dependent \citet{Schechter} function
\begin{eqnarray}
\label{eqn:schechter}
\phi_{UV}(M_{UV}, z) &= 0.4 \log 10 \phi_{\star}(z) [10^{0.4(M^\star-M_{UV})}]^{\alpha+1} \nonumber\\
&\times \exp[-10^{0.4(M^\star-M_{UV})}]
\end{eqnarray}
\noindent
where the redshift-dependent normalization $\phi_{\star}(z)$ can
be further parameterized. Our fiducial choice for the
normalization evolution is
\begin{equation}
\label{eqn:linear_phistar}
\log_{10} \phi_\star^{l}(z) = \log_{10} \phi_{\star,0} + 
\eta (z-z_0).
\end{equation}
We will refer to $z_0$ as the reference redshift, which we will
take fixed at $z_0=12$ unless otherwise noted. The
default parameters of the model then include $\vec{\theta} = [M^\star,\alpha,\phi_{\star,0},\eta]$, or the
characteristic magnitude $M^\star$, the faint-end slope $\alpha$,
the normalization at the reference redshift $\phi_{\star,0}$,
and the log-linear rate of change with redshift $\eta$.
In practice, we fit in maggies $l = -0.4M_{UV}$ and then 
convert to absolute magnitudes after inference.
We adopt log-uniform priors for $\phi_{\star,0}$ and 
$\eta$, a uniform prior in magnitude, and 
a uniform prior in $\alpha$.
The priors are reported in Table \ref{tab:lf_model}, along
with our inferred constraints on the parameters. 
We emphasize again that information from all redshifts where
the selection function has non-negligible support is included
by our inference procedure, which accounts both for regions
of redshift and magnitude space with detections and those
absent samples that could have been detected if present.
The effective redshift range where our model is informative
for the luminosity function is mostly set by the selection
completeness (Figure \ref{fig:selection_completeness}),
or roughly $z\sim11-20$. We present the full posterior distributions on the parameters in Figure \ref{fig:posteriors}. We here emphasize 
that the clear covariance between
$\phi_\star$ and $M^\star$ mostly acts to keep 
the luminosity density $\rho_{UV}\propto L^\star \phi_\star$
roughly constant at a given redshift. This feature
is reflected in our constraints on $\rho_{UV}$ 
shown in Figure \ref{fig:uvdensity}.

Since we constrain the abundance of galaxies at all
selected and detectable redshifts and magnitudes simultaneously,
evaluating the luminosity function at any one redshift
requires computing the marginal distribution of the
luminosity function equation \ref{eqn:schechter} 
over the posterior distribution of parameters
for a given redshift and range of absolute magnitudes.
At each $z$ and $M_{UV}$, equation \ref{eqn:schechter} is
evaluated for all posterior samples, and the cumulative distribution
of $\rho_{UV}$ weighted by the sample weights $w_k$ constructed.
Figure \ref{fig:lf} shows the marginal constraint on the
UV luminosity function at redshift $z=12$, with the 16-84\%
of $\phi_{UV}$ shown as a shaded region and the median $\phi_{UV}$
shown as a white line. We also show the median inferred $\phi_{UV}$
at $z=14$ as a light gray line.
Note that none of these $\phi_{UV}$ 
percentiles are guaranteed to follow equation \ref{eqn:schechter}
individually, but we do report the marginalized constraints on the
luminosity function parameters in Table \ref{tab:lf_model}.
We also show our step-wise luminosity function estimates
computed in redshift bins of $11.5<z<13.5$ and $13.5<z<15$. These step-wise luminosity function measures are reported in Table \ref{tab:lf_stepwise}.

In Figure \ref{fig:lf}, we also show $z\sim12-14$
luminosity function determinations
reported in the literature. These measurements
include
the $z\sim12$ data from \citet{harikane23_uvlf}, \citet{harikane2023b},
\citet{perez-gonzalez2023a} and
\citet{willott2023a}, the \citet{adams2023a} constraints
at $z\sim12.5$,
$z\sim13$ measurements from \citet{donnan2023b} and
\cite{mcleod2023a}, and the $z\sim14$ determinations
from \citet{finkelstein23}.
The median luminosity function constraints
inferred from our sample and our forward modeling approach
agree with the available observations to within about
$1\sigma$, excepting the $z\sim14$ constraints from
\citet{finkelstein23} that lie above our inference.
We note here that the $z\sim11$ luminosity function constraints from
\citet{donnan2023b}, 
\cite{mcleod2023a}, and \citet{finkelstein23}
lie above our 84\% inference of the $z=12$ luminosity
function, and that our selection function (Figure \ref{fig:selection_completeness})
by design removes $z\sim11$ galaxies from our sample.
We also emphasize that our results
are completely independent of the other data shown 
in Figure \ref{fig:lf}.

\subsubsection{Luminosity Density Evolution}

Given the evolving luminosity function parameters inferred
given the sample properties, the UV luminosity density
evolution $\rho_{UV}(z)$ can be computed. Figure \ref{fig:uvdensity}
presents the marginalized constraints on the
UV luminosity density evolution. Shown are 16-84\%
(jade shaded region) and median $\rho_{UV}$ (white line)
integrated to $M_{UV}<-17$, along with measured (left panel)
or extrapolated (right panel) constraints to $M_{UV}<-17$ from the literature.
Our measurements have sensitivity to objects at
redshifts $11\lesssim z\lesssim 20$, and we indicate the luminosity
density evolution inferred for the model represented
by equations \ref{eqn:schechter} and \ref{eqn:linear_phistar}.
As the Figure shows, we infer that 
the UV luminosity density declines at high-redshift at a
rate of $\eta\equiv d\log \phi_\star/dz\approx-0.2$ per unit redshift. Between $z=12$ and $z=14$,
we therefore infer that the luminosity
density declines by a factor of $10^{-0.2(14-12)}\approx2.5$.
Within our statistical uncertainties,
this inference agrees with almost all the literature
determinations including \citet{ishigaki18},
\citet{bouwens22},
\citet{mcleod2023a}, \citet{donnan23},
\citet{harikane23_uvlf}, \citet{harikane2023b}, \citet{adams2023a},
\citet{perez-gonzalez2023a}, \citet{leung23}, and \citet{willott2023a}.
The constraints at $z\sim11$ from \citet{finkelstein23} agree
with our results, but their $z\sim14$ point lies above our constraints
albeit with large uncertainties. If we extrapolate the UV luminosity
evolution inferred by our model, we find good agreement with
the literature measurements back to $z\sim8$
\citep[e.g.,][]{ishigaki18,bouwens22,perez-gonzalez2023a, willott2023a,adams2023a}. Also shown in 
Figure \ref{fig:uvdensity} is the corresponding evolution in the
cosmic star formation rate density $\rho_{SFR}$, using the approximate
conversion from
$\rho_{UV}$ of $\kappa_{UV} = 1.15\times10^{-28}$ $M_\sun$ yr$^{-1}$ erg$^{-1}$ s Hz from \citet{madau14}.
For comparison, we also show the \citet{madau14} model for the evolving
cosmic star formation
rate density.

\subsection{Caveats}
\label{sec:caveats}

Of course, with only nine objects at these extreme distances and depths, there are important caveats to consider about the LF measurement.
First, most of our objects are photometric candidates, and despite the closer spacing of the medium bands and our care in selection, we consider it possible that some might be lower redshift interlopers.  A Lyman-$\alpha$ break at $z=14$ falls at the same wavelength as a Balmer break around $z\approx4$.  
We stress that false positives would likely have a redshift distribution that falls less slowly than the true Lyman-$\alpha$ break population, so a population of false positives will typically cause the LF to appear to evolve more shallowly at extreme redshifts. However, the success of our selection method in 
providing a photometric redshift for 183348 of $z=14.32$ that was confirmed
by Carniani et al. (submitted) provides some evidence that our highest
redshift candidates could bear out.

\begin{figure*}[t]
\noindent
\begin{minipage}{\textwidth}
\begin{minipage}{0.5\textwidth}
\includegraphics[width=\linewidth]{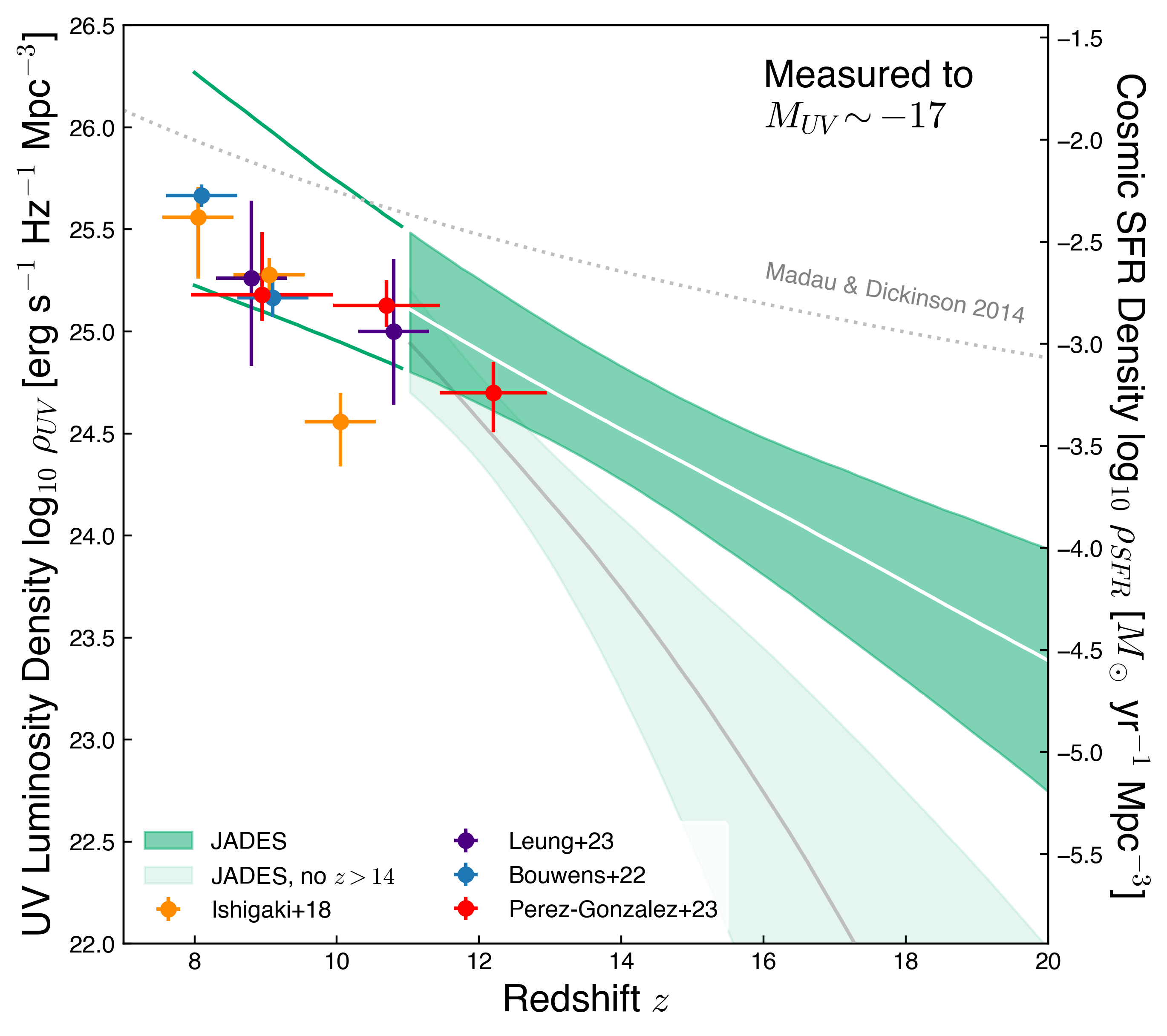}
\end{minipage}
\begin{minipage}{0.5\textwidth}
\includegraphics[width=\linewidth]{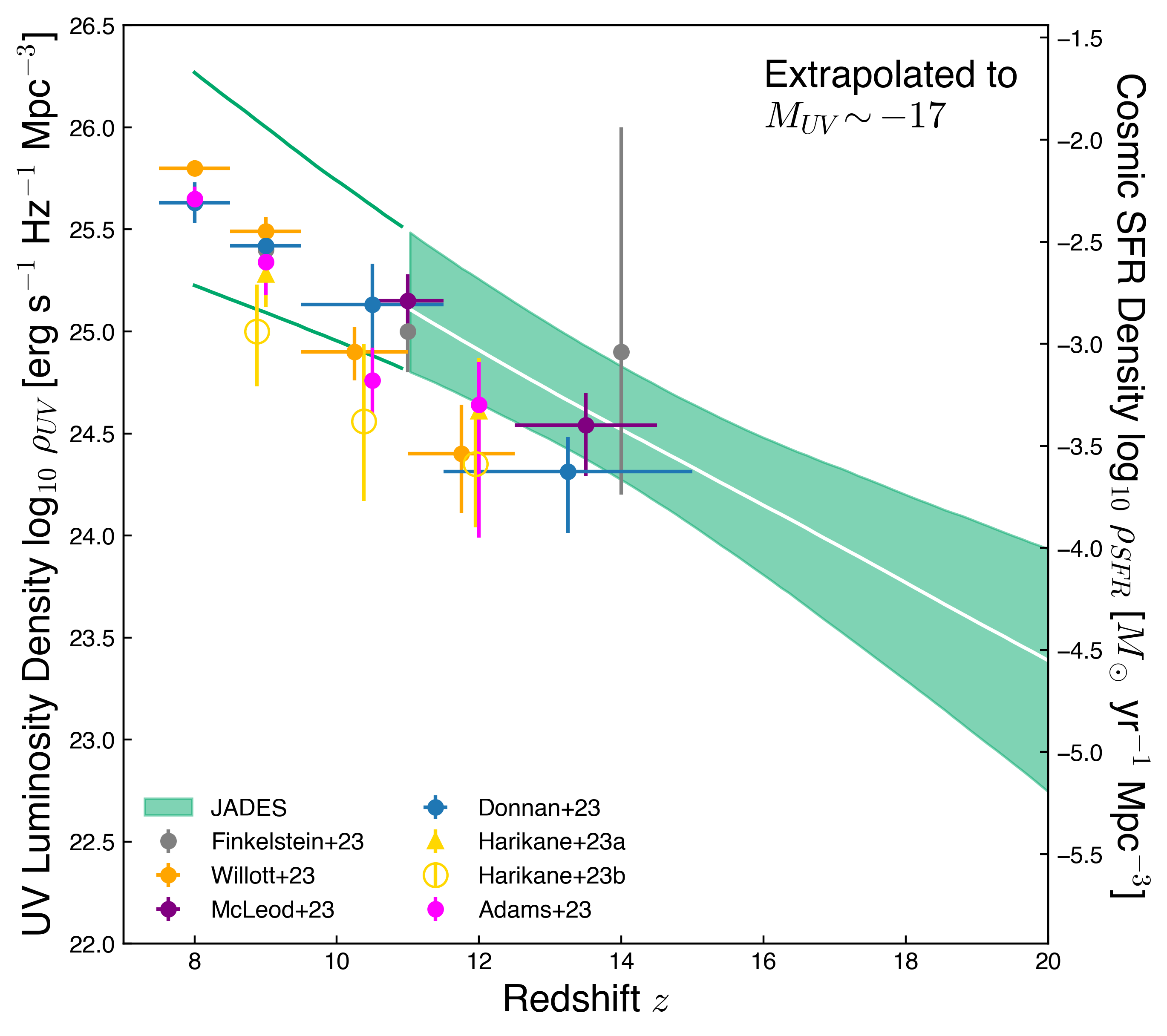}
\end{minipage}
\end{minipage}
\caption{Evolution of the UV luminosity density $\rho_{UV}(M_{UV}<-17)$ with redshift derived from the JOF sample. 
Shown are literature values for $\rho_{UV}(z)$
measured (left panel) or extrapolated (right panel) to $M_{UV}<-17$.
In both panels, 
the shaded jade region shows the 16\% and 84\% marginal constraints on the luminosity density computed from the posterior samples of the evolving luminosity function inference, as well as the median luminosity density with redshift (white line). These constraints model a linear evolution in $\log_{10} \phi_{\star}$ and include a permissive prior on the faint-end slope $\alpha$. 
Overall, our constraints agree well with prior literature results even as
our inference is completely independent.
The dark green lines extending to $z\sim8$ show the
low-redshift extrapolation of the inferred $\rho_{UV}(z)$ evolution, while the
shaded region indicates the redshift range where our detection and selection completeness is non-negligible.
 We also indicate an approximate cosmic star formation rate density (right axis; $M_\sun$ yr$^{-1}$ Mpc$^{-3}$) using the conversion $\kappa_{UV} = 1.15\times10^{-28}$ $M_\sun$ yr$^{-1}$ erg$^{-1}$ s Hz, and show the  \citet{madau14} model (left panel; dotted line).
For comparison, inn the left panel, we show the corresponding
constraint if the JOF high-redshift galaxies and candidates at $z>14$ are excluded and $\log_{10} \phi_{\star}$ is fit with an exponential evolution. In this case, we would infer the light jade region (16\%-84\% marginal constraint) with gray line (median).
\label{fig:uvdensity}}
\end{figure*}

Since the remaining candidates at $z>13.5$ have some imperfection in their cases, as discussed in \S~\ref{sec:sample}, and to illustrate the
relative impact of the highest-redshift objects on our inferences, we consider the impact on the LF estimate if we were to ignore the $z>14$ objects.
Removing these objects makes the inferred evolution of the LF notably steeper, which we show through the UV luminosity density
evolution in Figure \ref{fig:uvdensity} where the light jade region and
gray line report
the marginalized 16-84\% credibility interval and median $\rho_{UV}$, respectively. Since the
fiducial model assumes an evolution $\phi_\star^l(z)$ that has a log-linear
dependence on redshift, the $\rho_{UV}$ inferred by the model beyond the
redshift of our observed sample can in principle be artificially inflated
by the inferred trend at $z\sim12-14$. Instead, when removing the $z>14$
objects, we explore a more rapid decline given by
\begin{equation}
\label{eqn:exp_phistar}
\log \phi_\star^e(z) = \log(\phi_{\star,0}) \times \exp\left[(z-z_0)/h_\phi\right].
\end{equation}
This model enables a log-exponential drop in the galaxy abundance.
Indeed, without the $z>14$ objects the inferred $\rho_{UV}$ would drop much
more rapidly than in the fiducial model based on the Main Sample. For reference,
by $z\sim16$ the difference between the two inferences is more than
an order of magnitude.
Of course, given the small number statistics, we are also sensitive to the impact of a single false negative.  If any of the remaining Auxiliary Sample objects in \S~\ref{sec:sample} were to prove out, the LF would surely rise.

\subsection{Comparison with Halo Abundance and Large-Scale Structure}
\label{sec:halos}

The large-scale structure of the Universe is expected to present a substantial cosmic variance uncertainty given the small size of this field.  High-redshift galaxies likely live in rare halos of high mass for their epoch, leading to a large clustering bias and substantial number density fluctuations.  
To investigate this, we utilize the halo catalog from a cold dark matter simulation performed by the Abacus N-body code as part of the AbacusSummit suite \citep{garrison2021a,maksimova21}.  This simulation used $6144^3$ particles in a $300h^{-1}$~Mpc box, resulting in a particle mass of $1.5\times 10^7$~M$_\odot$, and a force softening of $21$ comoving kpc.  Halos were found using the CompaSO algorithm \citep{hadzhiyska21}.  While this simulation has high accuracy, we caution that the measurement of halo mass always depends on the halo-finding algorithm; we focus here on the relative trends across redshift and on the clustering.

\begin{figure}[t]
\noindent
\includegraphics[width=\linewidth]{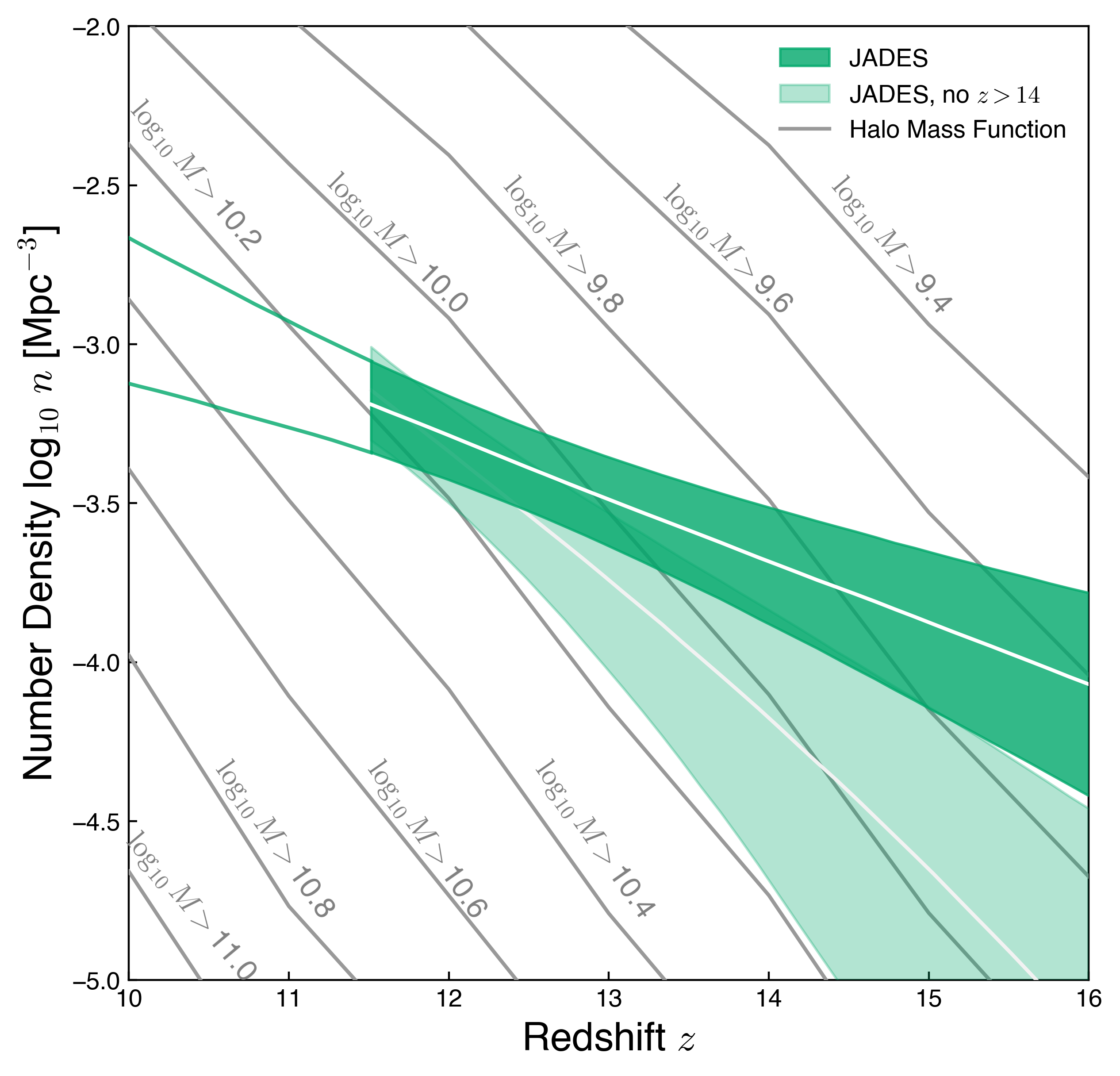}
\caption{\label{fig:mhalo}Comparison of the inferred evolution of the JOF galaxy number density $n(z)$ and the abundance of dark matter halos in cosmological simulations. Shown are the inferred number density constraints (dark jade region 16-84\%, white line 50\%) for model with a linear evolution in $\log_{10} \phi_{\star}$ with redshift $z$. The grid of gray lines show the abundance of dark matter halos with masses greater than $\log_{10} M \sim9.4-11$ computed from the AbacusSummit simulation suite \citep{maksimova21}. In the inferred JOF $n(z)$, if simply matched by abundance the halo mass of the
typical galaxy would vary by roughly a factor of $\sim10$. If instead we were to discard the $z>14$ objects and fit an exponential evolution to $\log_{10}\phi_{\star}$, the typical galaxy would mostly track a halo mass = $\log_{10} M \sim 10$ (light jade region). For reference, we indicate the extrapolation of the inferred number density constraints to lower redshifts with jade lines.}
\end{figure}

In Figure \ref{fig:mhalo}, we compare our LF measurements to
the cumulative halo mass function as a function of redshift.  One sees that if the shallow LF is correct, then matching the abundance of these galaxies to the abundance of the most massive halos would require a strongly evolving halo mass.  On the other hand, if one were to discard
the objects at $z>14$, then the result is more similar to the abundance of a constant mass, roughly of $10^{10}$~\Msun.  
Of course, the galaxies may live in less massive halos, with a scatter between luminosity and mass \citep[e.g.,][]{shen23_thesan,sun23_bursty}; indeed, some scatter is inevitable \citep{pan2023a}.  In what follows, we therefore
consider the properties of halos with virial masses of 
 $10^{9.7}$~\Msun, about 340 particles, which has comparable abundance
 to our galaxy sample at $z\sim12-14$.

We then calculate
the variation within the simulation of regions similar in size to the JOF.  We use pencil-shaped regions of $6h^{-1}$~Mpc square, roughly $3'$ at $z\sim12$, with a depth appropriate to $\Delta z=1$.  We find that at $z=12$ (11.5--12.5), there are an average of 8.3 halos above our mass threshold in a region, but with a standard deviation of 5.6.  At $z=13$, this abundance drops to $2.3\pm2.3$; at $z=14$, the abundance drops further to $0.7_{-0.7}^{+1}$.  The distribution of halo number counts becomes noticeably skewed, and by $z=14$ we find that 6\% of regions have $\ge3$ halos.
Hence, we find that unless the host halos are much less massive (and their luminosity much more variable), the large-scale structure contributes an error at least as large as the Poisson error. We caution that this uncertainty could impact the observed rate of
decline of the UV luminosity density, given our area, and motivates further studies
over larger fields. However, to combat other systematics such studies should also leverage the depth and filter coverage comparable to that afforded by the JOF, which is
challenging given the necessary exposure time.

Finally, we note that we have neglected the effect of magnification by gravitational lensing in our inference of $M_{UV}$.  While none of our candidates show obvious lens morphology, the high-luminosity tail of the high-redshift luminosity function will likely be enhanced by lensing
\citep[e.g.,][]{wyithe11,mason2015a,ferrami2023a}, which might affect interpretations of the luminosity function in the context of theories of galaxy formation.

\begin{figure}[t]
\noindent
\includegraphics[width=\linewidth]{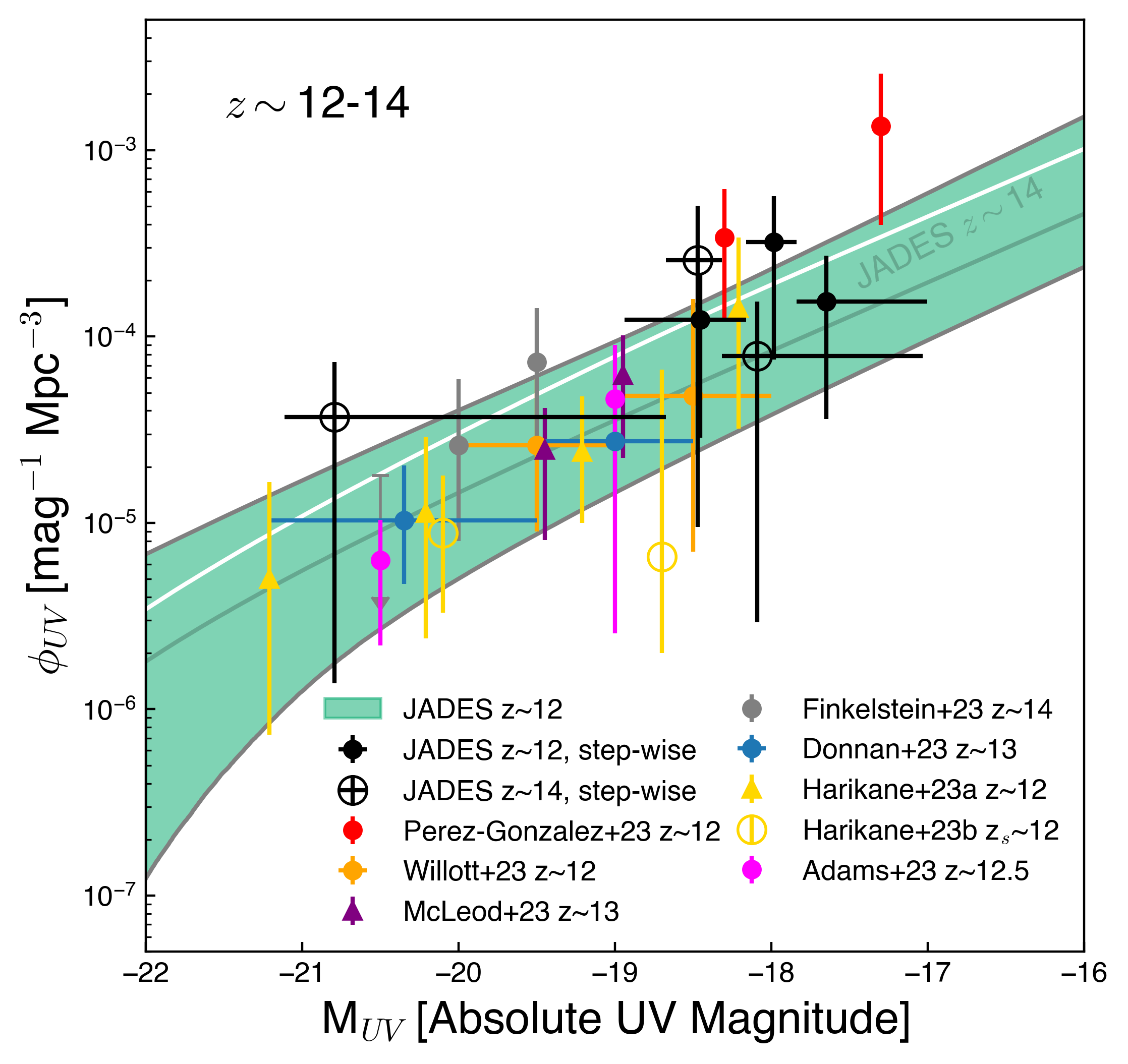}
\caption{UV luminosity at $z\sim12$ inferred from the JADES Origins Field (JOF). Using the method described in \S \ref{sec:lf_inference}, we compute the marginalized constraints on the UV luminosity function inferred from galaxies discovered in the JOF with photometric redshift distributions that overlap the redshift range $11.5<z<13.5$. We account for photometric scatter, the photometric redshift distribution of each object, the selection completeness for each object, and potential contamination from proximate redshifts. The 16\%-84\% marginal constraints on the abundance $\phi_{UV}$ as a function of absolute UV magnitude $M_{UV}$ are shown as a jade-shaded area and the median $\phi_{UV}(M_{UV})$ is shown as a white line. For comparison, we also compute step-wise luminosity function constraints as described in \S \ref{sec:lf_stepwise} at $z\sim12$ (solid black points) and at $z\sim14$ (open black circles). These step-wise estimates agree with the inferred $\phi_{UV}$, but the continuous constraints represent our results for the UV LF. We also show a variety of constraints from the literature at comparable redshifts (colored points), and note that none of these data were used to aid our inference of the UV LF.
\label{fig:lf}}
\end{figure}

\section{Physical Properties of the High-Redshift Population}
\label{sec:props}

\begin{deluxetable*}{ccccccc}
\centerwidetable
\tabletypesize{\footnotesize}
\tablecaption{Sample physical properties, assuming best-fit redshift.\label{tab:physical}}
\tablehead{
	\colhead{Name} & \colhead{NIRCam ID} & \colhead{$z_{\rm phot}$}  & \colhead{$M_{UV}$} & \colhead{$\beta$}&  \colhead{$\log_{10}~M_{\star}$ [$M_\sun$]} & \colhead{$\mathrm{SFR}$ [$M_\sun$~$\mathrm{yr}^{-1}$]}
}
\startdata
JADES+53.09731-27.84714 & 74977 & 11.53 & $-17.66\pm0.14$ & $-2.09\pm0.28$ & $7.63_{-0.53}^{+0.79}$ & $0.47_{-0.42}^{+0.47}$\\
JADES+53.02618-27.88716 & 16699 & 11.56 & $-17.94\pm0.15$ & $-2.91\pm0.35$ & $7.08_{-0.27}^{+0.20}$ & $0.29_{-0.14}^{+0.20}$\\
JADES+53.04017-27.87603 & 33309 & 12.1 & $-17.73\pm0.10$ & $-2.46\pm0.24$ & $7.62_{-0.20}^{+0.21}$ & $0.02_{-0.02}^{+0.08}$\\
JADES+53.03547-27.90037 & 160071 & 12.38 & $-18.16\pm0.11$ & $-2.43\pm0.27$ &$7.81_{-0.54}^{+0.28}$ &$0.20_{-0.19}^{+0.52}$\\
JADES+53.06475-27.89024 & 13731 & 12.93 & $-18.78\pm0.04$ & $-2.73\pm0.13$ &  $7.90_{-0.20}^{+0.19}$ & $0.18_{-0.18}^{+0.52}$\\
JADES+53.02868-27.89301 & 11457 & 13.52 & $-18.55\pm0.11$ & $-2.46\pm0.30$ &  $7.08_{-0.03}^{+0.13}$ & $1.14_{-0.13}^{+1.15}$\\
JADES+53.07557-27.87268 & 376946 & 14.38 & $-18.30\pm0.22$ & $-2.42\pm0.56$ & $7.38_{-0.21}^{+0.84}$ & $0.96_{-0.79}^{+1.23}$\\
JADES+53.08294-27.85563 & 183348 &14.39 & $-20.93\pm0.04$ & $-2.40\pm0.12$ & $8.86_{-0.03}^{+0.35}$ & $6.45_{-4.53}^{+2.18}$\\
JADES+53.10762-27.86013 & 55733 &14.63 & $-18.54\pm0.13$ & $-2.52\pm0.36$ & $7.80_{-0.05}^{+0.58}$ & $0.78_{-0.66}^{+0.82}$\\
\enddata
\tablecomments{The UV absolute magnitude $M_{UV}$ and rest-frame UV slope $\beta$ are jointly fit to common-PSF Kron photometry for each object. We report here the mean and standard deviation of other posterior distributions for each parameter. The star formation rates are averaged over the last 10 Myr of the inferred star formation histories.}
\end{deluxetable*}

Beyond the abundance and UV luminosity 
of these $z\gtrsim12$ galaxies, the physical properties
of the galaxies are of particular interest for understanding
the process of galaxy formation at the earliest epochs.
With the high-quality space-based optical-infrared photometry available in the JOF, physical properties of the
high-redshift galaxy stellar populations can be inferred.

\subsection{Rest-frame UV Magnitude and Spectral Slope}
\label{sec:muv_beta}

\begin{figure}[t]
\noindent
\includegraphics[width=\linewidth]{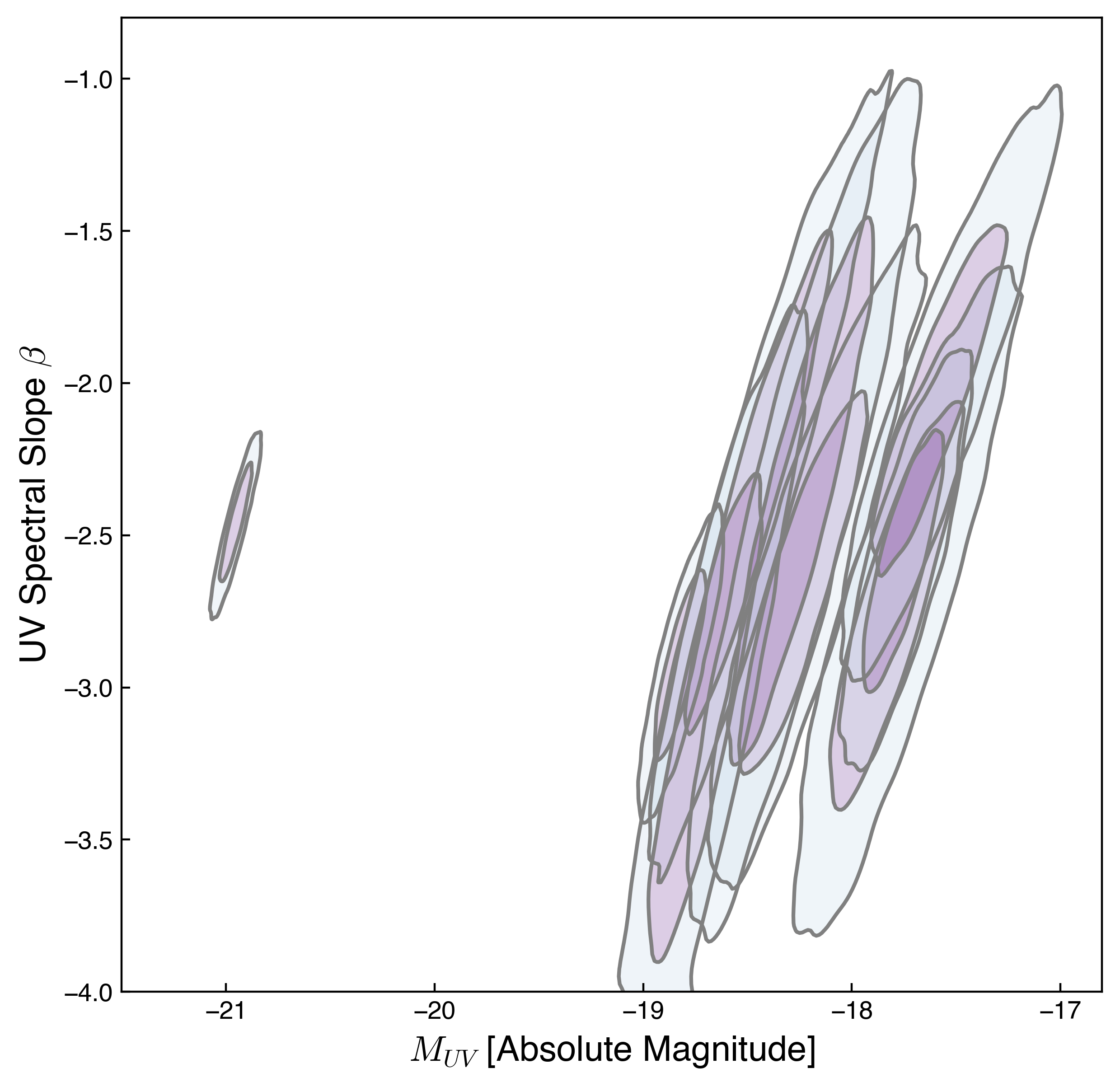}
\caption{\label{fig:muv_beta}Posterior distributions of
rest-frame UV absolute magnitude $M_{UV}$ and spectral
slope $\beta$ for candidate galaxies in our Main Sample at $z>11.5$.
Shown as kernel-density-estimated contours 
are the 68\% and 95\% credibility intervals on the
joint posterior distributions for each object. The
maximum likelihood values for the UV spectral slope are
$-2\gtrsim \beta \gtrsim-3$.} {The outlier at $M_{UV}\approx-21$
is 183348, spectroscopically confirmed at $z=14.32$ (Carniani et al., submitted).}
\end{figure}

Given the dramatic distances to these objects, the
photometry obtained in the JOF primarily probes only their
rest-frame UV spectra. Using common-PSF images and
aperture-corrected Kron photometry as a proxy for the
total fluxes, we can fit the rest-frame UV photometry
with a power law $f_\nu\propto \lambda^{2+\beta}$ and
jointly constrain $M_{UV}$ and $\beta$ given the object
redshifts. Figure \ref{fig:muv_beta} shows the posterior
distribution of $M_{UV}$ and $\beta$ for the candidate
galaxies in our Main Sample at $z>11.5$. The posterior
mean
and standard deviation for each parameter are reported
in Table \ref{tab:physical}, and for convenience
we also report $M_{UV}$ in Table \ref{tab:properties}.
The maximum likelihood values for the rest-frame
spectral slope are $-2\gtrsim\beta\gtrsim-3$. These
values are comparable to the rest-frame spectral
properties of high-redshift photometric samples
\citep[e.g.,][]{cullen2023a,topping23}, although
not quite blue enough to suggest completely dust-free
objects \citep[e.g.,][]{cullen2023b}.

\subsection{Morphology and Size}
\label{sec:sizes}

As expected, these galaxies show small angular sizes.  As described in \ref{sec:forcepho}, we fit single S\'ersic profiles to the individual exposures in the F200W and F277W filters, reporting the half-light radii in Table \ref{tab:properties}.  The posterior distributions are often non-Gaussian and asymmetric.  Unsurprisingly, most of the objects are small, with half-light radii below 50 mas, excepting the
unusual $z=14.32$ galaxy 183348.

To characterize the limiting angular resolution of our images, 
we have also fit S\'ersic profiles to the exposures (separated by epoch of observation) in the same bands for known brown dwarfs of similar flux levels in the JOF and wider GOODS-S areas \citep{hainline23_bd}. As in our past work \citep{robertson23}, we find that brown dwarfs in the JADES Deep imaging are recovered with 95\% upper limits on sizes of 20 mas in F200W, so we regard objects with 95\% lower limits above 20mas as inconsistent with a point source.  
As such, candidates 16699, 160071, and 55733 are resolved, with half-light angular sizes up to 50 mas and half-light physical sizes of 132, 118, and 142 pc, respectively. 
The galaxy 183348 spectroscopically-confirmed at $z=14.32$ by
Carniani et al. (submitted) shows a size of 76 mas, or about 240 pc.
The remaining sources are consistent with a point source, though many have non-negligible probability of having larger sizes. 
We note that objects 13731 and 376946 are both constrained to be very small.
In addition to the multiband \emph{Forcepho} fit reported in Table \ref{tab:properties}, independent
single-band \emph{Forcepho} 
fits to the 13731 infer its size be
less than 10 and 16 mas (95th percentile) in F200W and F277W respectively.
While 376946 appears unresolved in F200W and F277W, it appears more extended in some medium band filters. Regardless,
the sizes of these objects are small enough that we expect their extents do not impact their detection completeness (e.g., Figure \ref{fig:detection_completeness}).

These results are similar to those found in \citet{robertson23}, where 2 of the 4 $z>10$ galaxies were resolved. One consequence of being resolved is that the light from these galaxies cannot be purely from an accreting massive black hole \citep{tacchella23}. Other spectroscopically-confirmed galaxies at $z>12$ have had size measurements inferred from scene modeling, and show sizes of $R_{1/2}\sim100-300$pc \citep[e.g.,][]{wang2023a}. 
Collectively, these results indicate that
compact sizes are
a common property of
many high-redshift galaxies and candidates.

\subsection{Star Formation Rate Histories}
\label{sec:sfh}

To perform detailed modeling of the SEDs in terms of stellar populations, we use the \texttt{Prospector}
code \citep{johnson21}, following the methods described
in \citet{tacchella22_highz, tacchella23}. Briefly, we assume
a variable star-formation history (SFH) with a bursty continuity
prior, with 8 time bins spanning $0-5$ Myr, $5-10$ Myr and 6 
logarithmically spaced up to $z = 25$. We allow the redshift 
to vary within the \texttt{EAZY} posterior. We adopt a single
metallicity for both stars and gas, assuming a truncated 
log-normal centered on $\log(Z/Z_{\odot}) =-1.5$ with width 
of 0.5, minimum of --2.0, and maximum of 0.0. We model dust 
attenuation using a two-component model 
with a flexible attenuation curve. For the stellar population 
synthesis, we adopt the MIST isochrones \citep{choi16}
that include effects of stellar rotation but not binaries, 
and assume a \citet{chabrier03} initial mass function (IMF) 
between 0.08 and 120 $M_{\odot}$. No Ly$\alpha$ emission 
line is added to the model to account for resonant absorption 
effects, while the IGM absorption model \citep{inoue14,madau95} is 
taken into account (normalization is a free parameter). We do not try to 
constrain independently the effects of possible additional Lyman-$\alpha$ damping-wing absorption.
For consistency with Figures \ref{fig:sed_74977}-\ref{fig:sed_183348}, we use the $r=0.1"$ aperture fluxes, but we note that using $r=0.3"$ aperture fluxes provide quantitatively similar results for these compact objects.
We put an error floor of 5\% on the photometry. 
The rest of the nebular emission (emission lines and continuum) 
is self-consistently modeled \citep{byler17} with two 
parameters, the gas-phase metallicity (tied to the stellar
metallicity), and the ionization parameter (uniform 
prior in $-4 < \log(U) < -1$). 
By combining these inferred stellar population properties with 
the size measurements from \texttt{ForcePho}, we can 
additionally infer the stellar mass and star formation
rate surface densities of the candidate galaxies.

\begin{figure*}[t]
\noindent
\includegraphics[width=\linewidth]{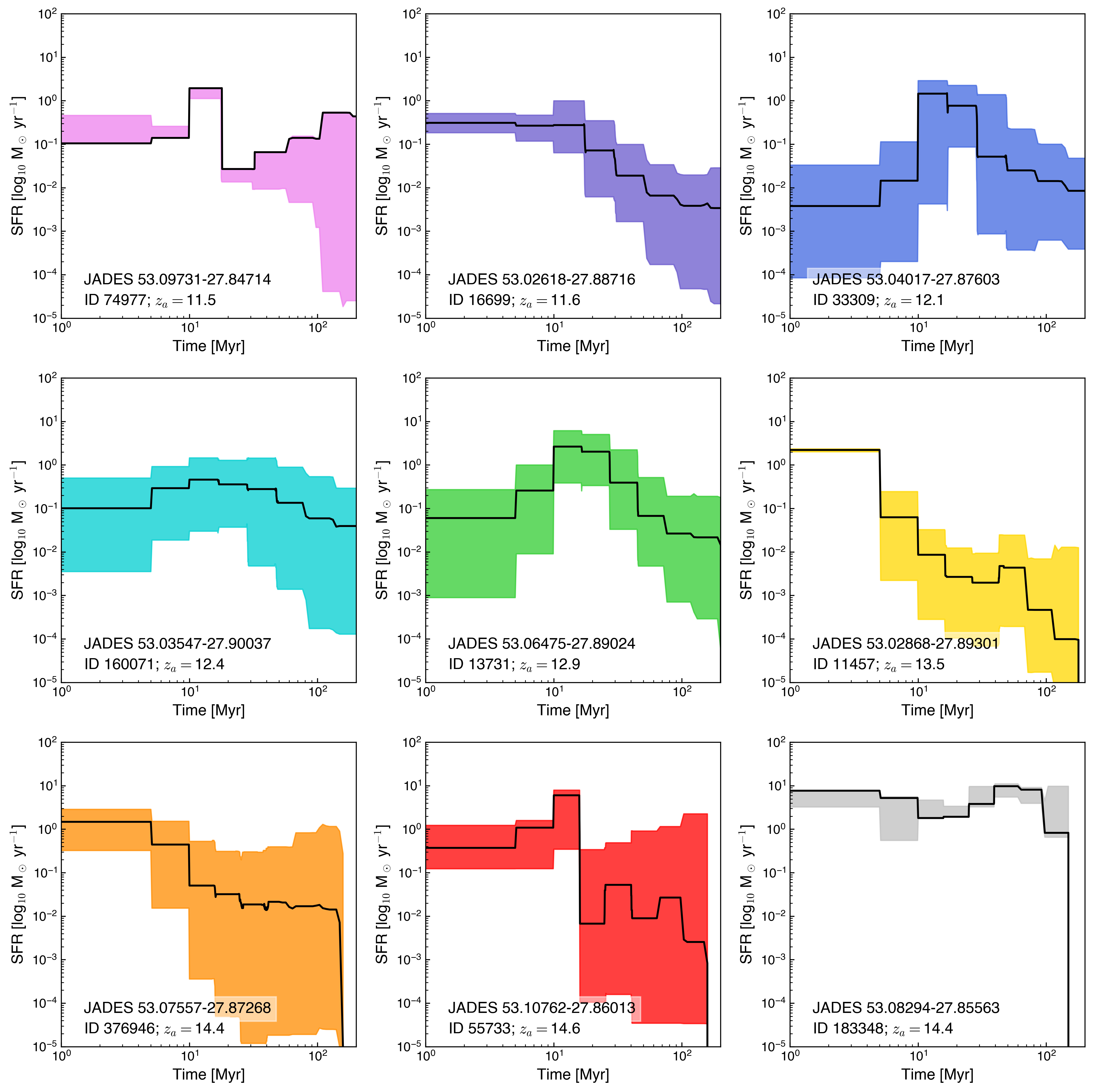}
\caption{Star formation histories (SFHs) inferred using the
\texttt{Prospector} code \citep{johnson21}, assuming a
continuity prior and following the methods described in
\citet{tacchella23}. The galaxy candidates
show star formation rates of $\SFR\approx0.1-1~\Msun~\yr^{-1}$
over the last $\sim10$ Myr, measured backward from the epoch of observation. Roughly half of the objects show
increasing star formation histories, while the others
indicate a peak or burst in their star formation rates roughly 
10 Myr before the observation epoch. This feature may indicate an
episode of ``mini-quenching'' \citep{looser23} in these objects.
Only one galaxy indicates a comparable or higher SFR $t\sim100$ Myr
before the observation epoch, such that no object indicates
evidence of substantial star formation before $z\sim15$.
Each galaxy is labeled
by both their [RA,Dec] designation, photometric redshift, and internal JADES NIRCam ID.
\label{fig:star_formation_histories}}
\end{figure*}

Figure \ref{fig:star_formation_histories}
shows the resulting star formation rate histories (SFHs) of the
eight galaxy candidates in our sample. The
average SFR over the last $10$ Myr is also
reported for each candidate galaxy in
Table \ref{tab:physical}. 
In
each case, the continuity prior on the 
star formation history was used to inform the
point-to-point star formation rate variations
in the galaxies. For each object, the
photometry listed in Tables \ref{tab:acs}-\ref{tab:nircam_lw}
were used, except for the faintest object 74977 $(f_{\nu}\sim2-3$nJy)
where the lower SNR Kron fluxes were used.
We find that the typical
star formation rate of these objects are
$\mathrm{SFR}\approx0.1-10~M_{\odot}~\yr^{-1}$ over the
last $t\sim10-30$ Myr. The galaxies formed
substantial fractions of their stars in the recent past,
and have characteristic ages of just a 
few tens of millions of years. A few of the
objects (NIRCam IDs 13731, 33309, 55733, 74977)
show features in their SFHs roughly $10-20$ Myr
before their observed epoch, with flat or even
falling SFR thereafter. We speculate
that these features may reflect
``mini-quenching'' events where star
formation shuts down briefly after exhausting
or removing fuel \citep{looser23}. For the other objects, the
SFHs appear to increase to the epoch of
observation, suggesting some upswing in the
star formation rate and luminosities of these
objects. 
In two cases (NIRCam ID 74977 and 183348)
the objects show evidence of comparable
or higher star formation rates 100 Myr before the
observed epoch. For 74977, this early star formation
would correspond to $z\sim14.2$. For 183348, the
early star formation would potentially
start at $z\sim20$.
The uncertainties
on the SFH are large, and we cannot constrain
well the star formation rate before
$z\sim15$ for most objects. Given the physical sizes of the objects
of $R_{1/2}\approx50-200$pc inferred from the
\texttt{ForcePho} analysis, the star formation rate
surface densities of these objects are $\Sigma_{SFR}\sim10-100~
\Msun~\yr^{-1}~\mathrm{kpc}^2$. Both the SFR and
SFR surface densities are comparable to those found by
\citet{robertson23} for spectroscopically-confirmed
galaxies at $z\sim12-13$, and consistent with 
being
from the same galaxy population.

The above analysis assumes no luminous contribution from an active galactic nucleus.  Of course, some of these galaxies may possibly host luminous AGN, as have been found or suspected in some other high-redshift galaxies \citep[e.g.,][]{goulding23,ubler23,kokorev23,maiolino23_bh,maiolino23_gnz11}.  AGN emission would decrease the inferred stellar emission and require a re-assessment of the star formation histories and stellar masses, and possibly the photometric redshifts.  We note that the fact that some of these galaxies are angularly resolved implies that some of the emission is stellar.

\subsection{Stellar Mass Distributions}
\label{sec:mstar}

\begin{figure}[t]
\noindent
\includegraphics[width=\linewidth]{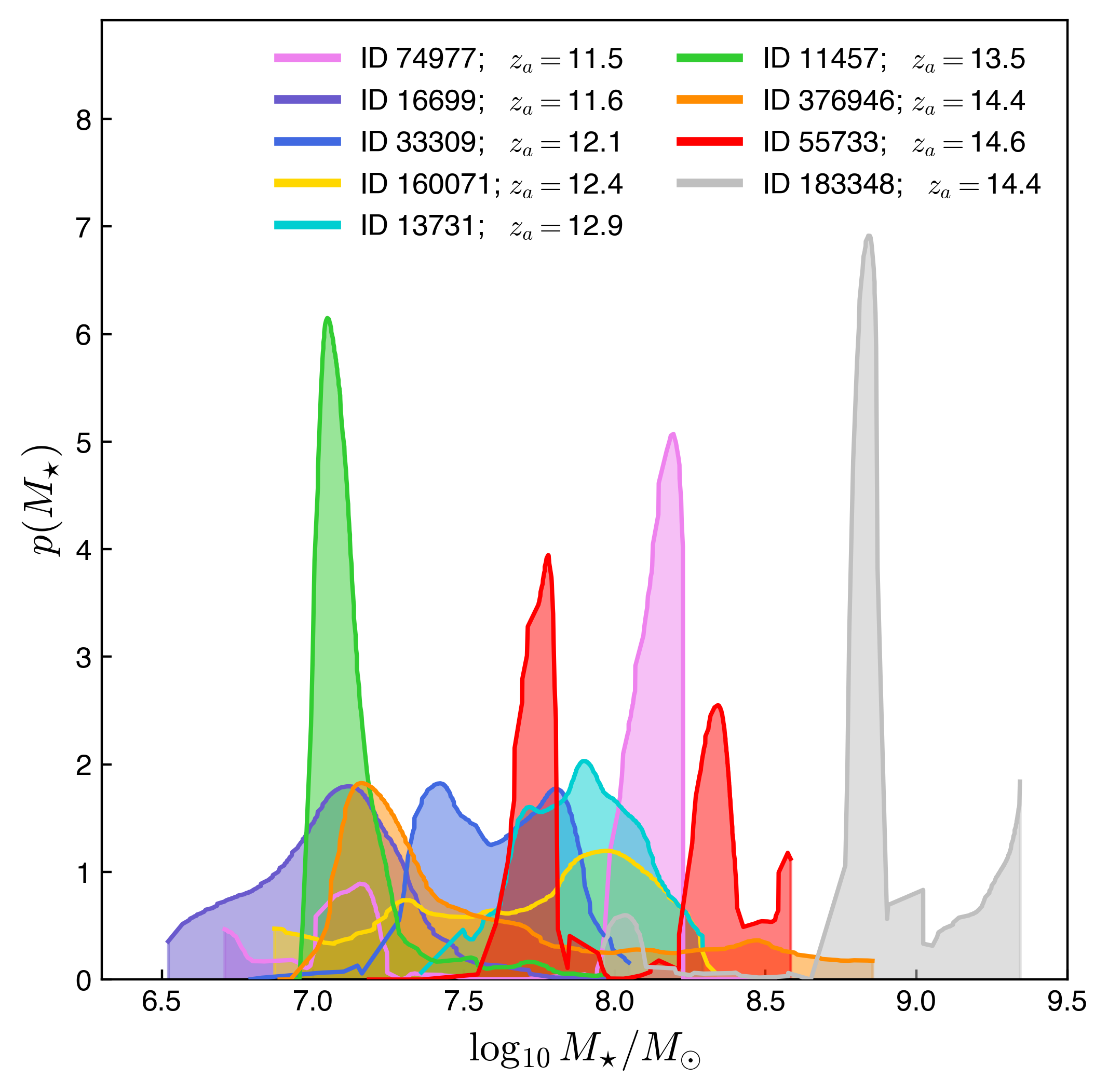}
\caption{Posterior distribution of stellar mass for candidate $z>11.5$ galaxies. Shown
are the stellar mass distributions constructed from posterior samples of
the \texttt{Prospector} code \citep{johnson21}. The objects have
inferred stellar masses of $\Mstar\sim10^7-10^8~\Msun$, comparable to that
inferred for the spectroscopically-confirmed $z\sim12-13$ analyzed by
\citet{robertson23}.
Each galaxy candidate is
labeled by its JADES NIRCam ID and photometric redshift, and color-coded the same in 
Figure \ref{fig:star_formation_histories}.
\label{fig:mass_marginal_distributions}}
\end{figure}

Figure \ref{fig:mass_marginal_distributions} presents
the marginal stellar mass distributions inferred from
\texttt{Prospector} fits to the observed photometry.
The posterior samples of the galaxy properties were
used to produce marginal distributions of the stellar
mass, following the procedure described in \citet{robertson23}.
In agreement with \citet{robertson23}, we find that the
stellar masses of these $z\sim12-15$ galaxies are
$\Mstar\sim10^7-10^9~\Msun$.
Given the
sizes of $R_{1/2}\sim50-200$pc we measure from the
surface brightness profiles, the stellar mass surface
densities of the objects are then $\Sigma_{\star}\sim10^3-10^4
\Msun~\mathrm{pc}^{-2}$.  For a self-gravitating system, the
dynamical timescale is then comparable to the star formation 
timescale inferred in \S \ref{sec:sfh}. Overall, in
agreement with our previous findings in \citet{robertson23},
these objects are consistent with rapidly star-forming, compact
galaxies with formation timescales comparable to 
a few dynamical times.
Using the simple abundance matching comparison with dark matter
halos discussed in \S \ref{sec:halos}, we note that matching to
number densities would place these objects in $M_h\sim10^{10}M_\odot$
dark matter halos, with $M_\star/M_h\sim10^{-1}-10^{-3}$, well
above the present-day stellar mass to halo mass relations
\citep[e.g.,][]{wechsler18}.

\section{Discussion}
\label{sec:discussion}

The luminosity function evolution remains the
best current indicator of the connection
between galaxies, dark matter halos, and
cosmic reionization at 
the highest redshifts \citep[for a review, see][]{robertson2022a}.
These results from the JADES Origins Field
provide some new insight into the process of
high-redshift galaxy formation.

The JOF provides the best currently available
data for probing faint galaxies at redshifts
$z>12$, given its depth and filter array.
Using an area twice the size of the Hubble
Ultra Deep Field, the JOF area reaches a 
deeper limit ($30.2-30.5$AB) and has 
fourteen \jwst{} filters including the
ultradeep JADES Program 1210. The inclusion
of deep F162M provides an essential check
on the reality of the highest-redshift
candidates.

Of our Main Sample, none of the galaxies
are brighter than $M_{UV}=-18.6$, and
many have $M_{UV}>-18$. The depth allows
us to constrain the UV luminosity function
to fainter limits at $z\sim14$ than previously
possible, while retaining tighter control
of systematics by having additional
medium band filters to probe the Lyman break
with more fidelity. Following the stellar population
modeling procedure of \citet{tacchella23}, we
find that the star formation
rate and stellar mass properties are comparable
to galaxies spectroscopically confirmed at $z\sim12-13$
\citep{robertson23,curtis-lake23,wang2023a}.
Using the \texttt{ForcePho} forward model for the
surface brightness distribution of these galaxies,
we find that they have compact sizes of $R_{1/2}\sim50-200$pc,
also in agreement with spectroscopically confirmed galaxies at these
redshifts \citep{robertson23,wang2023a}.

In agreement with
previous determinations of UV luminosity
function in extragalactic JWST fields
\citep{mcleod2023a,donnan23,adams2023a,harikane23_uvlf,harikane2023b,perez-gonzalez2023a,willott2023a,finkelstein23}, we
find that the luminosity function of 
galaxies has smoothly declined from $z\sim8$,
as first established by HST observations
\citep[e.g.,][]{mclure2013a}, to $z\sim12$.
Our results for the abundance of galaxies
at $z\sim12$ are in broad agreement with
the literature values, as shown in Figures \ref{fig:lf} and \ref{fig:uvdensity}.
We do note that our inferred UV luminosity density at $z\sim14$ is
lower than that reported by \citet{finkelstein23}, but the uncertainties are large.

However, our 
selection completeness using the JOF observations
is sensitive to galaxies out to $z\sim20$ when the Lyman-$\alpha$ break
enters F250M.
With a suitable revision to our
selection, we would be sensitive to bright galaxies
at even greater distances. Our work presents
a new method for modeling the redshift-dependent
UV luminosity function incorporating both
detections and non-detections to constrain its
evolution over the
redshift range $z\approx11-20$ where our completeness is high. 
From the lack of
galaxy candidates at $z>15$, we find that
the decline to $z>14$ continues at $d\log\phi_\star/dz\sim-0.2$
with our nominal Main Sample presented in Tables \ref{tab:properties}-\ref{tab:nircam_lw}. 
We note that uncertainties owing to cosmic variance are clearly non-negligible for the JOF, and a larger sample
of galaxies at $z>11.5$ is needed to confirm this decline.
Nonetheless,
we now know that the $M_{UV}\sim-21$
object NIRCam ID 183348 selected
by our JOF Medium band photometry to be at a photometric
redshift of $z\approx14.4$ has been spectroscopically confirmed
at $z=14.32$ by Carniani et al. (submitted).
As Figure \ref{fig:mhalo} shows, the evolving
luminosity density at $z>14$ we infer from 183348
and our photometric candidates, while declining,
still requires
a constant remapping between galaxy and halo abundance,
with increasing efficiency in low-mass halos 
at higher redshifts. This evolution is in contrast
to the possibility that $z>14$ galaxies
were not abundant, where a 
rapid drop in the
UV luminosity density would track more closely the abundance
of $M_{\mathrm{vir}}\sim10^{10}M_\sun$ halos and
the galaxy efficiency could stabilize at early
times. Given the confirmation of 183348, we
see no evidence for such a stabilization in the
efficiency of galaxy formation out to $z\sim14$
or beyond.

Lastly, since our results are consistent with
prior literature results at $z\sim12$, theoretical
models that match those observations also match
ours. For instance, the feedback-free models of \citet{dekel23}
and \citet{li23} agree with our $z\sim12$
observations for an efficiency of $\epsilon_{\max}\approx0.2$.
Models for the evolving
number counts of high-redshift galaxies
based on dust-free populations \citep[e.g.,][]{ferrara23}
also predict a star formation rate density
evolution to $z\sim15$ in agreement
with our inferences, assuming all our candidates
are really high-redshift sources \citep{ferrara2023b}.

\section{Summary and Conclusions}
\label{sec:conclusion}

Using ultra-deep \jwst\
observations of the JADES Origins Field (JOF), we 
search for the most distant galaxies in the universe.
With fourteen JWST and up to nine Hubble Space Telescope 
filters covering the JOF, we can carefully select galaxies
at $z>12$ by identifying dropouts in NIRCam F162M and 
bluer filters using SED template-based
photometric redshift fitting. Our findings include:

\begin{itemize}

\item We select nine galaxy candidates at $z\sim12-15$ and no galaxy candidates at $z\gtrsim15$. These objects include the most distant
candidates detected in more than five filters and displaying a
dropout in more than 10 filters. Our sample selection includes
a galaxy at $z=14.32$ since spectroscopically confirmed.
Simulations of our detection and photometry methods and our prior spectroscopic confirmations of high-redshift JADES sources suggest that the other candidates
without spectroscopic confirmation are robust.
Several of our candidates have been identified in previous analyses, including \citet{hainline23} and \citet{williams23}.

\item These objects show apparent total magnitudes of
$m_{AB}\sim29.5-30.5$ in the rest-frame UV
and blue rest-UV spectral slopes $-2\gtrsim\beta\gtrsim-3$.

\item Performing detailed structural modeling 
with ForcePho and stellar population inference 
using Prospector, we find that the
galaxies have star-formation rates of 
$SFR\approx0.1-10~\Msun~\yr^{-1}$, stellar 
masses of $\Mstar\sim10^{7}-10^{9}\Msun$, 
sizes of {$R\sim50-200~\pc$}, and 
stellar ages of $\tstar\approx30-50~\Myr$. 
The properties of our low-mass candidates
are comparable to the 
properties of $z\sim12-13$ galaxies with
confirmed redshifts, as first identified by 
the JADES collaboration.

\item We develop a new forward modeling method to infer constraints on the evolving UV luminosity function without binning in redshift or luminosity while marginalizing over the
photometric redshift posterior distribution
of candidates in our sample. This method
allows for an accounting of potential contamination by adjacent redshifts and
includes the impact of non-detections on 
the inferred galaxy luminosity function evolution.

\item With the population of $z>12$ galaxy 
candidates newly discovered in JOF, we provide 
an inference on the $z\sim15$ luminosity 
function and a refined measure of the
luminosity function at $z\sim12$ in agreement
with literature values. 
At $z\sim15$, 
we infer a continued decline from $z\sim12$.
Over the redshift range $z\sim12-14$, where we have detected galaxies, we infer a factor of $2.5$
decline in the luminosity function normalization $\phi_\star$ and a corresponding decline in the luminosity density $\rho_{UV}$. We note that cosmic variance uncertainties for the high-redshift JOF sample are not negligible, and this decline should be confirmed with a larger sample over a wider area.

\end{itemize}

This demonstrates the immediate impact new \jwst{}
observations can have on our knowledge of the distant universe.
With high-redshift galaxy populations now established fewer than
300 million years after the Big Bang, we have extended our reach
into the cosmic past by 40\% during the first eighteen months of \jwst{} operations.

\vspace*{\fill}
\vspace*{\fill}

\begin{acknowledgments}
The JADES Collaboration thanks the Instrument Development
Teams and the instrument teams at the European
Space Agency and the Space Telescope Science
Institute for the support that made this program possible.
The authors acknowledge use of the lux supercomputer at UC Santa Cruz, funded by NSF MRI grant AST 1828315.
\end{acknowledgments}\vspace*{-15pt}
\begin{acknowledgments}
BER, BDJ, DJE, PAC, EE, MR, FS, \& CNAW acknowledge support from the 
JWST/NIRCam contract to the University of Arizona, NAS5-02015. 
BER acknowledges support from JWST Program 3215. 
DJE is supported as a Simons Investigator.
SA acknowledges support from Grant PID2021-127718NB-I00 funded by the Spanish Ministry of Science and Innovation/State Agency of Research (MICIN/AEI/ 10.13039/501100011033). 
WB, FDE, RM, \& JW acknowledge support by the Science and Technology Facilities Council (STFC), ERC Advanced Grant 695671 ``QUENCH".
AJB, JC, \& GCJ acknowledge funding from the ``FirstGalaxies" Advanced Grant from the European Research Council (ERC) under the European Union’s Horizon 2020 research and innovation programme (Grant agreement No. 789056).
\end{acknowledgments}
\begin{acknowledgments}
SC acknowledges support by European Union’s HE ERC Starting Grant No. 101040227 - WINGS.
ECL acknowledges support of an STFC Webb Fellowship (ST/W001438/1).
FDE, RM, \& JW acknowledge support by UKRI Frontier Research grant RISEandFALL.
Funding for this research was provided by the Johns Hopkins University, Institute for Data Intensive Engineering and Science (IDIES).
RM also acknowledges funding from a research professorship from the Royal Society.
The Cosmic Dawn Center (DAWN) is funded by the Danish National Research Foundation under grant DNRF140.
PGP-G acknowledges support from grant PID2022-139567NB-I00 funded by Spanish Ministerio de Ciencia e Innovaci\'on MCIN/AEI/10.13039/501100011033, FEDER, UE.
DP acknowledges support by the Huo Family Foundation through a P.C. Ho PhD Studentship.
RS acknowledges support from a STFC Ernest Rutherford Fellowship (ST/S004831/1).
H{\"U} gratefully acknowledges support by the Isaac Newton Trust and by the Kavli Foundation through a Newton-Kavli Junior Fellowship.
LW acknowledges support from the National Science Foundation Graduate Research Fellowship under Grant No. DGE-2137419.
The research of CCW is supported by NOIRLab, which is managed by the Association of Universities for Research in Astronomy (AURA) under a cooperative agreement with the National Science Foundation. 
\end{acknowledgments}

\facilities{HST(ACS,WFC3), JWST(NIRCam)}

\software{astropy \citep{astropycollaboration18, astropy-collaboration22}, EAZY \citep{brammer08}, Source Extractor \citep{bertin96}, photutils \citep{bradley23}, nautilus \citep{lange2023a}
          }


\end{document}